\documentclass[a4paper,11pt]{article}
\pdfoutput=1 
\usepackage{jheppub}  
\usepackage{graphicx,color}
\usepackage{amsmath}
\usepackage{autobreak}
\allowdisplaybreaks
\usepackage{floatrow}
\DeclareFloatFont{tiny}{\tiny}
\newcommand{\ben}{\begin{enumerate}}
\newcommand{\een}{\end{enumerate}}
\newcommand{\beq}{\begin{equation}}
\newcommand{\eeq}{\end{equation}}
\newcommand{\bal}{\begin{align}}
\newcommand{\eal}{\end{align}}

\newcommand{\bea}{\begin{eqnarray}}
\newcommand{\eea}{\end{eqnarray}}
\newcommand{\nn}{\nonumber}
\def\gZ#1{g_{#1}}
\def\Ap#1{\tilde{A}_{#1}} 
\def\Dp#1{\tilde{D}_{#1}}
\def\btp#1{\tilde{\beta}_{#1}}
\def\btz{\beta_{0}}

\newcommand{\WpbWpp}{\frac{\lambda}{\lambda^{'}}}
\def\wp{{\lambda}}
\def\fo{{(\ln \lambda^{'})}}
\def\onemWpi{{\bigg(\frac{1}{\lambda^{'}}\bigg)}}
\def\varo{{\frac{\lambda (2-\lambda)}{(\lambda^{'})^2}}}
\def\varz{{\frac{\lambda^2}{(\lambda^{'})^2}}}

\newcommand*\oline[1]{%
   \vbox{%
     \hrule height 0.5pt
     \kern0.25ex
     \hbox{%
       \kern-0.2em
       \ifmmode#1\else\ensuremath{#1}\fi
       \kern-0.05em
     }
   }
}

\newcommand{\Nb}{\oline{N}}
\newcommand{\gt}{\tilde{G}}

\def\gdSVN#1{{\Delta_N^{\rm sv,(#1)}}}
\newcommand{\LNb}{{\bf\oline{L}}}

\newcommand{\F}{\hat{\cal F}}
\newcommand{\ph}{{\cal \varPhi}}
\newcommand{\Z}{{\cal Z}}
\newcommand{\K}{{\cal K}}
\newcommand{\G}{{\cal G}}
\newcommand{\eps}{\epsilon}
\newcommand{\ash}{\hat{a}_s}
\newcommand{\mur}{\mu_r}
\newcommand{\muf}{\mu_f}

\newcommand{\KB}{\overline{{\cal K}}}
\newcommand{\GB}{\overline{{\cal G}}}
\newcommand{\gb}{\overline g}

\def\wp{\lambda}
\def\gZ#1{g_{0#1}}
\def\gN#1{g_{#1}}
\def\Ap#1{\tilde{A}_{#1}} 
\def\Dp#1{\tilde{D}_{#1}}
\def\btp#1{\tilde{\beta}_{#1}}
\def\btz{\beta_{0}}

\newcommand{\Apo}{\tilde{A}_1}
\newcommand{\Dpo}{\tilde{D}_1}

\newcommand{\GE}{\gamma_E}

\newcommand{\dFAoNA }{\frac{d_F^{abcd}d_A^{abcd}}{N_A}}
\newcommand{\dFFoNA }{\frac{d_F^{abcd}d_F^{abcd}}{N_A}}

\def\A#1{{A_{#1}}}
\def\B#1{{{ B}_{#1}}}
\def\f#1{{f_{#1}}}

\def\GGO#1{G_{1#1}}
\def\GGTW#1{G_{2#1}}
\def\GGTH#1{G_{3#1}}
\def\GtGO#1{\tilde{G}_{1#1}}
\def\GtGTW#1{\tilde{G}_{2#1}}
\def\GtGTH#1{\tilde{G}_{3#1}}

\newcommand{\as}{a_s}

\def\gzNt#1{{\bar{g}_{0#1}}}
\def\gNS#1{{\Delta^{\rm Soft}_{g_{#1}}}}
\def\gSOFT#1{{\Delta^{\rm Soft}_{g_{0#1}}}}

\def\Gone#1{G_{1#1}}
\def\Gtwo#1{G_{2#1}}
\def\Gthree#1{G_{3#1}}

\def\Gtone#1{\tilde{G}_{1#1}}
\def\Gttwo#1{\tilde{G}_{2#1}}
\def\Gtthree#1{\tilde{G}_{3#1}}

\def\bt#1{\beta_{#1}}

\def\A#1{A_{#1}}
\def\B#1{{ B}_{#1}}
\def\f#1{f_{#1}}

\def\z#1{\zeta_{#1}}

\newcommand{\Ca}{C_A}
\newcommand{\Cf}{C_F}
\newcommand{\nf}{n_F}
\newcommand{\CA}{C_A}
\newcommand{\CF}{C_F}
\newcommand{\NF}{n_f}
\newcommand{\nfv}{n_{fv}}
\newcommand{\Nfour}{N_{4}}

\def\Dm1{{{\delta(1-z)}}}

\def\nf{{n^{}_{\! f}}}

\def\g0#1DY{{g_{0#1}^{DY}}}

\newcommand{\Lqr}{L_{qr}}
\newcommand{\Lfr}{L_{fr}}

\def\LogmW1{{{\ln (1-\omega)}}}

\def\btz{{\beta_0}}

\newcommand{\eq}[1]{eq.\ (\ref{#1})}
\newcommand{\fig}[1]{fig.\ (\ref{#1})}

\newcommand{\bg}{\bar{g}}

\title{Resummed Drell-Yan cross-section at N$^3$LL}
\author[a]{ Ajjath A H,}
\author[b,c]{Goutam Das,}
\author[d]{M. C. Kumar,}
\author[a]{Pooja Mukherjee,}
\author[a]{V. Ravindran,}
\author[d]{Kajal Samanta }

\affiliation[a]{The Institute of Mathematical Sciences, HBNI, IV Cross Road, Taramani, Chennai 600113, India}
\affiliation[b]{Theory Group, Deutsches Elektronen-Synchrotron (DESY), Notkestrasse 85, D-22607 Hamburg, Germany}
\affiliation[c]{Theoretische Physik 1, Naturwissenschaftlich-Technische Fakult{\"a}t, Universit{\"a}t Siegen,
Walter-Flex-Strasse 3, 57068 Siegen, Germany}
\affiliation[d]{Department of Physics, Indian Institute of Technology Guwahati,  Guwahati-781039, India}
\emailAdd{ajjathah@imsc.res.in}
\emailAdd{goutam.das@uni-siegen.de}
\emailAdd{mckumar@iitg.ac.in}
\emailAdd{poojamukherjee@imsc.res.in}
\emailAdd{ravindra@imsc.res.in}
\emailAdd{kajal.samanta@iitg.ac.in }
\abstract{We present the resummed predictions for inclusive cross-section for  Drell-Yan (DY) 
production as well as onshell $Z,W^\pm$ productions at next-to-next-to-next-to leading logarithmic (N$^{3}$LL) accuracy. Using the standard techniques, we derive the $N$-dependent coefficients in the Mellin-$N$ space as well as the $N$-independent constants and match the resummed result through the minimal prescription matching procedure with that of existing next-to next-to leading order (NNLO). In addition to the standard $\ln N$ exponentiation, we study
the numerical impacts of exponentiating $N$-independent part of the soft function 
and the complete $\gb_0$ that appears in the resummed predictions in $N$ space. 
All the analytical pieces needed in these different approaches are extracted from the soft-virtual part
of the inclusive cross section known to next-to-next-to-next-to leading order (N$^3$LO).
We perform a detailed analysis on the scale and parton distribution function (PDF) variations and present predictions for the 13 TeV LHC for the neutral Drell-Yan process as well as onshell charged and neutral vector boson productions.
}

\begin{document} 
\preprint{ IMSc/2019/12/16, DESY 19-191, SI-HEP-2019-23 }
\maketitle
\section{Introduction} \label{intro}

Standard Model(SM) has been very successful so far in describing the physics of 
elementary particles. Precision study has played an important role in establishing the 
SM through the latest discovery of Higgs boson at the Large Hadron Collider (LHC). 
The properties of the Higgs boson is being studied with higher accuracy. 
Recent observations at the LHC demonstrate that the systematic precision study is essential 
to look for any deviation from the SM in search of new physics beyond the SM (BSM). 
While there is no promising sign of new physics signature so far at the LHC, 
it is extremely important to know the SM predictions for the standard processes 
like Higgs and DY or $Z,W^\pm$ productions to utmost accuracy. 
Not only this could help in BSM searches but also help to understand the 
perturbative structure of the underlying gauge theory.

Drell-Yan production has been a standard candle at the hadron colliders and is extremely important 
for luminosity monitoring. This is one of the hadronic processes which is well understood theoretically.
For example, next to next to leading order (NNLO) quantum chromodynamics (QCD) 
correction \cite{Hamberg:1990np,Matsuura:1990ba,Harlander:2002wh} to this
process was computed three decades ago.
DY is also an important process experimentally for several BSM searches.  
Experimentally, one has a very clean environment for precise measurements in terms of 
the kinematics of the final state lepton pairs. 
Higher order perturbative QCD corrections to DY provides ample opportunity 
to explore the structure of the perturbation series. Thus DY serves as an important process 
in collider experiments.  At the LHC, the strong interaction dynamics dominates over the others and hence 
There have been attempts to go beyond NNLO accuracy in order to improve the precision from the theoretical 
side.  

The calculation of complete N$^3$LO cross-section is extremely 
difficult due to increasing number of subprocesses involved, however there have been significant 
progress to obtain third order contribution to this process in QCD. Very recently the first result at complete N$^3$LO from only virtual photon mediator has been calculated in \cite{Duhr:2020seh}.  From the theory side, 
DY is seen to be extremely stable with respect to factorization and renormalisation scales 
already at NNLO. The scale uncertainty has been seen to be reduced to $2\%$ for a canonical 
variation of factorization and renormalisation scales compared to NLO where uncertainty is about $9.2\%$, whereas the K-factor seem to improve marginally from $1.25$ at NLO to $1.28$ at NNLO. However keeping in mind the importance of this process, it is worth studying the results from next orders and devise methods 
to incorporate more and more higher order corrections.
Since a complete calculation beyond NNLO level is difficult,  the soft-virtual (SV)            
contributions is often computed as first step. 
In addition, the later constitutes a significant part of the cross-section 
in the region where the partonic scaling variable $z\to1$, called 
the threshold region.   
The SV cross-sections are known for many SM processes {\it e.g.} 
Higgs production \cite{Anastasiou:2014vaa,Moch:2005ky,Laenen:2005uz,Ravindran:2005vv,
Ravindran:2006cg,Idilbi:2005ni,Li:2014afw}, associated production \cite{Kumar:2014uwa}, 
bottom quark annihilation  \cite{Ahmed:2014cha}, pseudo-scalar Higgs production  \cite{Ahmed:2015qda}.For DY production, using the three loop quark form factor \cite{Gehrmann:2010ue}, exploring the
universal structure of the soft part \cite{Ravindran:2006bu} of SV cross 
section to Higgs production \cite{Anastasiou:2014vaa},
the dominant soft-virtual (SV) corrections for DY at third order was obtained
\cite{Ahmed:2014cla} and later it was confirmed in \cite{Li:2014bfa}.

The SV contributions dominate at every order in the perturbation theory
through large logarithms spoiling the reliability of
the fixed order predictions.   The resolution to this is to resum these large logarithms to all orders.
Resummation of these large logarithms is thus very important to correctly describe the cross 
section in the threshold region.
In \cite{Sterman:1986aj,Catani:1989ne,Catani:1990rp}, a systematic
approach was proposed to resum these logarithms to all orders.
The large logarithms arise in the hard partonic cross section when the total available center-of-mass energy ($\hat{s}$) becomes close to the invariant mass ($Q$) of the  final state, in other words the partonic scaling
variable $z=q^2/\hat{s} \to 1$.
This results from the soft gluon emissions, as a consequence of which the cross-section is enhanced 
by the large logarithms that appear as distributions namely Dirac delta $\delta(1-z)$ and $+$ distributions: 
\begin{eqnarray}
\label{eq:calD}
{\cal D}_j(z) = \left( { \log^j(1-z) \over 1-z }\right)_+
\end{eqnarray}
In Mellin space these singular
terms are transformed into powers of logarithms of
the Mellin variable $N$. In Mellin $N$ space, these contributions can be systematically resummed 
to all orders and they display a nontrivial pattern of exponentiation. In the threshold region, the fixed order predictions often fail to describe the cross-section well and hence the resummation of these large logarithms becomes very important to correctly describe the region. Moreover, it has been very well established that the resummed contributions give a sizeable contributions to the cross-section. In fact many SM fixed order calculations have been improved with the corresponding resummed results, for example, the inclusive scalar Higgs production in gluon fusion \cite{Catani:2003zt,Moch:2005ky,Catani:2014uta,Bonvini:2014joa,Bonvini:2016frm} (see  also \cite{Ahmed:2015sna} for renormalisation group improved prediction) as well as in bottom quark annihilation \cite{H:2019dcl}, deep inelastic scattering \cite{Moch:2005ba,Das:2019btv}, DY production \cite{Moch:2005ky,Idilbi:2006dg,Catani:2014uta}, pseudoscalar Higgs production \cite{Ahmed:2016otz,Schmidt:2015cea,deFlorian:2007sr}, spin-2 production \cite{Das:2019bxi,das:rs} etc.
Threshold resummation not only improves the inclusive fixed order results but also differential observables like rapidity \cite{Catani:1989ne,Westmark:2017uig,Banerjee:2017cfc,Banerjee:2018vvb,Lustermans:2019cau} and in the context of LHC precision measurements, it is important to include these corrections and they are 
shown to improve the fixed order results.  

In the resummed predictions for the cross-sections, there is an intrinsic ambiguity on what is 
exponentiated and what is not. In the standard approach, one exponentiates only large-$N$ pieces coming 
from the soft function which are enhanced in the threshold region. However one can also define large logarithms in terms of a new variable $\overline{N} = N \exp(\gamma_E)$, $\gamma_E$ being the {\it Euler-Mascheroni} constant. Theoretically this is allowed, since $\gamma_E$ arises as a mathematical artifact due to dimensional regularization in $d$-space time dimensions. Moreover, this does not spoil the fact that the large-$N$ pieces are exponentiated in the threshold region. In this terminology, one exponentiates $\overline{N}$ instead of $N$. Numerically, however this makes a difference already at the leading logarithmic accuracy. It has been already seen in \cite{Das:2019btv} that the perturbative convergence is improved if one exponentiates the large-$\overline{N}$ terms. Apart from the standard threshold exponentiation, one can in fact exponentiate the complete soft function \textit{i.e.,} all the large-$\overline{N}$ terms as well as the $\delta(1-z)$ terms  arising from the soft function. We call this `Soft exponentiation' which renders some part of the $N$-independent constant ($\gb_0$) for the standard $\overline{N}$-threshold. 
%
In addition to these procedures, one can also exponentiate the complete form factor along with the soft function. This was studied in the context of the SM Higgs production \cite{Bonvini:2014joa,Bonvini:2016frm} and was shown to improve the scale uncertainty better than the standard threshold approach. 
The form factor is process dependent and therefore is non-universal unlike the soft function. However, the form factor as well as the soft function both satisfy the similar Sudakov K+G type equation \cite{Sudakov:1954sw, Mueller:1979ih, Collins:1980ih, Sen:1981sd, Ravindran:2005vv,Ravindran:2006cg}.  Hence the solution to K+G equation for the form factor is 
an exponential $N$ independent constant justifying the exponentiation. 
The numerical impact of this has already been studied in the past for the DY production in \cite{Eynck:2003fn} where the authors show that both in DIS scheme and in $\overline{\rm MS}$ scheme the complete form factor exponentiates to the orders currently known. 


The goal of the present article is to study the effect of threshold logarithms at N$^3$LL accuracy and match it to the known NNLO results. We perform this study for the neutral DY production as well as for onshell $Z$ and $W^\pm$ productions. The paper is organized as follows. In  Sec.\ \ref{sec:theory}, we collect the useful formulae required for the invariant mass distribution for DY and total cross-section for  $Z, W^\pm$ productions at the LHC.  Next we discuss the theoretical set up in the context of resummation. Here we describe in detail the factorization of soft-virtual coefficient in Sec.\ \ref{sec:softvir}.  Next we set up in Sec.\ \ref{sec:resummation} different resummation prescriptions as well as derive some useful formulas needed.  Sec.\ \ref{sec:numerical} we study in detail the effect of threshold logarithms for different prescriptions and present our results along with the estimation of uncertainties. We finally conclude in Sec.\ \ref{sec:conclusion}.
\section{Theoretical Framework} 

\subsection{Drell-Yan and $Z,W^\pm$ production}\label{sec:theory}
The hadronic cross-section for DY or onshell $Z,W^\pm$ production at the LHC can be written as
%

\begin{align}\label{eq:hadronic-xsect}
\sigma
&=\sigma^{(0)}\sum_{ab={q,\overline q,g}} \int_0^1 dx_1
\int_0^1 dx_2~ f_a(x_1,\mu_f^2) ~
f_b(x_2,\mu_f^2)
\int_0^1 dz \,\,
\Delta_{ab}(z,q^2,\mu_f^2)
\delta(\tau-z x_1 x_2)\,,
\end{align}
where $\sigma =  {d \sigma \over d Q}\left(\tau,q^2\right)$ for DY production, 
with $Q$ being the invariant mass of the di-lepton pair.  
Here $f_a(x_1,\mu_f^2)$ and $f_b(x_2,\mu_f^2)$ 
are the non-perturbative parton distribution functions (PDFs) of the partons $a,b$ 
carrying momentum fractions $x_1,x_2$ of the incoming protons at the factorization scale
$\mu_f$. These PDFs are appropriately 
convoluted with perturbatively calculable partonic coefficients $\Delta_{ab}(z,q^2,\mu_f^2)$.  
For the on-shell $Z,W^\pm$ production, $\sigma = \sigma_V, V=Z/W^\pm$ and $Q=M_V$, 
the mass of the vector boson. 
The partonic coefficients are obtain from the partonic cross section using perturbation theory. 
For the DY production $V$ we include contributions from $\gamma$ and $Z$ as well as their interference. 

The partonic cross-section can be decomposed as 
\begin{align}\label{eq:sv-decompose}
\Delta_{ab}(z,q^2,\mu_f^2) &=  
\Delta^{{\rm(sv)}}_{ab}(z,q^2,\mu_f^2)
+ \Delta^{{\rm(reg)}}_{ab}(z,q^2,\mu_f^2) \,.
\end{align}
The first term $\Delta^{{\rm(sv)}}$ is called the SV partonic coefficient and it contains 
distributions such as $\delta(1-z)$ and ${\cal D}_+$, whereas the second term $\Delta^{{\rm(reg)}}$ 
contains those terms that are regular in the scaling variable $z$.
The prefactors for DY and $Z,W^\pm$ production  are given as below:
\begin{align}
\sigma_{DY}^{(0)} &= \frac{2\pi }{n_c } \bigg[ \frac{Q}{S} {\cal F}^{(0)} \bigg]\,, \nn\\
\sigma_Z^{(0)} &=  \frac{2\pi }{n_c } \bigg[ \frac{\pi\alpha}{8s_w^2 c_w^2 S}  \bigg]\,, \nn\\
\sigma_{W^\pm}^{(0)} &= \frac{2\pi }{n_c } \bigg[ \frac{\pi\alpha}{4 s_w^2 S}  \bigg]\,, 
\end{align}
where $S$ is the hadronic centre-of-mass energy and $n_c=3$ in QCD. For DY production, the factor ${\cal F}^{(0)}$ is found to be,
\begin{align}
\begin{autobreak}
{\cal F}^{(0)} =
{4 \alpha^2 \over 3 q^2} \Bigg[Q_q^2 
- {2 q^2 (q^2-M_Z^2) \over  \left((q^2-M_Z^2)^2
+ M_Z^2 \Gamma_Z^2\right) c_w^2 s_w^2} Q_q g_e^V g_q^V 
+ {Q^4 \over  \left((q^2-M_Z^2)^2+M_Z^2 \Gamma_Z^2\right) c_w^4 s_w^4}\Big((g_e^V)^2
+ (g_e^A)^2\Big)\Big((g_q^V)^2+(g_q^A)^2\Big) \Bigg]\,.
\end{autobreak}
\end{align}
Here $\alpha$ is the fine structure constant, $c_w,s_w$ are sine and cosine of Weinberg angle respectively. $M_Z$ and $\Gamma_Z$ are the mass and the decay width of the $Z$-boson.
\begin{align}
 g_a^A = -\frac{1}{2} T_a^3 \,, \qquad g_a^V = \frac{1}{2} T_a^3  - s_w^2 Q_a \,,
\end{align}
$Q_a$ being electric charge and $T_a^3$ is the weak isospin of the electron or quarks.

In the threshold region, the SV terms which consist of distributions 
contribute significantly at the hadron level.  After mass factorization, the partonic coefficient in the threshold region experiences further factorization in terms of the form factor and soft-collinear function. In the next section we will discuss in detail on the structure of distributions in the SV coefficient which will form the basis for the resummation.


\subsection{Soft-virtual cross-section}\label{sec:softvir}
In the following, we briefly describe the theoretical 
set up that is required to study the impact of 
threshold corrections within the framework of resummation a la 
Sterman, Catani and Trentedue \cite{Sterman:1986aj,Catani:1989ne}.
We do this in order to understand the role of various pieces that
contribute to the resummed result.
Exploiting the factorization of infrared sensitive contributions 
and gauge and renormalisation group invariances,   
inclusive cross section for the DY and on-shell $Z/W^\pm$ productions in the threshold limit can be
expressed in terms of 
form factor of the neutral/charged current, soft distribution function and Altarell-Parisi kernels
(see \cite{Ravindran:2005vv,Ravindran:2006cg}).
The resulting expression expressed in $z$ space is free of both ultraviolet and infrared divergences and
captures the  
distributions ${\cal D}_j$ with given logarithmic accuracy to all orders in perturbation theory. 
In the Mellin $N$ space, we can achieve the same and in addition, one has the advantage to 
reorganize the series in such a way that order one contributions of the form $a_s \beta_0 \log(N)$ can
be resummed systematically to all orders in the large $N$ limit.  
Here, $a_s$ is defined by 
$a_s = g_s^2(\mu_r^2)/16 \pi^2$ with $g_s$ begin the strong coupling constant and $\mu_r$ the
renormalisation scale and $\beta_0$ is the first coefficient of the coupling constant 
beta function.  Note that in Mellin N space, the convolutions in $z$ space become
simple products.  The $z$ space result can be used to compute the soft-virtual contributions       
in power series expansion of strong coupling constant $a_s$.  

In $d=4+\epsilon$ space time dimensions, the threshold enhanced partonic 
soft-virtual cross-section to all orders in perturbation theory 
in $z$ space can be written \cite{Ravindran:2005vv,Ravindran:2006cg} as
\begin{align}
  \label{eq:sigma}
  \Delta^{{\rm{(sv)}}} (z, q^2, \mu_r^{2}, \mu_f^2) = 
  {\cal C} \exp \Big( \Psi \left(z, q^2, \mu_r^2, \mu_f^2,
  \epsilon \right)  \Big)  \Big|_{\epsilon = 0} \,.
 \end{align}
Here $\Psi$ is a distribution function which is finite in the limit $\epsilon \to 0$. 
The symbol $ {\cal C}$ denotes the Mellin convolution (denoted below as $ \otimes$) which in the above expression should be treated as
\begin{align}
  \label{eq:conv}
  {\cal C} \exp \Big( {f(z)} \Big) = \delta(1-z) + \frac{1}{1!} f(z) + \frac{1}{2!} f(z) \otimes f(z) + \cdots \,,
\end{align}
with $f(z)$ being a function containing only $\delta(1-z)$ and plus distributions. The finite exponent in the above is refactorized in the threshold limit and  gets contribution from the form factor \big(${\F} (\hat{a}_s, Q^2, \mu^2, \epsilon)$\big) with $q^2=-Q^2$, soft-collinear function \big($\ph (\hat{a}_s, z, q^2, \mu^2, \epsilon)$\big) (later called as soft function) as well as mass factorization kernels  \big($\Gamma (\hat{a}_s, z, \mu_f^2, \mu^2, \epsilon)$\big) and takes the following form in dimensional regularization: 
\begin{align}\label{sv:psi}
  \Psi \Big(z, q^2, \mu_r^2, \mu_f^2, \epsilon \Big)  
  = &\bigg( \ln \Big[ \Z (\hat{a}_s, \mu_r^2, \mu^2, \epsilon) \Big]^2 
      + \ln \Big|  {\F} (\hat{a}_s, q^2, \mu^2, \epsilon)   \Big|^2 \bigg) \delta(1-z) 
      \nonumber\\
    & + 2 \ph (\hat{a}_s, z, q^2, \mu^2, \epsilon) 
      - 2 {\cal C} \ln \Gamma (\hat{a}_s, z, \mu_f^2, \mu^2, \epsilon) \, .
\end{align}
 $\mu$ keeps the strong coupling ($\ash$) dimensionless in the $d=4+\eps$ dimensions. $\Z (\hat{a}_s, \mu_r^2, \mu^2, \epsilon)$ denotes the overall UV renormalization constant which for the processes considered here is unity due to conserved current.

The bare quark form factor satisfies the Sudakov K+G equation \cite{Sudakov:1954sw, Mueller:1979ih, Collins:1980ih, Sen:1981sd, Ravindran:2005vv,Ravindran:2006cg} which follows as a consequence of the gauge invariance as well as renormalisation group invariance,
\begin{align}\label{eq:FFKG}
\frac{d \ln \F}{d\ln q^2} =   \frac{1}{2} \bigg[ {\K}(\ash, \frac{\mur^2}{\mu^2},\eps) + {\G}(\ash, \frac{q^2}{\mur^2},\frac{\mur^2}{\mu^2},\eps)\bigg] \,.
\end{align}
The function $\K$ contains all the infrared poles in $\eps$ whereas the function $\G$ is finite in the limit $\eps \to 0$. The renormalisation group invariance leads to the following solutions of these functions in terms of cusp anomalous dimensions ($A$):
\begin{align}
\frac{d\K}{d\ln\mur^2} = - \frac{d\G}{d\ln\mur^2}  = A(\as(\mur)) = \sum_{i=1}^{\infty} \as^i(\mur) A_{i} \,.
\end{align}
The cusp anomalous dimensions are known to fourth order \cite{Moch:2004pa,Lee:2016ixa,Moch:2017uml,Grozin:2018vdn,Henn:2019rmi,Davies:2016jie,Lee:2017mip,Gracey:1994nn,Beneke:1995pq,Moch:2017uml,Moch:2018wjh,Henn:2019rmi,Bruser:2019auj,Lee:2019zop} and are collected in Appendix \ref{app:anodim}. The $\mur$ independent piece of the $\G$ can be written in perturbative series as
\begin{align}
\G(\as(q),\eps) = \sum_{j=1}^{\infty} \as^j(q) \G_{j}(\eps)\,,
\end{align} 
where the coefficients $\G^{(j)}(\eps)$  can be decomposed as 
\begin{align} \label{eq:G}
 \G_i(\eps) = 2 B_{i} + f_{i} + C_{i}  + \sum_{k=1}^{\infty} \epsilon^k G_{ik}    \,,
\end{align}
where 
\begin{align}
C_1  &=   0 \,, \nn\\
C_2  &= - 2 \beta_0 G_{11} \,, \nn\\
C_3  &= - 2 \beta_1 G_{11} - 2 \beta_0 \Big( G_{21} + 2 \beta_0  G_{12} \Big) \,.
\end{align}
The coefficients $G_{ik}$ are the finite coefficients found in terms of QCD color factors and can be extracted from explicit calculation of quark form factor. Note that up to the third order one also needs coefficients $G_{22},G_{31}$ and thereby needs the three-loop calculation of the form factor \cite{Gehrmann:2010ue}. We have collected them in the Appendix \ref{app:anodim}.
%
Similar to the cusp anomalous dimension, the coefficients $f_{i}$ have been found to be maximally non-abelian to third order in strong coupling {\it i.e.} they satisfy 
\begin{align}
f^g_{i} = \frac{C_F}{C_A} f^{q}_{i} \,.
\end{align}

The initial state collinear singularities are removed using the Altarelli-Parisi (AP) splitting kernels $\Gamma (\hat{a}_s, \mu_f^2, \mu^2, z, \epsilon)$. They satisfy the well-known DGLAP evolution given as,
 \begin{align}
\label{eq:APRGE}
 \frac{d\Gamma (  z, \muf^2, \eps)}{d\ln \muf^2} = \frac{1}{2} P(z,\muf^2) \otimes \Gamma (  z, \muf^2, \eps)\,,
 \end{align}
 where $P(z,\muf^2) $ is the AP splitting functions. The perturbative expansion for these splitting functions has the following form:
 \begin{align}
 P(z,\muf^2) = \sum_{i=0}^{\infty} \as^{i+1}(\muf) P^{(i)}(z) \,.
 \end{align}
 As already discussed, only the $q\bar{q}$ channel contributes to the SV cross-section and thus we find that, only the diagonal terms of the splitting functions contribute to the SV cross-section. The diagonal part of the splitting functions is known to contain the $\delta(1-z)$ and distributions and can be written as,
\begin{align}
P_{II}^{(i)} = 2 \Big[B_{i+1} \delta(1-z) + A_{i+1} {\cal D}_0 \Big] + P_{II}^{(reg,i)}(z) \,.
\end{align}
The splitting functions are known exactly to four loops \cite{Vogt:2004mw,Moch:2004pa,Moch:2014sna,Henn:2019swt}.

The finiteness of the soft-virtual cross-section demands that the soft-collinear function $\ph$ will also satisfy similar Sudakov type equation like the form factor {\it i.e.} one can write
\begin{align}\label{eq:PHIKG}
\frac{d\ph}{d \ln q^2} =   \frac{1}{2} \bigg[ \KB(\ash, z, \frac{\mur^2}{\mu^2},\eps) + \GB(\ash, z, \frac{q^2}{\mur^2},\frac{\mur^2}{\mu^2},\eps)\bigg] \,,
\end{align}
where $\KB(\ash, z, \frac{\mur^2}{\mu^2},\eps)$ contains all the poles and $\GB(\ash, z, \frac{q^2}{\mur^2},\frac{\mur^2}{\mu^2},\eps)$ is finite in the dimensional regularization such that $\Psi$ becomes finite as $\eps \to 0$. The solution to the above equation has been found \cite{Ravindran:2005vv,Ravindran:2006cg} to be 
\begin{align}\label{eq:phi-sol}
\ph =  \sum_{j=1}^{\infty} \ash^j       \frac{j\eps}{1-z}  \bigg(  \frac{q^2(1-z)^2}{\mu^2}  \bigg)^{j\eps/2}   {\cal S}_{\eps}^j ~\hat{\ph}^{(j)}(\eps) \,.
\end{align}
$\hat{\ph}^{(j)}$ can be found from the solution of the form factor by the replacement  as $A \to -A, \G(\eps) \to \GB(\eps) $. Notice that $ \GB(\eps) $ are now new finite $z$-independent coefficients coming from the soft function whereas the $z$ dependence has been taken out in Eq.\ (\ref{eq:phi-sol}). This can be found by comparing the poles and non-pole terms in $\hat{\ph}^{(j)}$ with those coming from the form factors, overall renormalisation constants, splitting kernel and the lower order SV terms.
The coefficient $\GB$ has same structure as the form factor in Eq.\ (\ref{eq:G}) after setting $f_{i} \to - f_{i}, B_{i} \to 0, \gamma_{i} \to 0 $,
\begin{align}\label{eq:Gt}
\GB_{i} = - f_{i}  +    \tilde{C}_{i}    + \sum_{k=1}^{\infty} \eps^k \gt_{ik} \,,
\end{align}
where 
\begin{align}
\tilde{C}_{1} &= 0, \nn\\
\tilde{C}_{2} &= -2 \beta_0 \gt_{11}, \nn\\
\tilde{C}_{3} &= -2 \beta_1 \gt_{11} - 2 \beta_0 \Big( \gt_{21} + 2 \beta_0 \gt_{12}    \Big) \,. 
\end{align}
The coefficients $f_{i}$ are same as those appear in the quark form factor in Eq.\ (\ref{eq:G}). 
The coefficients $\gt_{ij}$ required up to three loops have been extracted in \cite{Ahmed:2015qia} and also collected in the Appendix \ref{app:anodim}. 
Note that one has to perform the following expansion in Eq.\ (\ref{eq:phi-sol}) in order to  get all the distributions and delta function coming from the soft function,
\begin{align}
\frac{1}{(1-z)} \Big[ (1-z)^2 \Big]^{j\eps/2} = \frac{1}{j\eps} \delta(1-z) + \sum_{k=0}^{\infty} \frac{(j\eps)^k}{k!} {\cal D}_k   \,.
\end{align}
It is worth noting that $\GB$ as well as the complete soft function $\ph_I$ satisfy the maximally non-abelian property up to three loops. Moreover $\ph_I$ is also universal in the sense that it only depends on the initial legs and is completely unaware of the color neutral final state. 
Expanding $\Delta^{(\rm{sv})}$ in powers of $a_s$ as
\begin{align}
\Delta^{{\rm(sv)}}_{ab}  = \delta_{a \overline b} \sum_{i=0}^{\infty} a_s^i \Delta^{(i)} \,,
\end{align}
with the born contribution being $\Delta^{(0)} = \delta(1-z) $.
The SV correction at the three loops are known \cite{Ahmed:2014cla} which we collect here for completeness in the Appendix \ref{app:svn}.

In the following, we will study the numerical impact of resummed result resulting from
$\Delta^{{\rm(sv)}}_{ab}$ after performing the Mellin transformation in the large $N$ limit. 
We start with $\Psi$ which is finite while the individual contributions to it contain 
UV and IR singularities.  Decomposing the later ones as sum of singular and finite parts as 
\begin{eqnarray}
\ln \big|  {\F} \big|^2(q^2) &=& {\cal L}_{\cal F}^{\rm{sing}}(q^2,\mu_r^2) + {\cal L}_{\cal F}^{\rm{fin}}(q^2,\mu_r^2)  \nonumber\\
\Phi(z,q^2) &=& \Phi^{\rm{sing}}(z,q^2,\mu_f^2,\mu_r^2) 
+ \Phi_{\cal D}^{\rm{fin}}(z,q^2,\mu_f^2,\mu_r^2)  
+ \Phi_{\delta}^{\rm{fin}}(q^2,\mu_f^2,\mu_r^2)  \delta(1-z)
\nonumber\\
{\cal C} \ln \Gamma (z,\mu_f^2)& =& {\cal L}_\Gamma^{\rm{sing}}(z,\mu_f^2,\mu_r^2) 
+ {\cal L}_{\Gamma {\cal D}_0}^{\rm{fin}} (\mu_f^2,\mu_r^2){\cal D}_0 
+ {\cal L}_{\Gamma\delta}^{\rm{fin}} (\mu_f^2,\mu_r^2) \delta(1-z) 
\end{eqnarray}
Substituting the above equations in Eq.(\ref{sv:psi}), we can easily show
that all the singular terms in the limit $\epsilon\rightarrow 0$ cancel among 
themselves.  In addition, ${\cal D}_0$ terms in finite part of ${\cal C} \ln\Gamma$ 
go away when added to $\Phi_{\cal D}^{\rm{fin}}$ resulting in a finite distribution.
Substituting the $\Psi$ in Eq.(\ref{eq:sigma}), we obtain
\begin{eqnarray}
\label{eq:resumz}
\Delta^{\rm{sv}}(z,q^2) = C_0(q^2) \otimes {\cal C} e^{
G_+\left(z,q^2\right)}
\end{eqnarray}
where (supressing dependence on $\mu_f$ and $\mu_r$), the $N$ independent constant $C_0$ is given by
\begin{eqnarray}
C_0(z,q^2)&=& \exp\left({\cal L}_{\cal F}^{\rm{fin}}
+ 2 \Phi^{\rm{fin}}_\delta-2 {\cal L}_{\Gamma \delta}^{\rm{fin}}\right) \delta(1-z)
\\
G_+(z,q^2) &=&\left({1 \over 1-z} \left[\int_{\muf^2}^{q^2(1-z)^2} \frac{d\mu^2}{\mu^2}2~ A(a_s(\mu^2)) + D(a_s(q^2(1-z)^2))\right] \right)_+
\end{eqnarray}
and $D$ in $G_+$ is related to $\overline {\cal G}$
by $D = 2 \overline {\cal G}$ 
and $\overline {\cal G} (a_s(q^2 (1-z)^2))$ is $\overline {\cal G}$ in Eq.(\ref{eq:PHIKG}) 
evaluated at $\mu_r^2=\mu_f^2=q^2$.  

So far, we showed how various collinear soft gluon emissions as well as 
the wide angle soft emissions can be systematically summed to all orders to obtain Eq.(\ref{eq:resumz}) 
in the $z$ space when partonic variable $z\rightarrow 1$. 
Note that $C_0$ is obtained by first collecting those terms that are proportional to
$\delta(1-z)$ terms of $\Psi$ and then expanding the exponential of them in powers of $a_s$. 
The remaining function $G_+$ contains only distributions ${\cal D}_j$.  
Hence, one can predict the following structure for $G_+$:
\begin{eqnarray}
\label{eq:expz}
G_+(z,q^2) = G_1 (q^2) \otimes {\cal D}_0 + G_2(z,q^2) + a_s G_3(z,q^2) + \cdots \,.
\end{eqnarray}
where each $G_1$ sums certain terms of the $a_s^i {\cal D}_{i-1}$ to all orders,
and $G_2$ sums $a_s^i {\cal D}_{i-2}$ terms to all orders, etc etc.
The result $\Delta^{\rm{sv}}$ expressed in terms of $C_0$ and the exponential of 
$G_+$ using Eq.(\ref{eq:expz}) systematically sums the distributions 
${\cal D}_j$ to all orders and hence can predict
these distributions to all orders provided $A$ and $D$ are known to desired accuracy in
$a_s$.  For example, knowing $A_{1}$, we can predict
all the terms $a_s^i {\cal D}_{i}$ with $i=1,2,...,\infty$ in $\Phi$, similarly given $A_{1}$ and
$D_{1}$, we can predict $a_s^i {\cal D}_{i-1}$ with $i=1,2,...,\infty$ etc.  
Hence, expression given
in Eq.(\ref{eq:resumz}) has the predictive power for $\Delta^{\rm{sv}}$ to all orders in
$a_s$ given the logarithmic accuracy in $z$ space, quantified by terms of the form $a_s^i {\cal D}_{j}$.  
Note that when the exponential of $\Phi$ is expanded
using convolution rules given in Eq.(\ref{conv}), we will get not only ${\cal D}_j$ but also $\delta(1-z)$.
In other words, $\delta(1-z)$ terms in $\Delta^{\rm{sv}}$ can come from both $exp(G_+)$ as well as $C_0$.    

Often in certain kinematic regions, these contributions can be enhanced when convoluted with the
parton distribution functions spoiling the reliability of the perturbation theory.  Hence we need to
include these potentially large terms to all orders in perturbation theory for any sensible predictions.
Such an exercise in the $z$ space is technically challenging due to the complexity involved in
computing the convolutions of ${\cal D}_j$.  However, in the Mellin $N$ space, the convolutions
become simple products allowing us to study the impact of these large logarithms to all orders
in a systematic fashion.  In the following, we will describe how this can be done in Mellin $N$ space. 

\subsection{Threshold Resummation}\label{sec:resummation}
In the last sub-section, we showed that threshold effects 
for partonic coefficients can be obtained 
near threshold as a product of well-defined functions, each organizing a class of infrared and 
collinear enhancements as can be seen from Eq.\ (\ref{sv:psi}). 
This refactorization is valid up to corrections which are nonsingular at threshold 
when partonic $z \to 1$.  While the $z$ space result captures the entire underlying
infrared dynamics in the threshold limit, it can be better described in the Mellin-$N$ space 
where the threshold limit $z\to 1$ translates into $N\to \infty$. 
We found that the form in Eq.\ (\ref{sv:psi}) was already suitable for all order study, 
however complications arise in performing the convolution.  On the other hand any 
such convolution becomes simple product in Mellin space and all the distributions coming 
from the soft function are thus translated into large logarithms in Mellin $N$.

Following \cite{Catani:1989ne}, the resummed partonic SV coefficient function can be organized as follows: 
\begin{align}\label{eq:res}
\hat{\sigma}_N (q^2)&= \int_0^1 dz z^{N-1} \Delta^{\rm{sv}}(z,q^2)
\nonumber\\
&= \gb_0(q^2) \exp \Big( G_{\Nb}(q^2) \Big) \,,
\end{align}
where $G_{\Nb}$ is obtained by computing the large $N$ limit of Mellin moment of $G_+$ and then 
by decomposing as  
\begin{eqnarray}
\label{eq:resGNb}
\lim_{N \to \infty} 
\int_0^1 dz z^{N-1} G_+(z,q^2)
= \overline G_0(q^2) + G_{\Nb}(q^2) 
\,, \quad {\rm {with}} \quad G_{\Nb}(q^2)\big|_{\Nb=1} = 0 
\end{eqnarray}
where $\Nb = N\exp(\gamma_{E})$ and $\gamma_E$ is E-M constant.
The $N$ independent constant $\overline g_0$ is given by
\begin{eqnarray}
\overline g_0(q^2) =\exp\left({\cal L}_{\cal F}^{\rm{fin}}
+ 2 \Phi^{\rm{fin}}_\delta-2 {\cal L}_{\Gamma \delta}^{\rm{fin}}+\overline G_0(q^2) \right)
\end{eqnarray}
$G_{\Nb}$ is function of the universal coefficients $A$ which are known to fourth order and 
$D$ known to third order in $a_s$.  $G_{\Nb}$ collects and resums all the large-$N$ 
logarithms to all orders and it can be expressed as a 
resummed perturbative series which takes the following form:  
\begin{align}\label{eq:gnb}
G_{\Nb}(q^2) = \ln \overline{N} ~\gb_1(\Nb,q^2) + \gb_2(\Nb,q^2) + a_s ~\gb_3(\Nb,q^2) + a_s^2~ \gb_4(\Nb,q^2) + \cdots \,.
\end{align}
Following \cite{Catani:2014uta,Catani:2003zt}, we computed $\gb_i$ up to $i=4$ 
(for $\gb_i$ up to $i=3$, see \cite{Idilbi:2006dg}) 
and they are given in Appendix \ref{app:NBres}. Note that $\gb_i$ coefficients are universal in the sense that it depends only on whether the born process is $q\bar{q}$ channel or $gg$ channel. 
In the Mellin $N$ space, the $\delta(1-z)$ in $z$-space directly translates into $N$ independent piece 
whereas the plus-distributions give rise to the $\ln(\Nb)$ as well as $N$ independent constants
in the large $N$ limit.  
Part of these constant pieces, namely $\overline G_0$, is absorbed into the 
coefficients $\gb_0$ in the standard resummation approach.
Hence, $\gb_0$ contains only $N$ independent pieces 
which come from the form factor, soft distribution function, AP kernels and $N$ independent part
of the Mellin moment of $G_+(z,q^2)$. 
The condition $G_{\Nb}=0$ for ${\Nb=1}$ allows the constants $\gb_i$ to contain $N$ independent terms.   
Note that the expressions for $\gb_0$ and $\gb_i$ obtained this way depend on the condition 
$G_{\Nb}=0$ for ${\Nb=1}$.  In other words, there is an ambiguity in
treating the $N$ independent terms in the resummed results.  Exploiting this, in \cite{Catani:1989ne}, 
the $N$ independent constants were defined by demanding $G_{\Nb}=1$ when $\Nb=1$.  
With this, $\gb_0$ has the following perturbative expansion:
\begin{align}\label{eq:gg0b}
\gb_0(q^2) = 1+\sum_{n=1}^{\infty} a_s^i \gb_{0i}(q^2) \,.
\end{align}
The successive terms in the resummed exponent Eq.\ (\ref{eq:gnb}) along with the corresponding terms in Eq.\ (\ref{eq:gg0b}) define the accuracies leading logarithmic (LL), next to LL (NLL), NNLL and N$^3$LL etc.
Terms independent of $N$ can be treated, in principle, by the same methods that resum terms enhanced by logarithms of $N$.

In summary, the resummed result will differ depending on how we treat the $N$-independent constants.
We define various 
schemes that differentiate how these constants are treated in our 
numerical implementation for the phenomenological 
studies.
This allows us to investigate numerical impact of the various resummed results in detail.  
\begin{itemize}
\item {\bf Standard $\Nb$ exponentiation:} This is the case we have discussed so far where we define 
large logarithms are functions of $\Nb = N\exp(\gamma_{E}^{})$, where $\gamma_E$ is E-M constant. 
The $N$ dependent functions 
$G_{\Nb}$ in this case can be computed by simply performing the Mellin moment of $G_+(z,q^2)$ in 
the large $\Nb$ limit and keeping only those terms that vanish when $\Nb=1$. 

\item {\bf Standard $N$ exponentiation:} This approach differs from the previous one in the 
definition of large-$N$ variable. In this case the large logarithm is simply $\ln N$ and these terms are exponentiated to all orders through the resummed exponent. It is evident that this only accounts for reshuffling of $\gamma_E^{}$ between $\gb_0$ and $G_{\Nb}$ in Eq.\ (\ref{eq:res}) which now takes the following form:
\begin{align}\label{eq:resN}
\hat{\sigma}_N(q^2) = g_0(q^2) \exp \Big( G_{N}(q^2) \Big) \,.
\end{align}
 The resummed exponent $G_N$ also takes a different form  compared to the standard $\Nb$ exponent,
\begin{align}
G_N(q^2) = \ln N ~g_1(N,q^2) + g_2(N,q^2) + a_s ~g_3(N,q^2) + a_s^2~ g_4(N,q^2) +\cdots \,.
\end{align}
The resummed coefficients $g_i$ in  the above equation which defines the resummed accuracy, 
differs from $\gb_i$ in Eq. (\ref{eq:gnb}).  The present scheme is defined by demanding $G_N=0$ when $N=1$.
\begin{eqnarray}
\label{eq:resGN}
\lim_{N \to \infty}
\int_0^1 dz z^{N-1} G_+(z,q^2)
= G_0(q^2) + G_{N}(q^2)
\,, \quad {\rm {with}} \quad G_{N}(q^2)\big|_{N=1} = 0
\end{eqnarray}
With this definition, the rest of the $N$ independent terms from the Mellin moment of $G_+$
is combined with finite parts of form factor, soft distribution function and the AP kernels
as 
The $N$ independent constant $g_0$ is given by
\begin{eqnarray}
g_0(q^2) =\exp\left({\cal L}_{\cal F}^{\rm{fin}}
+ 2 \Phi^{\rm{fin}}_\delta-2 {\cal L}_{\Gamma \delta}^{\rm{fin}}+G_0(q^2) \right)
\end{eqnarray}
and the above result is expanded in powers of $a_s$:  
\begin{align}\label{eq:gg0b}
g_0(q^2) = 1+\sum_{n=1}^{\infty} a_s^i g_{0i}(q^2) \,.
\end{align}
Numerically this can make a difference and it was seen in the context of DIS previously. 
In case of DY also we find such differences which will be discussed in the next section. 
Up to N$^3$LL accuracy, the resummed exponents $g_i, i=1,..,4$ for both quark as well as
for gluon initiated process in $N$ exponentiation scheme can be found in
\cite{Catani:2003zt,Moch:2005ba} and we computed the results for
the $g_{0i}$ coefficients up to $i=3$ which are listed in Appendix \ref{app:Nres} along with $g_i$.

\item {\bf Soft exponentiation:} In the standard $\Nb$ ($N$) exponentiation, one exponentiates 
$\ln \Nb$ ($\ln N$) and certain $\Nb$ ($N$) independent terms which arise from $G_+$, subjected
to the condition $G_{\Nb}=0$ $(G_N=0)$ when $\Nb=1$ ($N=1$).  The remaining $\Nb$ ($N$) independent
terms in the Mellin moment of $G_+$ along with $C_0$ give  
the coefficient $\gb_0$ ($g_{0})$.  In principle, we can define a scheme 
wherein entire $\Nb$ ($N$) independent terms of $G_+$ can be
kept in the exponent.  More specifically, we define the scheme (relaxing $G_{\Nb}=1$ ($G_N=0$) for 
$\Nb=1$ ($N=1$)) wherein we exponentiate all the terms coming 
from the finite part of soft distribution function and those from the AP kernels.  That is,
the exponential contains
\begin{eqnarray}
\label{GNsoft}
G_{\Nb}^{\rm{Soft}} = G_{\Nb} + 2 \Psi_\delta^{\rm{fin}} - 2 {\cal L}_{\Gamma \delta}^{\rm {fin}} 
\end{eqnarray}  
that is,
\begin{align}\label{eq:resN}
\hat{\sigma}_N(q^2) = g^{\rm {Soft}}_0(q^2) \exp \Big( G^{\rm {Soft}}_{\Nb}(q^2) \Big) \,.
\end{align}
with
\begin{align}\label{eq:gnb}
G_{\Nb}^{\rm {Soft}}(q^2) = \ln \overline{N} ~g^{\rm Soft}_1(\Nb,q^2) + g^{\rm Soft}_2(\Nb,q^2) + a_s ~
g^{\rm Soft}_3(\Nb,q^2) + a_s^2~ g^{\rm Soft}_4(\Nb,q^2) + \cdots \,.
\end{align}
The remaining $N$ independent terms define $g^{\rm{Soft}}_0$ that is obtained by
expanding $\exp({\cal L}_{\cal F \delta}^{\rm{fin}})$ in power series expansion
in $a_s$:  
\begin{align}\label{eq:gs0b}
g^{\rm{Soft}}_0(q^2) = 1+\sum_{n=1}^{\infty} a_s^i g^{\rm{Soft}}_{0i}(q^2) \,.
\end{align}
$G_{\Nb}^{\rm Soft}$ and $ \gb_0^{\rm Soft}$ have similar expansion as Eq.\ (\ref{eq:resGNb}) 
and Eq.\ (\ref{eq:gnb}) respectively and the corresponding coefficients are calculated and 
presented in Appendix \ref{app:softres}. 

\item {\bf All exponentiation:}  The soft function and the form factor 
satisfy K+G type Sudakov integro-differential equations given in 
Eqs.\ (\ref{eq:FFKG}),  and (\ref{eq:PHIKG}) and the AP kernels satisfy renormalisation
group equation Eq. (\ref{eq:APRGE}) governed by AP splitting functions. 
Hence, their solutions given the boundary conditions demonstrate exponential.  The $z$ space solutions
that we obtained carry all order information on the distribution ${\cal D}_j$ in terms of
universal cusp $A$, soft $f$ and collinear $B$ anomalous dimensions and certain process dependent
constants resulting from the form factor.  
Hence it is natural to study the numerical impact of the entire contribution
in the Mellin space without imposing any condition on the $N$ dependent terms.   
This can be easily achieved and the result for $\hat \sigma_N$ takes the following form
\begin{align}\label{eq:resall}
\hat{\sigma}_N = \exp \Big( G_{\Nb}^{\rm All} \Big) \,,
\end{align}
where 
\begin{eqnarray}
\label{eq:resGAb}
G_{\Nb}^{\rm All}(q^2)= {\cal L}_{\cal F}^{\rm{fin}}(q^2)
+ 2 \Phi^{\rm{fin}}_\delta(q^2)-2 {\cal L}_{\Gamma \delta}^{\rm{fin}}
+\overline G_0(q^2) + G_{\Nb}(q^2) 
\end{eqnarray}
where $G_{\Nb}^{\rm All}$ is expanded as  
\begin{align}
G_{\Nb}^{\rm All}(q^2) = \ln \Nb ~g_1^{\rm All}(q^2) + g_2^{\rm All}(q^2) + a_s ~g_3^{\rm All}(q^2) + a_s^2~ g_4^{\rm All}(q^2) \,.
\end{align}
The present scheme was already explored in \cite{Bonvini:2014joa,Bonvini:2016frm}
for studying inclusive cross section for the production of Higgs boson at the LHC. 
For similar study for the DY in DIS and $\overline{\rm MS}$ schemes, see \cite{Eynck:2003fn}. 
Here we will extend it to the N$^3$LL accuracy. The relevant resummed exponent has been 
provided in Appendix \ref{app:allres}.
\end{itemize}

Note that a detailed comparison between the $N$-exponentiation and 
$\overline{N}$-exponentiation has been done in \cite{Das:2019btv} for the charge and neutral DIS processes. 
There, one finds that the $\overline{N}$-exponentiation shows a faster convergence compared to 
the $N$-exponentiation. In fact, the convergence has already been achieved at NLO+NLL 
order in the threshold region in the case of $\overline{N}$-exponentiation, 
whereas in $N$-exponentiation, this occurs after the NLO+NLL order. 
Notice that the leading logarithmic term also differs between these two approaches. 
In the case of $\overline{N}$-exponentiation, all the $\gamma_{E}^{}$ terms 
are exponentiated through the variable $\overline{N} = N \exp(\gamma_E^{})$; 
but in the $N$-exponentiation these $\gamma_E^{}$ terms are distributed among the exponent 
and the $N$ independent term $g_0$. As a result the deviation starts already at the LL accuracy. 
In the next section, we will discuss how various schemes discussed so far can affect the predictions. 
Note that they all give same result at the LL accuracy, however from NLL they differ. 
At NNLO level, we have the contributions from all the channels and at N${}^3$LO only SV contribution
is known so far.  Hence, our numerical predictions will be based on fixed order N$^3$LO${}_{\rm {sv}}$
results for the parton coefficients and on parton distribution functions known to NNLO accuracy.  
Note that the resummed result has to be matched to the fixed order result 
in order to avoid any double counting of threshold logarithms. 
Hence, the matched result which is usually denoted by N$^n$LL is
computed by by taking the difference between the resummed result and the same truncated 
up to order $a_s^n$. Hence, it contains contributions from the threshold
logarithms to all orders in perturbation theory starting from $a_s^{n+1}$:   
\begin{align}\label{eq:matched}
\sigma_V^{N^nLO+N^nLL} &= 
\sigma_V^{N^nLO} +
\sigma_V^{(0)} 
\sum_{ab\in\{q,\bar{q}\}}
  \int_{c-i\infty}^{c+i\infty} \frac{dN}{2\pi i} (\tau)^{-N} \delta_{a\overline b}f_{a,N}(\muf^2) f_{b,N}(\muf^2) \nn\\
&\times \bigg( \hat{\sigma}_N \bigg|_{N^nLL} - \hat{\sigma}_N\bigg|_{trN^nLO}     \bigg)  \,.
\end{align}
The Mellin space PDF ($f_{i,N}$) can be evolved using QCD-PEGASUS \cite{Vogt:2004ns}. 
Alternatively they can be related to the derivative of $z$-space PDF as prescribed 
in \cite{Catani:1989ne,Catani:2003zt}. The contour $c$ in the Mellin 
inverse integration can be chosen according to {\it Minimal prescription} \cite{Catani:1996yz} procedure. 
Notice that the second term in Eq.\ (\ref{eq:matched}) represents the resummed result truncated 
to N$^n$LO order, i.e. the same order to which singular SV results are available. 
In the next section we present the numerical results for the DY production as well as 
on-shell $Z,W^\pm$ production for LHC where we match the existing N$^3$LO fixed order SV results with the 
N$^3$LL resummation derived in this article.

\section{Numerical Results}\label{sec:numerical}
In this section, we present the numerical impact of resummed threshold corrections 
for neutral DY production as well as on-shell $Z/W^\pm$ production at the LHC.
For neutral DY production we consider all the partonic channels at the FO up to 
NNLO with off-shell $\gamma^{*}, Z$ intermediate states. 
Detailed analysis is done for 13 TeV LHC, 
however it can be extended to other energies as well as to other colliders.

\subsection{Soft-virtual correction for neutral DY invariant mass}
We start our discussion by examining the SV corrections at N$^3$LO. For our numerical study, 
we use the following electro-weak parameters for the vector boson masses and widths, Weinberg 
angle ($\theta_w$) and the fine structure constant ($\alpha$):
\begin{align}\label{eq:parameters}
m_Z &= 91.1876 ~ \text{GeV},  ~~ \Gamma_Z=2.4952~ \text{GeV} \,,\nn\\
m_W &= 80.379  ~ \text{GeV}, ~~ \text{sin}^2\theta_w=0.2311 ~~ \alpha=1/128 \,.
\end{align}
We present our results for the default choice of hadronic center of mass energy $13$ TeV at the LHC. The parton distribution functions (PDFs) are directly taken from the $\tt{lhapdf}$ \cite{Buckley:2014ana} routine. Except for studying the PDF uncertainties, we use $\tt{MMHT2014}$ \cite{Harland-Lang:2014zoa} parton densities throughout.  
The $(n+1)$-loop strong coupling constant is used for computing N$^n$LO order cross sections with $\alpha_s=0.120 (0.117)$ at NLO(NNLO) respectively.

\begin{figure}[ht]
\centerline{
\includegraphics[width=7.5cm, height=7.0cm]{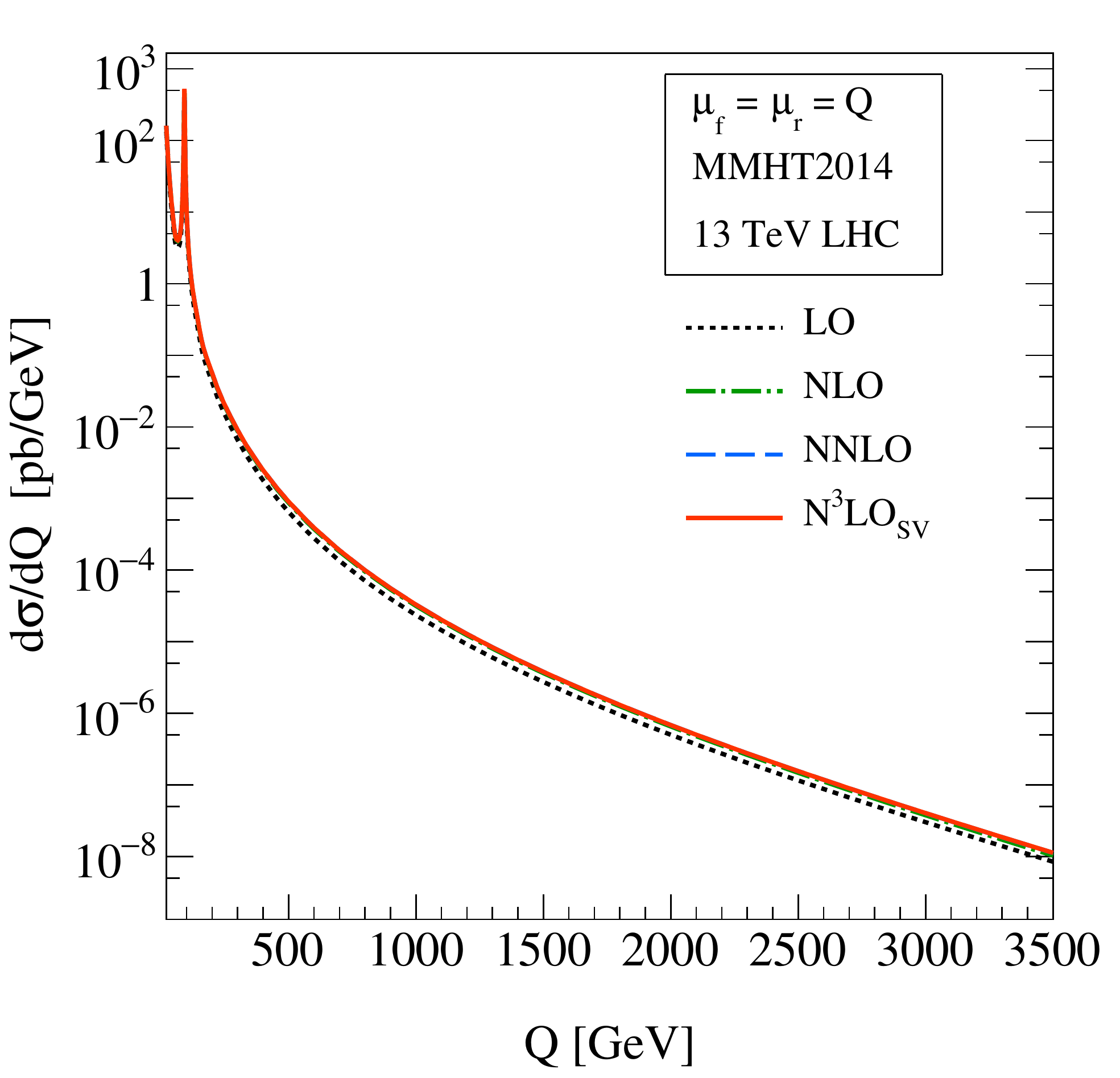}
\includegraphics[width=7.5cm, height=7.0cm]{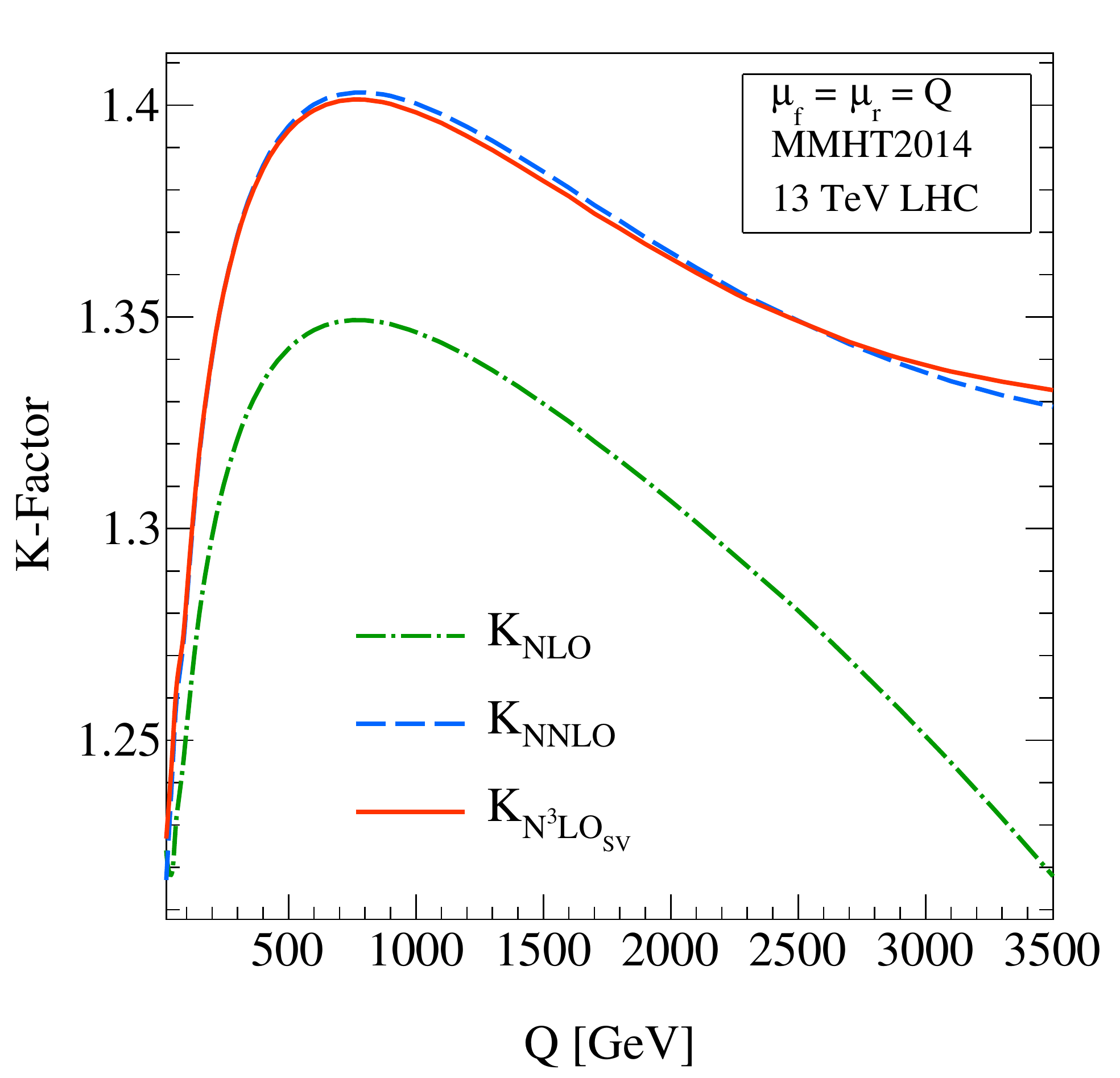}
}
\vspace{-2mm}
\caption{\small{Invariant mass distribution (left panel) of di-lepton pair for $13$ TeV LHC 
and the corresponding $K$-factors (right panel) to N$^3$LO$_{sv}$ level. The renormalization and factorization scales are set to be same as the di-lepton invariant mass.}}
\label{fig:foinv_kfac}
\end{figure}
In \fig{fig:foinv_kfac}, we present  the invariant mass distribution (left panel) of the 
di-lepton production for the neutral case to N$^3$LO$_{sv}$ in QCD for $13$ TeV LHC as well as the corresponding  K-factors  (right panel). It is worth noting here that at ${\cal O}(\alpha_s^3)$ level  the $\delta(1-z)$ contribution is comparable but opposite in sign to the sum of  logarithmic contributions as is mentioned in \cite{Ahmed:2014cla}.  The $3$-loop SV corrections are found to  be positive up to around $Q=400$ GeV and remain negative for $400\text{ GeV} < Q < 2200 \text{ GeV}$ and become positive thereafter as threshold logarithms dominate in the high Q region. At around 3500 GeV, the $3$-loop SV corrections contribute by about 2\%. The observed values of $Q$ where this change in the sign happens are not fixed but can change with the center of mass energy of the hadrons.

\begin{figure}[ht]
\centerline{
\includegraphics[width=12.5cm, height=7.0cm]{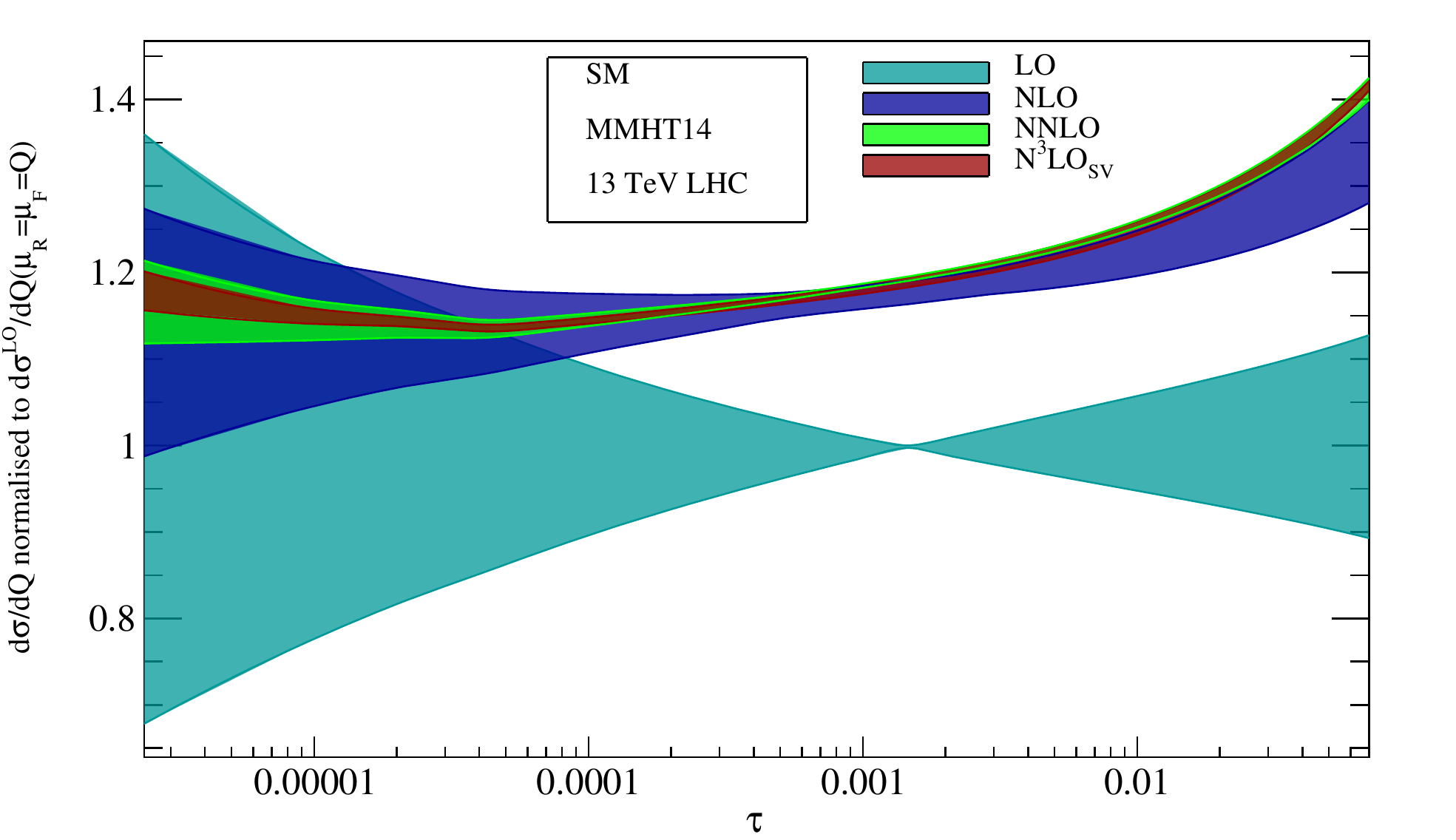}
}
\vspace{-2mm}
\caption{\small{Sevel-point scale variation is plotted against hadronic $\tau$ variable up to N$^3$LO$_{sv}$ order. All the figures are normalised to LO contribution evaluated at the central scale $\mur=\muf=Q$.}}
\label{fig:scalevarFO}
\end{figure}
While the perturbation series is asymptotic and the higher orders terms are very small, the reliability of the theory predictions depends somewhat on the uncertainties due to the unphysical factorization ($\mu_F$) and renormalization ($\mu_R$) scales as well as those due to choice of PDFs. To this end, we estimate the $7$-point scale uncertainties in the invariant mass distribution at various orders in the perturbation theory by varying the scales $\mu = \{ \mu_f, \mu_r \}$ in the range $\frac{1}{2} \leq \frac{\mu}{Q} \leq 2$. The scale uncertainties are conveniently presented in terms of the invariant mass distribution at higher orders normalized with respect to LO ones. In \fig{fig:scalevarFO} we present these normalized distributions up to N$^3$LO$_{sv}$ as a function of $\tau = Q^2/S$. At LO, there is no dependence on $\mu_r$, hence the observation that these scale uncertainties are minimum around $\tau=0.001$ (corresponding to about $Q=400$ GeV) can be directly related to the behavior of the corresponding quark fluxes. At higher orders, the dependence on $\mu_r$ and $\mu_f$ is known and the scale uncertainties are found to increase with Q in the region $Q > 400$ GeV.  For $Q=1500$ GeV, they are found to be $12.55\%$, $6.23\%$, $1.50\%$ and $1.91\%$ respectively at LO, NLO, NNLO and N$^3$LO$_{sv}$. For the $3$-loop SV case, the scale uncertainties are expected to get further reduced only after including the regular terms that are yet to be computed in the fixed order perturbation theory.  However, as we increase Q value, even N$^3$LO$_{sv}$ show reasonable reduction in scale uncertainty as threshold logarithms  dominate over the regular terms for larger Q values. For completeness, we note that the scale uncertainties for $Q=3000$ GeV are found to be 21.39\%, 10.95\%, 3.04\% and 2.16\% at LO, NLO, NNLO and N$^3$LO$_{sv}$ respectively. 

\subsection{Resummed prediction for neutral DY invariant mass}

\begin{figure}[ht]
\centerline{
\includegraphics[width=12.5cm, height=6.5cm]{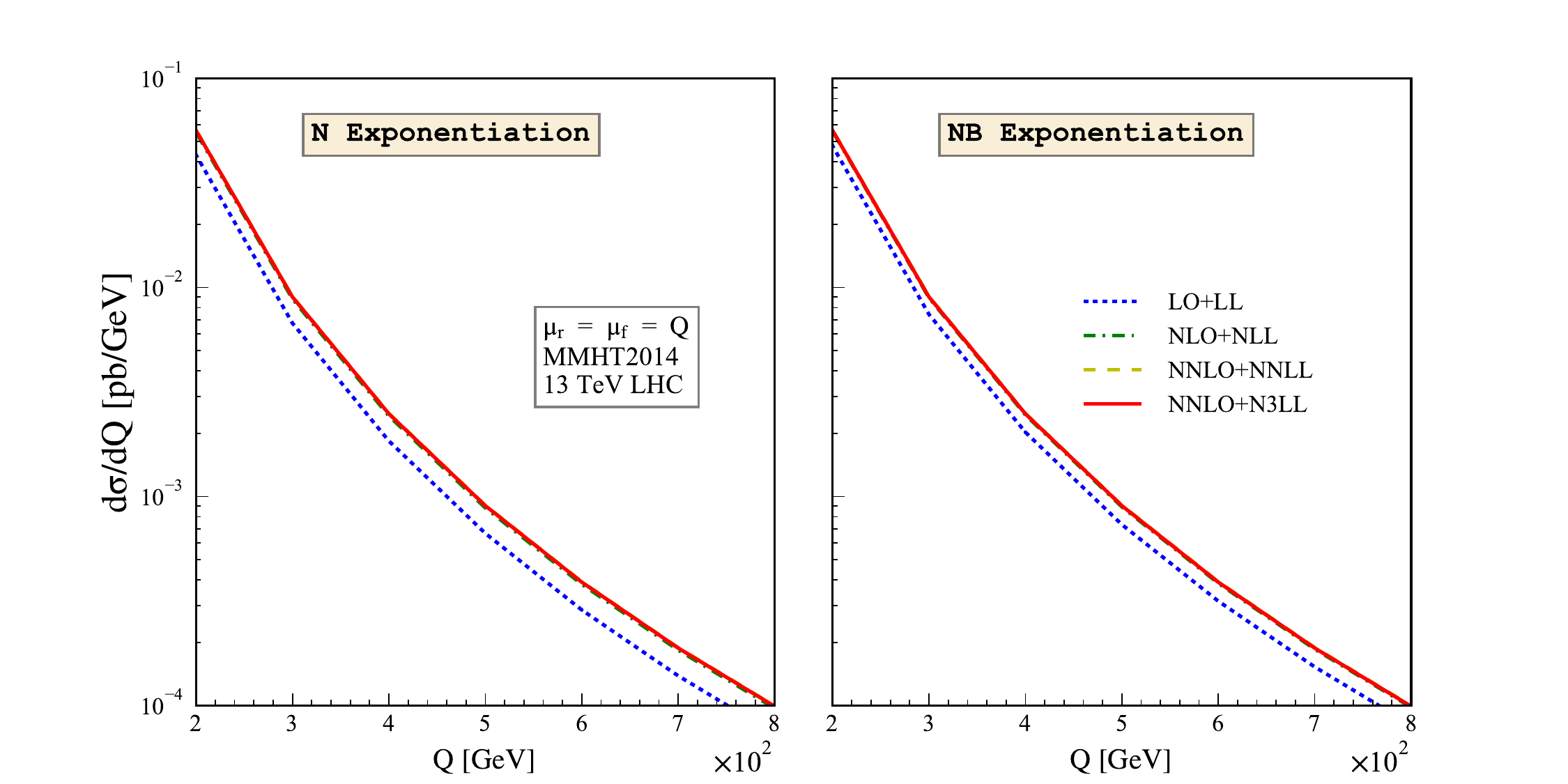}
}
\vspace{-2mm}
\caption{\small{
The comparison between Standard $N$ and $\bar{N}$ approaches are presented up to N$^3$LL setting $\mur=\muf=Q$.}}
\label{fig:NNbcompare1}
\end{figure}
We have studied the impact of different resummation schemes as described in the previous section. First we compare the resummed results between two approaches: the Standard $\Nb$ and Standard $N$ prescriptions. We find that the perturbative convergence is better in the case of $\Nb$ exponentiation for the scale choice $\mur=\muf=Q$. This can be clearly seen from \fig{fig:NNbcompare1} where the convergence is already achieved at NLO+NLL whereas in $N$ exponentiation it happens only after NLO+NLL order. At $Q=2500$ GeV, we see the corrections received in Standard $N$ exponentiation is $21.6\%$ at NLO+NLL, $2.2\%$ at NNLO+NNLL whereas in the Standard $\Nb$ exponentiation these are $6.7\%$ and $2.3\%$ respectively. This observation is also true for different scale choices. This is expected since naively one can expect that as we exponentiate more and more terms the convergence becomes better.  In the rest of the discussion we will mention `Standard' only in the context of $\Nb$ exponentiation unless otherwise stated.

\begin{figure}[ht]
\centerline{
\includegraphics[width=14.5cm, height=7.0cm]{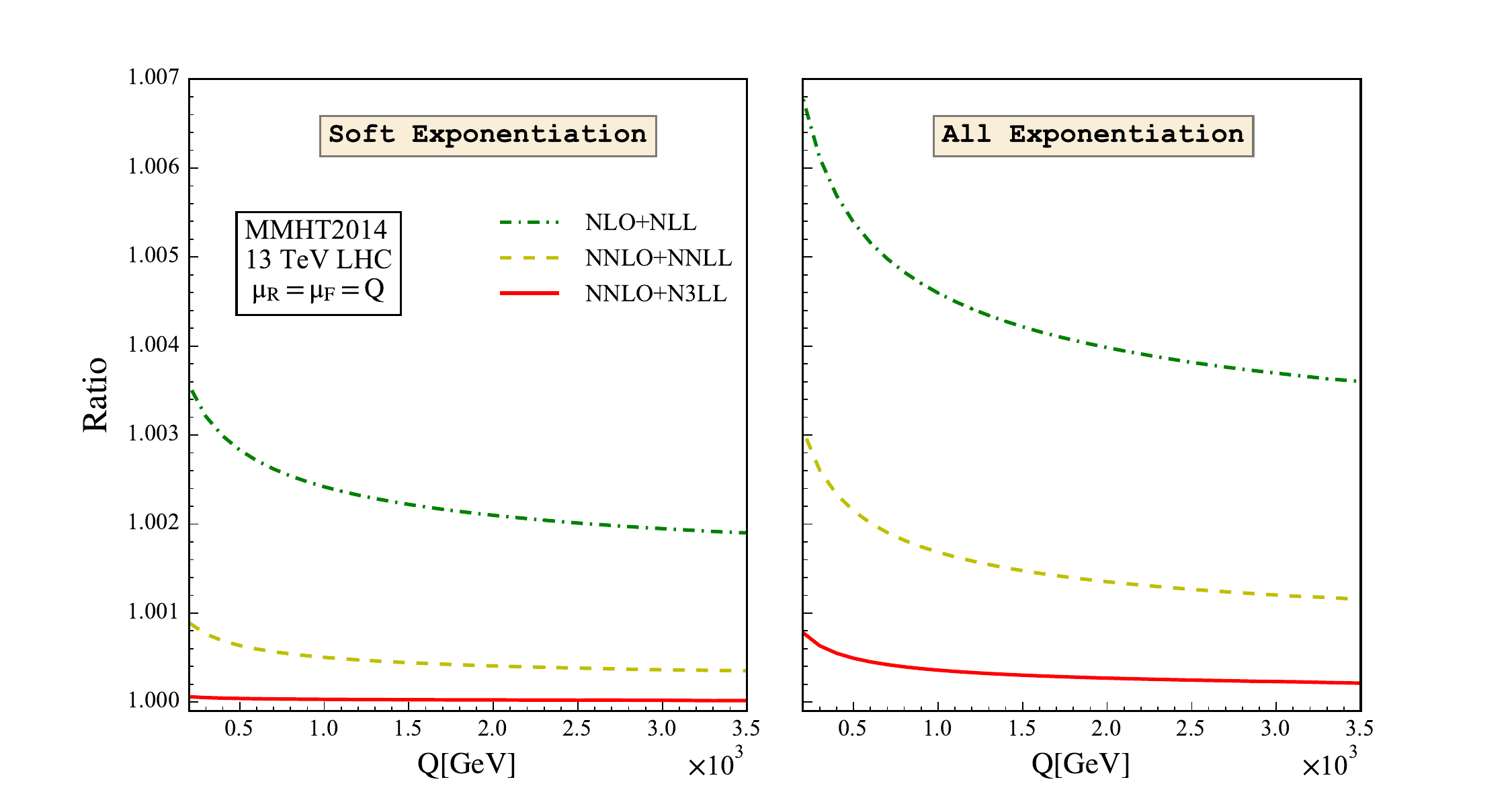}
}
\vspace{-2mm}
\caption{\small{Comparison between the Soft (left panel) and All exponentiation (right panel) with Standard $\Nb$ approach. The renormalisation and factorization scales are set to $Q$. }}
\label{fig:SAcompare}
\end{figure}
We now investigate the differences resulting from 
two approaches {\it viz.} the Soft exponentiation and All exponentiation to study their perturbative behavior.   To illustrate this, we show \fig{fig:SAcompare} where we took the ratio with respect to the Standard $\Nb$ results at each order. Notice that LO+LL results are same for all these three approaches by construction. To this end one sees that at lower orders the resummed cross-sections are improved over $\Nb$ exponentiations. At NNLL the Soft exponentiation gets additional $0.12\%$ corrections compared to the Standard $\Nb$ approach at $Q=100$ GeV. However at N$^3$LL level the Soft exponentiation does not improve over the Standard $\Nb$ results and both approaches provide almost same results. On the other hand,  All exponentiation still gets some contribution from higher orders through the exponentiation of complete $\gb_0$ even at N$^3$LL order.  The increment is however very small giving only $0.12\%$ corrections over the Standard $\Nb$ scenario.

\begin{figure}[ht]
\centerline{
\includegraphics[width=14.5cm, height=12.0cm]{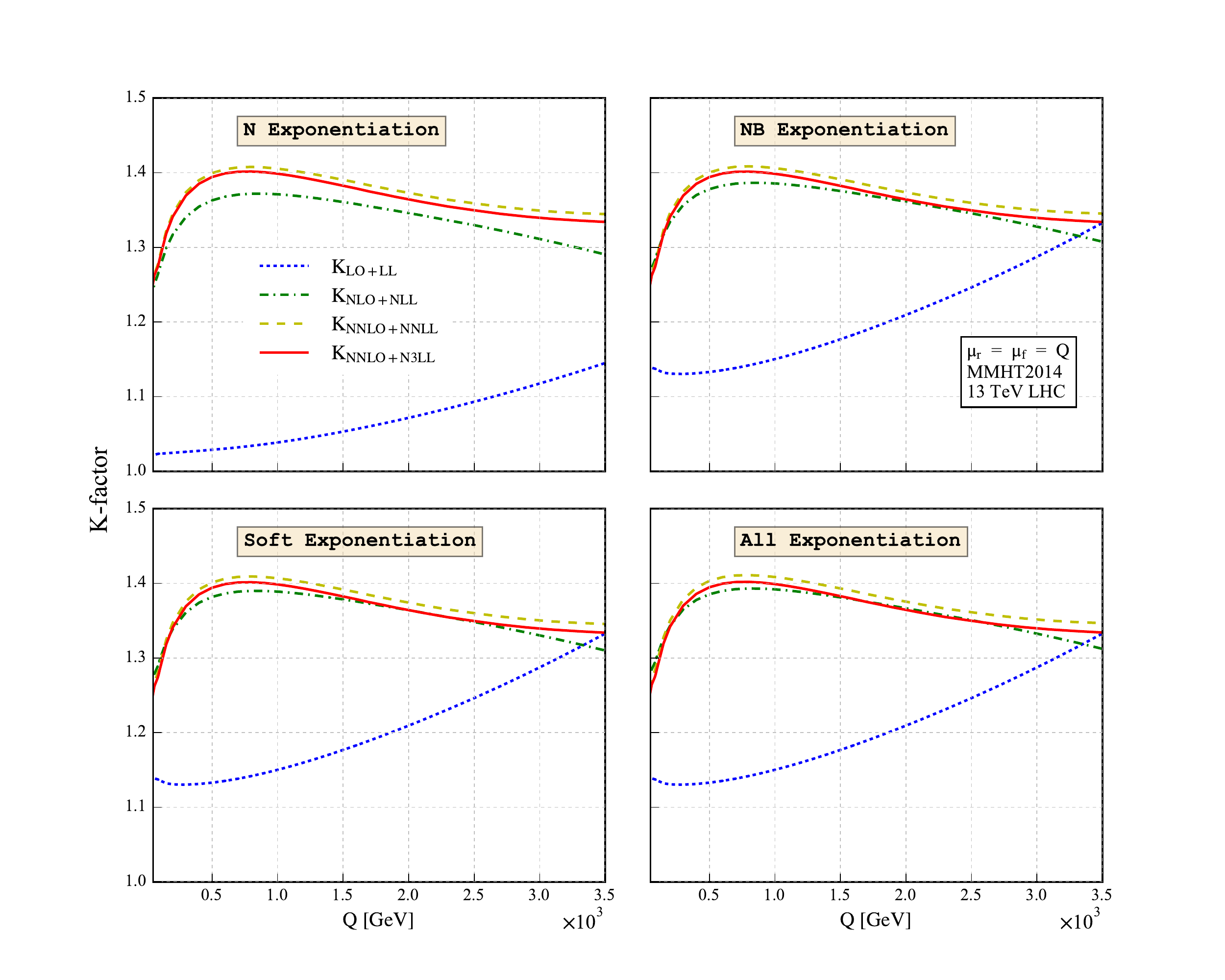}
}
\vspace{-2mm}
\caption{\small{The K-factors are shown for resummed results up to N$^3$LL level for all different resummed approaches consider. (see text) }}
\label{fig:resumKfactor}
\end{figure}
We have quantified the impact of resummed results through $K$-factor.
In \fig{fig:resumKfactor} we present the resummed $K$-factors (K$_{NLO+NLL}$, K$_{NNLO+NNLL}$, K$_{NNLO+N3LL}$) up to order N$^3$LL. We define the $K$-factor as $\frac{d\sigma^{resum}}{dQ}/\frac{d\sigma^{LO}}{dQ}$, where $resum$ represents all the resummed corrections up to NNLO+N$^3$LL. One observes that the perturbative convergence is improved in the case of All exponentiation compared to others although marginally. The $K$ factor defined this way will be useful to directly compare against the experimental results. For All exponentiation case, we find that the $K$-factor is $1.294$ at $Q=100$ at NNLL which changes to $1.286$ at N$^3$LL. The $K$-factor increases with $Q$. At higher $Q=2500$ GeV the $K$-factors  become $1.362$ at NNLL and $1.350$ at N$^3$LL.

\begin{figure}[ht]
\centerline{
\includegraphics[width=16.5cm, height=12.0cm]{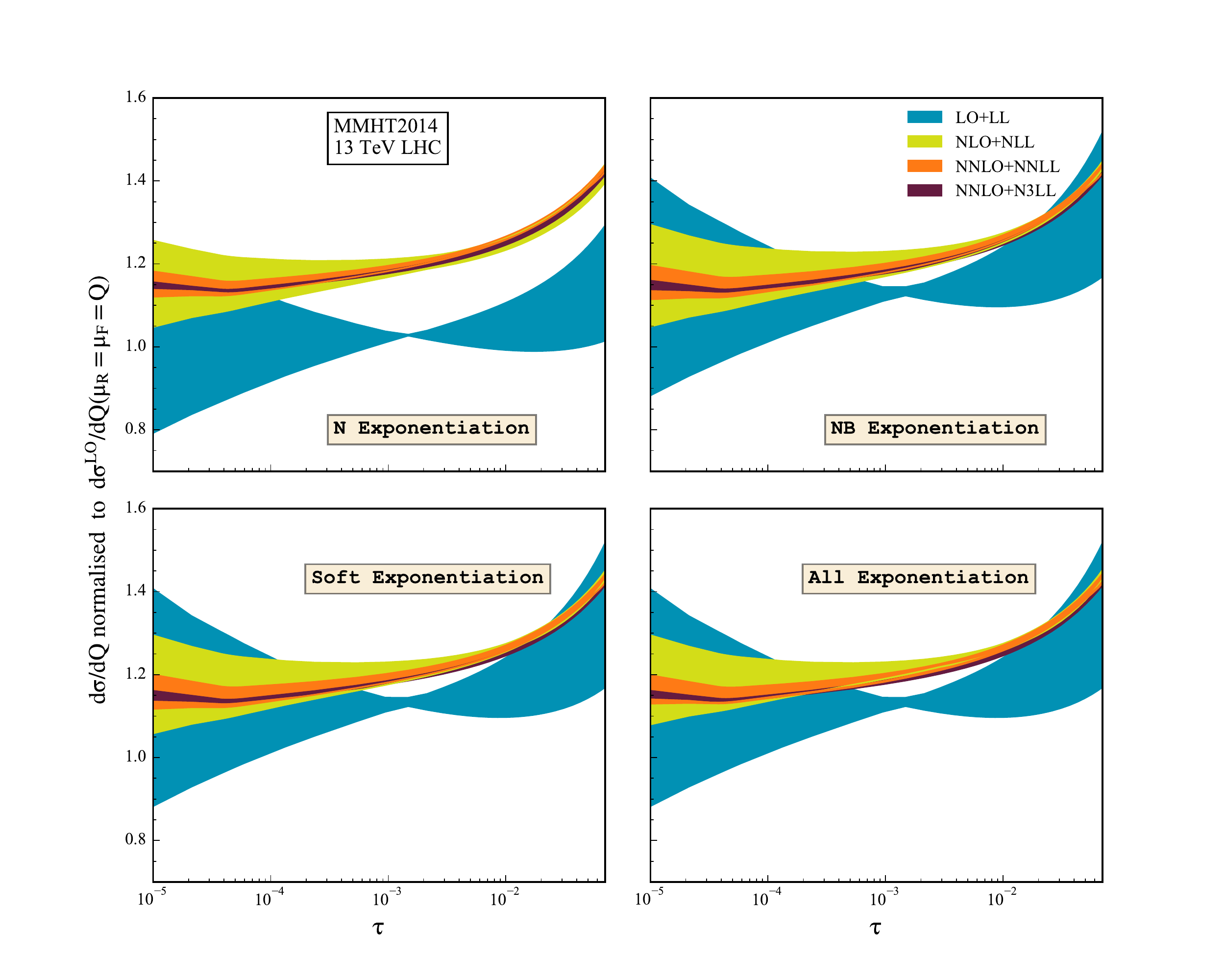}
}
\vspace{-2mm}
\caption{\small{Renormalisation and factorization scale uncertainties have been estimated through 7-point scale variation around the central scale choice $(\mur,\muf) = (1,1)Q$}.}
\label{fig:scaleresum}
\end{figure}
Next we study the uncertainties resulting from unphysical scale in these approaches. We follow the canonical variation of $\mu_f$ and $\mu_r$ around the final state invariant mass $Q$ within $[1/2,2]Q$ imposing additional constraint $1/2 \leq \mur/\muf \leq 2$ as was done in the third order SV prediction in the previous section.  We notice that different approaches for resummation provide a systematic scale reduction at lower invariant mass of the di-lepton pair. For example, in the Standard $\Nb$ case, the scale uncertainty reduces from $13.37\%$ at NLO+NLL to $1.99\%$ at NNLO+NNLL and $0.56\%$ at N$^3$LO$_{sv}$+ N$^3$LL.  Similar pattern is seen for the Soft and All exponentiation as well as seen in \fig{fig:scaleresum}. However, when we compare among themselves, we see that in the case of All exponentiation the scale uncertainty is reduced to $1.65\%$ at NNLO+NNLL compared to $1.99\%$ for $\Nb$ exponentiation and $2.09\%$ for Soft exponentiation at the same order. At the N$^3$LO$_{sv}$+ N$^3$LL, however All exponentiation gives relatively larger scale uncertainty compared to the other two approaches.
At some high invariant mass (say $Q=2500$ GeV), we see a better scale estimate at order NNLO+NNLL where we observe that the scale uncertainty systematically improved from $\Nb$ exponentiation from $0.53\%$ to $0.51\%$ for Soft exponentiation and $0.43\%$ for All exponentiation. However at N$^3$LO$_{sv}$+ N$^3$LL order we observe an over-estimation of scale uncertainty which gets larger for different approaches and can reach the size of NLO scale uncertainty. This shows that the sub-leading regular pieces are also important to capture the scale dependence properly.  This behavior is unlike the Higgs case where one sees a certain scale improvement for exponentiation of complete $\gb_0$.   We will again come back on this discussion at the end of this section.

\begin{figure}[ht]
\centerline{
\includegraphics[width=8.0cm, height=7.0cm]{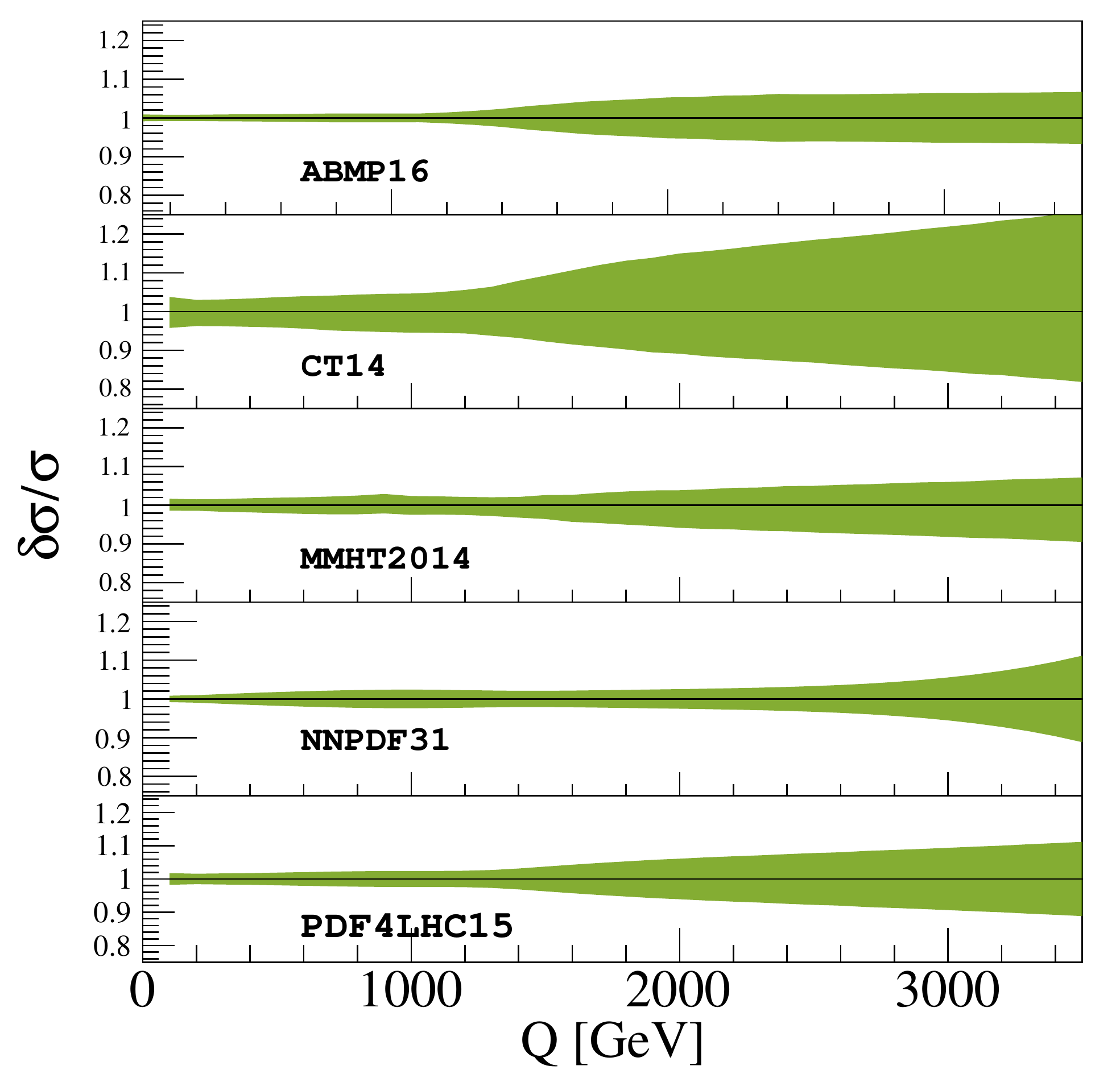}
}
\vspace{-2mm}
\caption{\small{PDF uncertainty has been estimated at NNLO+NNLL level taking $\mur=\muf=Q$. }.}
\label{fig:pdfresum}
\end{figure}
We have also estimated in our resummed predictions the uncertainties from the non-perturbative PDFs.  We convolute the resummed coefficient at N$^3$LL level with $n$ different  sets of a given PDF group and estimate the uncertainty from the {\tt lhapdf} routines.  We use the PDFs provided by {\tt ABMP16} (n= 30) \cite{Alekhin:2016uxn} , {\tt CT14} (n=57) \cite{Dulat:2015mca}, {\tt MMHT2014} (n=51) \cite{Harland-Lang:2014zoa}, {\tt NNPDF31} (n=101) \cite{Ball:2017nwa} and {\tt PDF4LHC15} (n=31) \cite{Butterworth:2015oua} groups. These results are shown in \fig{fig:pdfresum} in terms of $\delta\sigma/\sigma$ where $\delta\sigma$ is the difference between the extrema obtained from $n$ different sets and $\sigma$ is the one obtained from central set $n=0$. These PDF uncertainties in general are found to increase with the invariant mass of the di-lepton pair and, for the range of $Q$ considered here,we find that they are smallest in the low $Q$-region for {\tt AMP16} and are largest for {\tt CT14} case. These uncertainties for $Q=1500$ GeV are found to be $6.14\%$ ({\tt AMBP16}), $16.99\%$ ({\tt CT14}), $6.17\%$ ({\tt MMHT2014}), $4.21\%$ ({\tt NNPDF31}) and $7.43\%$ ({\tt PDF4LHC15}).

\begin{figure}[ht]
\centerline{
\includegraphics[width=7.5cm, height=7.5cm]{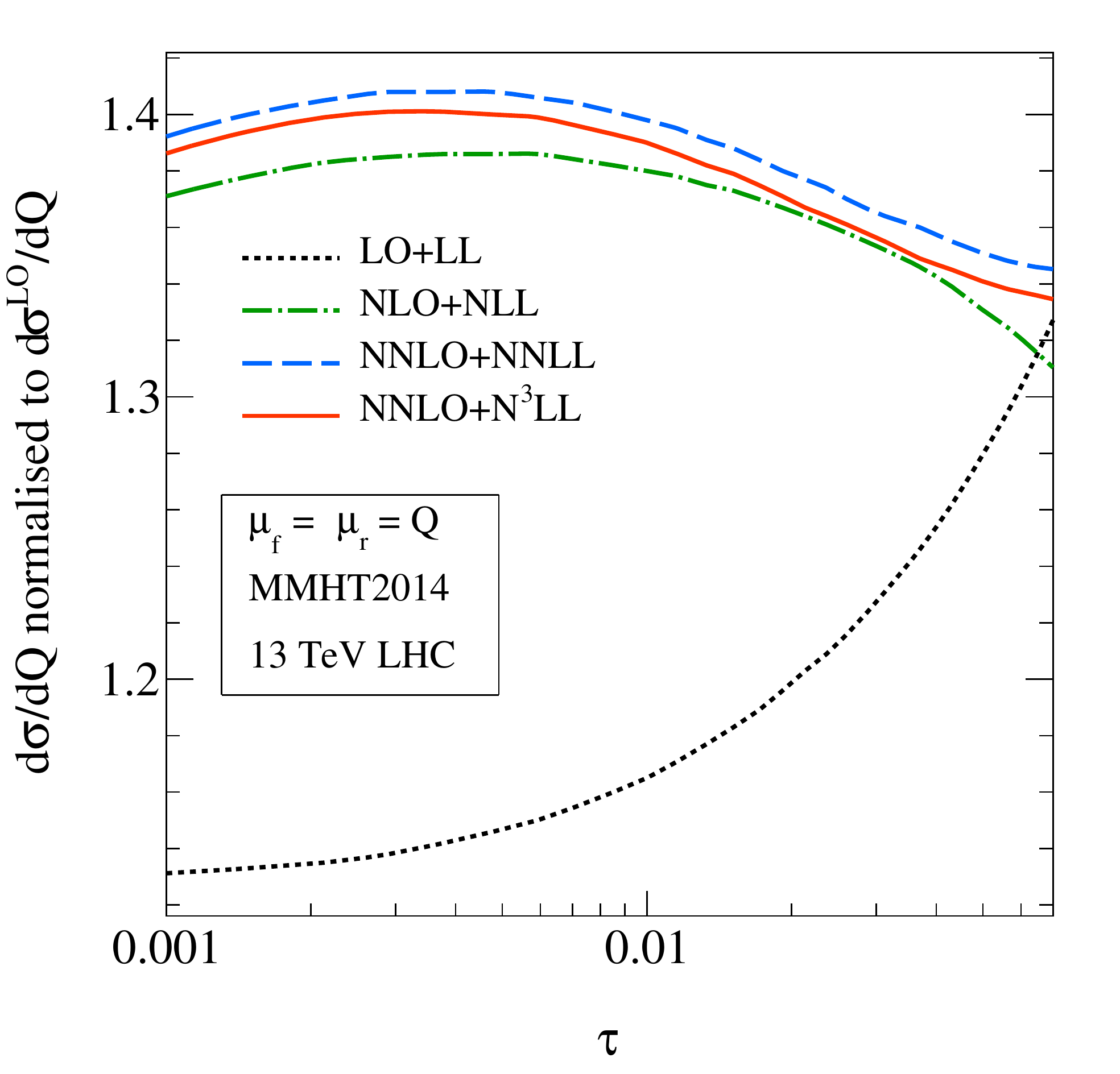}
\includegraphics[width=7.5cm, height=7.5cm]{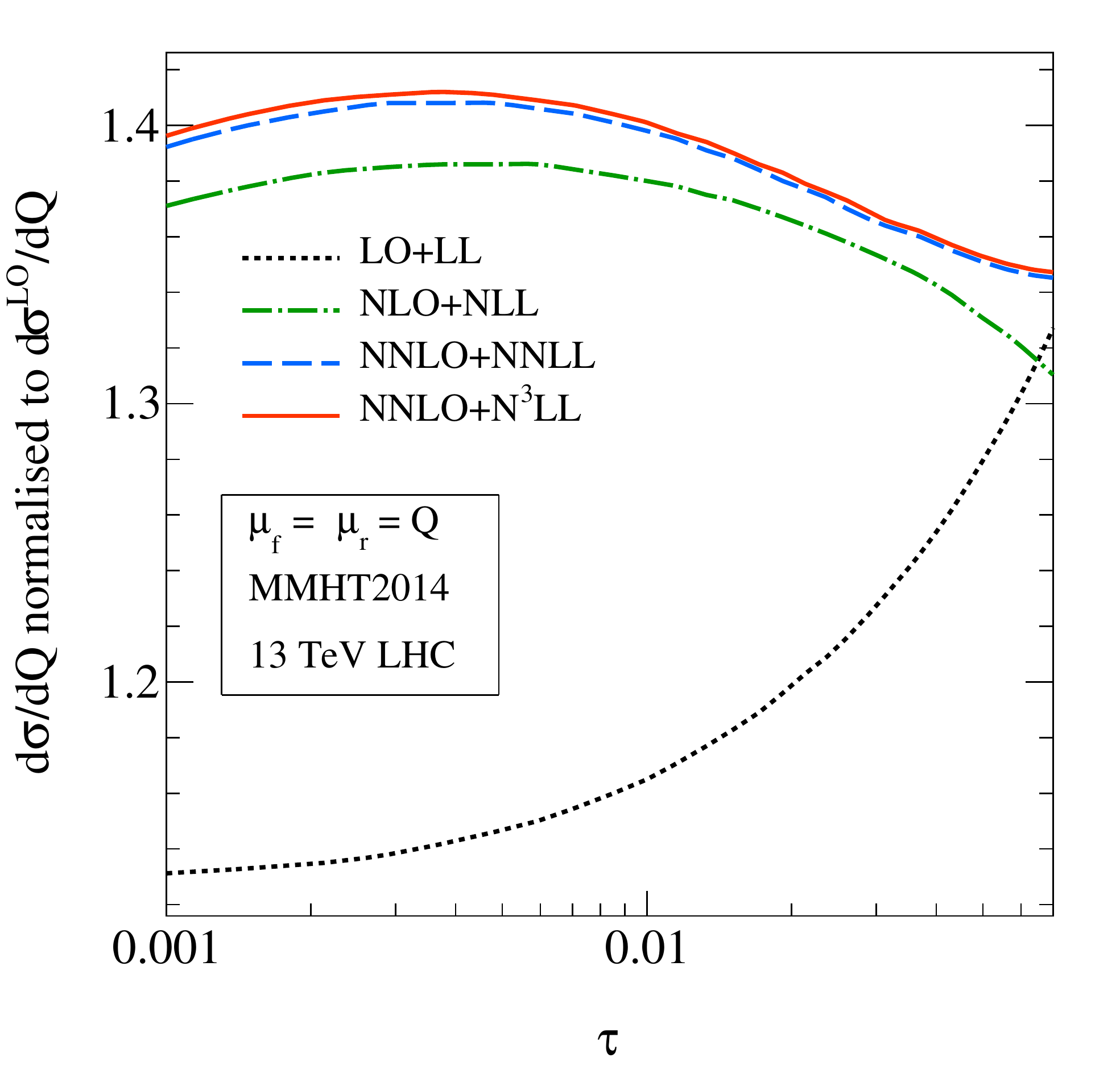}
}
\vspace{-2mm}
\caption{\small{
Comparison between two different way of matching with the fixed order. All scales are set same as $Q$. Left: the matching is done with threshold logarithms kept in distribution space. Right: matching is done with threshold logarithms in Mellin-$N$ space.  }}
\label{fig:matchcompare}
\end{figure}
Finally, we discuss the matching relation presented in \eq{eq:matched}. We notice that the matching relation \eq{eq:matched} can be interpreted in two ways. One can match the N$^3$LO$_{sv}$ fixed order results (with $n=3$) with the resummed results subtracted up to ${\cal O}(a_s^3)$ (with $n=3$) in order to avoid any double counting from the fixed order. So far, we have followed this approach.  Instead we can match the complete NNLO fixed order result  with the resummed result subtracted up to ${\cal O}(a_s^2)$, which also avoids double counting and retains the threshold terms at ${\cal O}(a_s^3)$ in $N$-space in the threshold limit $N\to \infty$. The difference in these two approaches is sub-leading and has to be related with the fact that $N$-space threshold results when transformed back into distribution space produces sub-leading logarithms in addition to the plus distributions. In \fig{fig:matchcompare} we compare these two approaches setting all the scales same as $Q$ in Standard $\Nb$ approach. We see that the threshold terms defined in Mellin-N space provide much better perturbative convergence compared to the $z$-space definition. This is a well-known observation which shows that the sub-leading pieces are also important at this order. As far as scale uncertainty is concerned, this approach gives better estimate of scale uncertainty at N$^3$LL level reducing in some cases by a factor of two, however the general behavior does not change much.

\subsection{Resummed prediction for $Z/W^\pm$ productions}
In this section we present the resummed results for on-shell $Z$ and $W^\pm$ productions to N$^3$LO$_{sv}$+N$^3$LL accuracy. We use 13 TeV as centre of mass energy at the LHC. We set all the parameters same as the previous section. For pdf, we chose the central value from {\tt MMHT2014} set at the corresponding order. At the LHC, the underlying parton fluxes for $W^+$ production are larger than for  $W^-$ case, consequently the production cross sections for the former case are larger than the latter one. This is true also for higher centre of mass energies. In tab.\ (\ref{table3}), (\ref{table1}), (\ref{table2}), we present the central predictions for on-shell $Z$, $W^+$ and $W^-$ respectively with the corresponding percentage scale uncertainties. Note that the scale uncertainties are calculated again using the same procedure i.e. the seven-point scale variation around the central scale which is now the vector boson mass i.e. the central scale has been chosen as $(\mur,\muf) = (1,1) M_V$, with $V=Z$ for $Z$ production and $V=W^\pm$ for $W$-boson production.  In all the cases we observed that the fixed order scale uncertainties are systematically reduced while going to higher orders, however at N$^3$LO$_{sv}$, it again increases which is due to the fact that, at this order there is still missing pieces which are essential to the scale uncertainty. Similar observation is also seen to the matched resummed prediction.
We see that compared to the fixed order, the resummed results provide better perturbative convergence. The scale uncertainty is also seen to improve starting from NNLO level compared to fixed order. 
The resummed $K$-factors as defined before, however increases from NNLO+NNLL to N$^3$LO$_{sv}$+N$^3$LL for all the cases. The absolute size of the perturbative corrections however decreases at N$^3$LO$_{sv}$+N$^3$LL compared to the previous orders confirming the reliability of perturbation theory.
\floatsetup[table]{font=tiny}
\begin{table}
	 \begin{tabular}{|c|c|c|c|c|c|c|c|c|}

\hline
	$\sqrt{S}$ (TeV) & LO & NLO  & NNLO & N$^3$LO$_{sv}$ & LO+LL & NLO+NLL & NNLO+NNLL& N$^3$LO$_{sv}$+N$^3$LL \\

\hline
		$13.00$ & $46.465$  & $57.958$    & $59.379$     & $59.840$
	& $52.829$  & $59.774$  &  $59.666$  & $60.008$\\
			 & ($\pm 13.84\%$) & ($\pm 4.91\%$)   & ($\pm 1.17\%$)    & ($\pm 2.04\%$)
	& ($\pm 14.31\%$) & ($\pm 7.35\%$)  & ($\pm 2.28\%$) & ($\pm 1.38\%$) \\
\hline
\hline
\end{tabular}
\caption{\small{Fixed order (up to N$^3$LO$_{sv}$ ) and resummed (up to N$^3$LO$_{sv}$ + N$^3$LL) cross section (in nb) for on-shell $Z$-boson production at 13 TeV LHC. The scale uncertainty has been estimated using seven-point scale variation around the central scale $(\mur,\muf)=(1,1)M_Z$.}}
\label{table3}
\end{table}

\floatsetup[table]{font=tiny}
\begin{table}
	 \begin{tabular}{|c|c|c|c|c|c|c|c|c|}
\hline
	$\sqrt{S}$ (TeV) & LO & NLO  & NNLO & N$^3$LO$_{sv}$ & LO+LL & NLO+NLL & NNLO+NNLL& N$^3$LO$_{sv}$+N$^3$LL \\

\hline
	$13.00$ & $86.542$  & $107.427$   & $109.454$  & $110.700$ 
	& $98.044$  & $110.700$   &  $109.967$  & $110.638$  \\
		 &  ($\pm 14.34\%$) &  ($\pm 4.41\%$)  &  ($\pm 1.43\%$) &  ($\pm 3.86\%$)
	&  ($\pm 14.8\%$) &  ($\pm 6.86\%$)  &   ($\pm 0.94\%$) &  ($\pm 1.4\%$) \\
\hline
\hline	
\end{tabular}
\caption{\small{The fixed order and resummed cross sections (in nb) for $W^+$ production at LHC for at 13 TeV with corresponding scale uncertainty.}}
\label{table1}
\end{table}

\floatsetup[table]{font=tiny}
\begin{table}
  	 	 \begin{tabular}{|c|c|c|c|c|c|c|c|c|}

\hline
	$\sqrt{S}$ (TeV) & LO & NLO  & NNLO & N$^3$LO$_{sv}$ & LO+LL & NLO+NLL & NNLO+NNLL& N$^3$LO$_{sv}$+N$^3$LL \\

\hline
	$13.00$ & $64.571$  & $79.089$    & $80.441$  & $81.358$  
	        & $73.646$   & $81.719$   &  $80.862$   & $81.365$  \\
	 & ($\pm 14.89\%$) & ($\pm 4.1\%$)   &  ($\pm 2.22\%$) &  ($\pm 4.66\%$) 
	        &  ($\pm 14.36\%$)  &  ($\pm 6.7\%$)   &   ($\pm 1.56\%$)  &  ($\pm 2.19\%$) \\
\hline
\hline
\end{tabular}
\caption{\small{Fixed order and resummed cross section (in nb) for $W^-$ production at LHC for 13 TeV centre of mass energy with corresponding scale uncertainties.}}
\label{table2}
\end{table}

%
\section{Conclusions}\label{sec:conclusion}
We have studied the Drell-Yan production as well as on-shell $Z,W^\pm$ productions in the context of threshold resummation. We have used all the necessary ingredients available to perform this task, in particular the threshold enhanced large-$N$ as well as the $N$-independent constants. The standard threshold resummation heavily reuses the results of the SV cross-section at a particular order. In particular we showed how the the large $N$-independent constants can be found at N$^3$LL level  using the existing SV results. We also explore other possibilities of doing resummation where we exponentiate the complete soft pieces coming from the soft distribution function and also exponentiate the complete $g_0$ coefficients including the form factor. All these different approaches show a systematic behavior of the resummed perturbative series which gets better when more and more terms are being exponentiated in terms of perturbative convergence. We have matched our resummed N$^3$LL results with the existing NNLO(N$^3$LO$_{sv}$) cross-section and presented results for 13 TeV LHC. We observe a systematic decrease of the size of the corrections at the third order.  At this accuracy however the missing regular pieces are also important to tame the scale uncertainty. The results for inclusive DY and $Z,W^\pm$ production demonstrate the ambiguity on exponentiation of 
$N$-independent terms in the resummed results.
\section*{Acknowledgements}
G.D. would like to thank G. Bell, C. Duhr, Y. Ji, L. Magnea, S. Moch and A. Vogt for useful discussions. The algebraic computations have been done with the latest version of the symbolic manipulation system {\sc Form}~\cite{Vermaseren:2000nd,Ruijl:2017dtg}. The research of G.D. is supported partially by DESY postdoctoral fellowship and the DFG within the Collaborative Research Center TRR 257 (``{\it Particle Physics Phenomenology after the Higgs Discovery}'').

\appendix
\section{Soft-Virtual coefficient in $N$-space}\label{app:svn}
The SV coefficient up to three loops are presented here (denoting $\LNb = \ln \Nb$), 

\begin{align} 
\begin{autobreak} 
\gdSVN1 = 
  \LNb^2    \bigg( 2  \A1 \bigg)      
+ \LNb    \bigg( 2  \A1  \Lfr
- 2  \A1  \Lqr
+ 2  \f1 \bigg)      
+ \gb_{01} ,   
\end{autobreak} 
\\ 
\begin{autobreak} 
\gdSVN2 = 
  \LNb^4    \bigg( 2  \A1^2 \bigg)      
+ \LNb^3    \bigg( 4  \A1^2  \Lfr
- 4  \A1^2  \Lqr
+ \frac{4}{3}  \A1  \bt0
+ 4  \A1  \f1 \bigg)      
+ \LNb^2    \bigg( 10  \A1^2  \z2
+ 2  \A1^2  \Lfr^2
- 4  \A1^2  \Lfr  \Lqr
+ 2  \A1^2  \Lqr^2
- 2  \A1  \bt0  \Lqr
- 4  \A1  \B1  \Lfr
+ 4  \A1  \B1  \Lqr
+ 4  \A1  \f1  \Lfr
- 4  \A1  \f1   \Lqr
+ 4  \A1  \Gone1
+ 4  \A1  \Gtone1
+ 2  \A2
+ 2  \bt0  \f1
+ 2  \f1^2 \bigg)      
+ \LNb    \bigg( 10  \A1^2  \z2  \Lfr
- 10  \A1^2  \z2  \Lqr
+ 4  \A1  \bt0  \z2
- \A1  \bt0  \Lfr^2
+ \A1  \bt0  \Lqr^2
+ 10  \A1  \z2  \f1
- 4  \A1  \B1  \Lfr^2
+ 8  \A1  \B1  \Lfr  \Lqr
- 4   \A1  \B1  \Lqr^2
+ 4  \A1  \Gone1  \Lfr
- 4  \A1  \Gone1  \Lqr
+ 4  \A1  \Gtone1  \Lfr
- 4  \A1  \Gtone1   \Lqr
+ 2  \A2  \Lfr
- 2  \A2  \Lqr
- 2  \bt0  \f1  \Lqr
+ 4  \bt0  \Gtone1
- 4  \B1  \f1  \Lfr
+ 4  \B1  \f1  \Lqr
+ 4  \f1  \Gone1
+ 4  \f1  \Gtone1
+ 2  \f2 \bigg)      
+ \gb_{02} ,   
\end{autobreak} 
\\ 
\begin{autobreak} 
\gdSVN3 = 
  \LNb^6    \bigg( \frac{4}{3}  \A1^3 \bigg)      
+ \LNb^5    \bigg( 4  \A1^3  \Lfr
- 4  \A1^3  \Lqr
+ \frac{8}{3}  \A1^2  \bt0
+ 4  \A1^2  \f1 \bigg)      
+ \LNb^4    \bigg( 10  \A1^3  \z2
+ 4  \A1^3  \Lfr^2
- 8  \A1^3  \Lfr  \Lqr
+ 4  \A1^3  \Lqr^2
+ \frac{8}{3}  \A1^2  \bt0  \Lfr
- \frac{20}{3}  \A1^2  \bt0  \Lqr
- 4  \A1^2  \B1  \Lfr
+ 4  \A1^2  \B1   \Lqr
+ 8  \A1^2  \f1  \Lfr
- 8  \A1^2  \f1  \Lqr
+ 4  \A1^2  \Gone1
+ 4  \A1^2  \Gtone1
+ 4  \A1   \A2
+ \frac{4}{3}  \A1  \bt0^2
+ \frac{20}{3}  \A1  \bt0  \f1
+ 4  \A1  \f1^2 \bigg)      
+ \LNb^3    \bigg( 20  \A1^3  \z2  \Lfr
- 20  \A1^3  \z2  \Lqr
+ \frac{4}{3}  \A1^3  \Lfr^3
- 4  \A1^3   \Lfr^2  \Lqr
+ 4  \A1^3  \Lfr  \Lqr^2
- \frac{4}{3}  \A1^3  \Lqr^3
+ \frac{44}{3}  \A1^2  \bt0  \z2
- 2   \A1^2  \bt0  \Lfr^2
- 4  \A1^2  \bt0  \Lfr  \Lqr
+ 6  \A1^2  \bt0  \Lqr^2
+ 20  \A1^2  \z2   \f1
- 8  \A1^2  \B1  \Lfr^2
+ 16  \A1^2  \B1  \Lfr  \Lqr
- 8  \A1^2  \B1  \Lqr^2
+ 4  \A1^2   \f1  \Lfr^2
- 8  \A1^2  \f1  \Lfr  \Lqr
+ 4  \A1^2  \f1  \Lqr^2
+ 8  \A1^2  \Gone1  \Lfr
- 8   \A1^2  \Gone1  \Lqr
+ 8  \A1^2  \Gtone1  \Lfr
- 8  \A1^2  \Gtone1  \Lqr
+ 8  \A1  \A2  \Lfr
- 8  \A1   \A2  \Lqr
- \frac{8}{3}  \A1  \bt0^2  \Lqr
- \frac{8}{3}  \A1  \bt0  \B1  \Lfr
+ \frac{8}{3}  \A1  \bt0  \B1  \Lqr
+ 4  \A1  \bt0  \f1  \Lfr
- 12  \A1  \bt0  \f1  \Lqr
+ \frac{8}{3}  \A1  \bt0  \Gone1
+ \frac{32}{3}  \A1  \bt0   \Gtone1
+ \frac{4}{3}  \A1  \bt1
- 8  \A1  \B1  \f1  \Lfr
+ 8  \A1  \B1  \f1  \Lqr
+ 4  \A1  \f1^2  \Lfr
- 4  \A1  \f1^2  \Lqr
+ 8  \A1  \f1  \Gone1
+ 8  \A1  \f1  \Gtone1
+ 4  \A1  \f2
+ \frac{8}{3}  \A2  \bt0
+ 4  \A2  \f1
+ \frac{8}{3}  \bt0^2  \f1
+ 4  \bt0  \f1^2
+ \frac{4}{3}  \f1^3 \bigg)      
+ \LNb^2    \bigg( 25  \A1^3  \z2^2
+ 10  \A1^3  \z2  \Lfr^2
- 20  \A1^3  \z2  \Lfr  \Lqr
+ 10   \A1^3  \z2  \Lqr^2
+ 8  \A1^2  \bt0  \z2  \Lfr
- 28  \A1^2  \bt0  \z2  \Lqr
+ \frac{16}{3}  \A1^2   \bt0  \z3
- 2  \A1^2  \bt0  \Lfr^3
+ 2  \A1^2  \bt0  \Lfr^2  \Lqr
+ 2  \A1^2  \bt0  \Lfr   \Lqr^2
- 2  \A1^2  \bt0  \Lqr^3
- 20  \A1^2  \z2  \B1  \Lfr
+ 20  \A1^2  \z2  \B1  \Lqr
+ 20   \A1^2  \z2  \f1  \Lfr
- 20  \A1^2  \z2  \f1  \Lqr
+ 20  \A1^2  \z2  \Gone1
+ 20  \A1^2  \z2   \Gtone1
- 4  \A1^2  \B1  \Lfr^3
+ 12  \A1^2  \B1  \Lfr^2  \Lqr
- 12  \A1^2  \B1  \Lfr  \Lqr^2
+ 4  \A1^2  \B1  \Lqr^3
+ 4  \A1^2  \Gone1  \Lfr^2
- 8  \A1^2  \Gone1  \Lfr  \Lqr
+ 4  \A1^2   \Gone1  \Lqr^2
+ 4  \A1^2  \Gtone1  \Lfr^2
- 8  \A1^2  \Gtone1  \Lfr  \Lqr
+ 4  \A1^2  \Gtone1   \Lqr^2
+ 20  \A1  \A2  \z2
+ 4  \A1  \A2  \Lfr^2
- 8  \A1  \A2  \Lfr  \Lqr
+ 4  \A1  \A2  \Lqr^2
+ 8  \A1  \bt0^2  \z2
+ 2  \A1  \bt0^2  \Lqr^2
+ 12  \A1  \bt0  \z2  \B1
+ 28  \A1  \bt0  \z2   \f1
+ 2  \A1  \bt0  \B1  \Lfr^2
+ 4  \A1  \bt0  \B1  \Lfr  \Lqr
- 6  \A1  \bt0  \B1  \Lqr^2
- 2   \A1  \bt0  \f1  \Lfr^2
- 4  \A1  \bt0  \f1  \Lfr  \Lqr
+ 6  \A1  \bt0  \f1  \Lqr^2
- 8  \A1  \bt0   \Gone1  \Lqr
+ 4  \A1  \bt0  \Gone2
+ 8  \A1  \bt0  \Gtone1  \Lfr
- 16  \A1  \bt0  \Gtone1  \Lqr
+ 4   \A1  \bt0  \Gtone2
- 2  \A1  \bt1  \Lqr
+ 10  \A1  \z2  \f1^2
+ 4  \A1  \B1^2  \Lfr^2
- 8  \A1   \B1^2  \Lfr  \Lqr
+ 4  \A1  \B1^2  \Lqr^2
- 8  \A1  \B1  \f1  \Lfr^2
+ 16  \A1  \B1  \f1  \Lfr   \Lqr
- 8  \A1  \B1  \f1  \Lqr^2
- 8  \A1  \B1  \Gone1  \Lfr
+ 8  \A1  \B1  \Gone1  \Lqr
- 8  \A1  \B1   \Gtone1  \Lfr
+ 8  \A1  \B1  \Gtone1  \Lqr
- 4  \A1  \B2  \Lfr
+ 4  \A1  \B2  \Lqr
+ 8  \A1  \f1  \Gone1   \Lfr
- 8  \A1  \f1  \Gone1  \Lqr
+ 8  \A1  \f1  \Gtone1  \Lfr
- 8  \A1  \f1  \Gtone1  \Lqr
+ 4  \A1   \f2  \Lfr
- 4  \A1  \f2  \Lqr
+ 4  \A1  \Gone1^2
+ 8  \A1  \Gone1  \Gtone1
+ 2  \A1  \Gtwo1
+ 4  \A1   \Gtone1^2
+ 2  \A1  \Gttwo1
- 4  \A2  \bt0  \Lqr
- 4  \A2  \B1  \Lfr
+ 4  \A2  \B1  \Lqr
+ 4  \A2   \f1  \Lfr
- 4  \A2  \f1  \Lqr
+ 4  \A2  \Gone1
+ 4  \A2  \Gtone1
+ 2  \A3
- 4  \bt0^2  \f1  \Lqr
+ 8  \bt0^2  \Gtone1
- 4  \bt0  \B1  \f1  \Lfr
+ 4  \bt0  \B1  \f1  \Lqr
- 4  \bt0  \f1^2  \Lqr
+ 4  \bt0  \f1  \Gone1
+ 12  \bt0  \f1  \Gtone1
+ 4  \bt0  \f2
+ 2  \bt1  \f1
- 4  \B1  \f1^2   \Lfr
+ 4  \B1  \f1^2  \Lqr
+ 4  \f1^2  \Gone1
+ 4  \f1^2  \Gtone1
+ 4  \f1  \f2 \bigg)      
+ \LNb    \bigg( 25  \A1^3  \z2^2  \Lfr
- 25  \A1^3  \z2^2  \Lqr
+ 20  \A1^2  \bt0  \z2^2
- 5   \A1^2  \bt0  \z2  \Lfr^2
- 10  \A1^2  \bt0  \z2  \Lfr  \Lqr
+ 15  \A1^2  \bt0  \z2  \Lqr^2
+ \frac{16}{3}  \A1^2  \bt0  \z3  \Lfr
- \frac{16}{3}  \A1^2  \bt0  \z3  \Lqr
+ 25  \A1^2  \z2^2  \f1
- 20   \A1^2  \z2  \B1  \Lfr^2
+ 40  \A1^2  \z2  \B1  \Lfr  \Lqr
- 20  \A1^2  \z2  \B1  \Lqr^2
+ 20   \A1^2  \z2  \Gone1  \Lfr
- 20  \A1^2  \z2  \Gone1  \Lqr
+ 20  \A1^2  \z2  \Gtone1  \Lfr
- 20  \A1^2   \z2  \Gtone1  \Lqr
+ 20  \A1  \A2  \z2  \Lfr
- 20  \A1  \A2  \z2  \Lqr
- 8  \A1  \bt0^2  \z2  \Lqr
+ \frac{32}{3}  \A1  \bt0^2  \z3
+ \frac{2}{3}  \A1  \bt0^2  \Lfr^3
- \frac{2}{3}  \A1  \bt0^2  \Lqr^3
+ 4  \A1   \bt0  \z2  \B1  \Lfr
- 4  \A1  \bt0  \z2  \B1  \Lqr
+ 10  \A1  \bt0  \z2  \f1  \Lfr
- 30  \A1  \bt0   \z2  \f1  \Lqr
+ 8  \A1  \bt0  \z2  \Gone1
+ 28  \A1  \bt0  \z2  \Gtone1
+ \frac{16}{3}  \A1  \bt0  \z3  \f1
+ 4  \A1  \bt0  \B1  \Lfr^3
- 4  \A1  \bt0  \B1  \Lfr^2  \Lqr
- 4  \A1  \bt0  \B1  \Lfr  \Lqr^2
+ 4  \A1  \bt0  \B1  \Lqr^3
- 2  \A1  \bt0  \Gone1  \Lfr^2
- 4  \A1  \bt0  \Gone1  \Lfr  \Lqr
+ 6   \A1  \bt0  \Gone1  \Lqr^2
+ 4  \A1  \bt0  \Gone2  \Lfr
- 4  \A1  \bt0  \Gone2  \Lqr
- 2  \A1  \bt0   \Gtone1  \Lfr^2
- 4  \A1  \bt0  \Gtone1  \Lfr  \Lqr
+ 6  \A1  \bt0  \Gtone1  \Lqr^2
+ 4  \A1  \bt0   \Gtone2  \Lfr
- 4  \A1  \bt0  \Gtone2  \Lqr
+ 4  \A1  \bt1  \z2
- \A1  \bt1  \Lfr^2
+ \A1  \bt1   \Lqr^2
- 20  \A1  \z2  \B1  \f1  \Lfr
+ 20  \A1  \z2  \B1  \f1  \Lqr
+ 20  \A1  \z2  \f1  \Gone1
+ 20  \A1  \z2  \f1  \Gtone1
+ 10  \A1  \z2  \f2
+ 4  \A1  \B1^2  \Lfr^3
- 12  \A1  \B1^2  \Lfr^2   \Lqr
+ 12  \A1  \B1^2  \Lfr  \Lqr^2
- 4  \A1  \B1^2  \Lqr^3
- 8  \A1  \B1  \Gone1  \Lfr^2
+ 16   \A1  \B1  \Gone1  \Lfr  \Lqr
- 8  \A1  \B1  \Gone1  \Lqr^2
- 8  \A1  \B1  \Gtone1  \Lfr^2
+ 16  \A1   \B1  \Gtone1  \Lfr  \Lqr
- 8  \A1  \B1  \Gtone1  \Lqr^2
- 4  \A1  \B2  \Lfr^2
+ 8  \A1  \B2  \Lfr   \Lqr
- 4  \A1  \B2  \Lqr^2
+ 4  \A1  \Gone1^2  \Lfr
- 4  \A1  \Gone1^2  \Lqr
+ 8  \A1  \Gone1  \Gtone1   \Lfr
- 8  \A1  \Gone1  \Gtone1  \Lqr
+ 2  \A1  \Gtwo1  \Lfr
- 2  \A1  \Gtwo1  \Lqr
+ 4  \A1  \Gtone1^2   \Lfr
- 4  \A1  \Gtone1^2  \Lqr
+ 2  \A1  \Gttwo1  \Lfr
- 2  \A1  \Gttwo1  \Lqr
+ 8  \A2  \bt0  \z2
- 2  \A2  \bt0  \Lfr^2
+ 2  \A2  \bt0  \Lqr^2
+ 10  \A2  \z2  \f1
- 4  \A2  \B1  \Lfr^2
+ 8   \A2  \B1  \Lfr  \Lqr
- 4  \A2  \B1  \Lqr^2
+ 4  \A2  \Gone1  \Lfr
- 4  \A2  \Gone1  \Lqr
+ 4  \A2   \Gtone1  \Lfr
- 4  \A2  \Gtone1  \Lqr
+ 2  \A3  \Lfr
- 2  \A3  \Lqr
+ 8  \bt0^2  \z2  \f1
+ 2   \bt0^2  \f1  \Lqr^2
- 8  \bt0^2  \Gtone1  \Lqr
+ 8  \bt0^2  \Gtone2
+ 12  \bt0  \z2  \B1  \f1
+ 10  \bt0  \z2  \f1^2
+ 2  \bt0  \B1  \f1  \Lfr^2
+ 4  \bt0  \B1  \f1  \Lfr  \Lqr
- 6  \bt0  \B1   \f1  \Lqr^2
- 8  \bt0  \B1  \Gtone1  \Lfr
+ 8  \bt0  \B1  \Gtone1  \Lqr
- 8  \bt0  \f1  \Gone1  \Lqr
+ 4  \bt0  \f1  \Gone2
- 8  \bt0  \f1  \Gtone1  \Lqr
+ 4  \bt0  \f1  \Gtone2
- 4  \bt0  \f2  \Lqr
+ 8  \bt0  \Gone1  \Gtone1
+ 8  \bt0  \Gtone1^2
+ 4  \bt0  \Gttwo1
- 2  \bt1  \f1  \Lqr
+ 4  \bt1   \Gtone1
+ 4  \B1^2  \f1  \Lfr^2
- 8  \B1^2  \f1  \Lfr  \Lqr
+ 4  \B1^2  \f1  \Lqr^2
- 8  \B1   \f1  \Gone1  \Lfr
+ 8  \B1  \f1  \Gone1  \Lqr
- 8  \B1  \f1  \Gtone1  \Lfr
+ 8  \B1  \f1  \Gtone1  \Lqr
- 4  \B1  \f2  \Lfr
+ 4  \B1  \f2  \Lqr
- 4  \B2  \f1  \Lfr
+ 4  \B2  \f1  \Lqr
+ 4  \f1  \Gone1^2
+ 8  \f1  \Gone1  \Gtone1
+ 2  \f1  \Gtwo1
+ 4  \f1  \Gtone1^2
+ 2  \f1  \Gttwo1
+ 4  \f2  \Gone1
+ 4  \f2   \Gtone1
+ 2  \f3 \bigg)      
+ \gb_{03}\,.
\end{autobreak} 
\end{align}

The coefficients $\gb_{0i}$ are given in Eq.\ (\ref{eq:g0b}).
\section{Resummed coeficients}
Here we collect $N$-dependent and $N$-independent coefficients for all different prescriptions for resummation.
 
\subsection{Resummation ingredients for the Standard $\overline{N}$ exponentiation}\label{app:NBres}
For the standard $\Nb$ exponentiation we present here the $\Nb$ independent coefficients $g_0$ to three loops in Eq.\ (\ref{eq:gnb}) below

\begin{align} 
\begin{autobreak} 
\gzNt1 =
\bigg[ \Gtone1  ~  \bigg( 2 \bigg)      
+ \Gone1  ~  \bigg( 2 \bigg)      
+ \B1  ~  \bigg( 2 ~ \Lqr
- 2 ~ \Lfr \bigg)      
+ \A1  ~  \bigg( 5 ~ \z2 \bigg)
\bigg],   
\end{autobreak} 
\\ 
\begin{autobreak} 
\gzNt2 =
\bigg[ \Gttwo1  ~  \bigg( 1 \bigg)      
+ \Gtone2  ~  \bigg( 2 ~ \bt0 \bigg)      
+ \Gtone1  ~  \bigg( 
- 2 ~ \bt0 ~ \Lqr \bigg)      
+ \Gtone1^2  ~  \bigg( 2 \bigg)      
+ \Gtwo1  ~  \bigg( 1 \bigg)      
+ \Gone2  ~  \bigg( 2 ~ \bt0 \bigg)      
+ \Gone1  ~  \bigg( 
- 2 ~ \bt0 ~ \Lqr \bigg)      
+ \Gone1 ~ \Gtone1  ~  \bigg( 4 \bigg)      
+ \Gone1^2  ~  \bigg( 2 \bigg)      
+ \f1  ~  \bigg( 5 ~ \bt0 ~ \z2 \bigg)      
+ \B2  ~  \bigg( 2 ~ \Lqr
- 2 ~ \Lfr \bigg)      
+ \B1  ~  \bigg( 
- \bt0 ~ \Lqr^2
+ \bt0 ~ \Lfr^2
+ 6 ~ \bt0 ~ \z2 \bigg)      
+ \B1 ~ \Gtone1  ~  \bigg( 4 ~ \Lqr
- 4 ~ \Lfr \bigg)      
+ \B1 ~ \Gone1  ~  \bigg( 4 ~ \Lqr
- 4 ~ \Lfr \bigg)      
+ \B1^2  ~  \bigg( 2 ~ \Lqr^2
- 4 ~ \Lfr ~ \Lqr
+ 2 ~ \Lfr^2 \bigg)      
+ \A2  ~  \bigg( 5 ~ \z2 \bigg)      
+ \A1  ~  \bigg( \frac{8}{3} ~ \bt0 ~ \z3
- 5 ~ \bt0 ~ \z2 ~ \Lqr \bigg)      
+ \A1 ~ \Gtone1  ~  \bigg( 10 ~ \z2 \bigg)      
+ \A1 ~ \Gone1  ~  \bigg( 10 ~ \z2 \bigg)      
+ \A1 ~ \B1  ~  \bigg( 10 ~ \z2 ~ \Lqr
- 10 ~ \z2 ~ \Lfr \bigg)      
+ \A1^2  ~  \bigg( \frac{25}{2} ~ \z2^2 \bigg)
\bigg],   
\end{autobreak} 
\\ 
\begin{autobreak} 
\gzNt3 =
\bigg[ \Gtthree1  ~  \bigg( \frac{2}{3} \bigg)      
+ \Gttwo2  ~  \bigg( \frac{4}{3} ~ \bt0 \bigg)      
+ \Gttwo1  ~  \bigg( 
- 2 ~ \bt0 ~ \Lqr \bigg)      
+ \Gtone3  ~  \bigg( \frac{8}{3} ~ \bt0^2 \bigg)      
+ \Gtone2  ~  \bigg( \frac{4}{3} ~ \bt1
- 4 ~ \bt0^2 ~ \Lqr \bigg)      
+ \Gtone1  ~  \bigg( 
- 2 ~ \bt1 ~ \Lqr
+ 2 ~ \bt0^2 ~ \Lqr^2
+ 8 ~ \bt0^2 ~ \z2 \bigg)      
+ \Gtone1 ~ \Gttwo1  ~  \bigg( 2 \bigg)      
+ \Gtone1 ~ \Gtone2  ~  \bigg( 4 ~ \bt0 \bigg)      
+ \Gtone1^2  ~  \bigg( 
- 4 ~ \bt0 ~ \Lqr \bigg)      
+ \Gtone1^3  ~  \bigg( \frac{4}{3} \bigg)      
+ \Gthree1  ~  \bigg( \frac{2}{3} \bigg)      
+ \Gtwo2  ~  \bigg( \frac{4}{3} ~ \bt0 \bigg)      
+ \Gtwo1  ~  \bigg( 
- 2 ~ \bt0 ~ \Lqr \bigg)      
+ \Gtwo1 ~ \Gtone1  ~  \bigg( 2 \bigg)      
+ \Gone3  ~  \bigg( \frac{8}{3} ~ \bt0^2 \bigg)      
+ \Gone2  ~  \bigg( \frac{4}{3} ~ \bt1
- 4 ~ \bt0^2 ~ \Lqr \bigg)      
+ \Gone2 ~ \Gtone1  ~  \bigg( 4 ~ \bt0 \bigg)      
+ \Gone1  ~  \bigg( 
- 2 ~ \bt1 ~ \Lqr
+ 2 ~ \bt0^2 ~ \Lqr^2
- 12 ~ \bt0^2 ~ \z2 \bigg)      
+ \Gone1 ~ \Gttwo1  ~  \bigg( 2 \bigg)      
+ \Gone1 ~ \Gtone2  ~  \bigg( 4 ~ \bt0 \bigg)      
+ \Gone1 ~ \Gtone1  ~  \bigg( 
- 8 ~ \bt0 ~ \Lqr \bigg)      
+ \Gone1 ~ \Gtone1^2  ~  \bigg( 4 \bigg)      
+ \Gone1 ~ \Gtwo1  ~  \bigg( 2 \bigg)      
+ \Gone1 ~ \Gone2  ~  \bigg( 4 ~ \bt0 \bigg)      
+ \Gone1^2  ~  \bigg( 
- 4 ~ \bt0 ~ \Lqr \bigg)      
+ \Gone1^2 ~ \Gtone1  ~  \bigg( 4 \bigg)      
+ \Gone1^3  ~  \bigg( \frac{4}{3} \bigg)      
+ \f2  ~  \bigg( 10 ~ \bt0 ~ \z2 \bigg)      
+ \f1  ~  \bigg( 5 ~ \bt1 ~ \z2
+ \frac{16}{3} ~ \bt0^2 ~ \z3
- 10 ~ \bt0^2 ~ \z2 ~ \Lqr \bigg)      
+ \f1 ~ \Gtone1  ~  \bigg( 10 ~ \bt0 ~ \z2 \bigg)      
+ \f1 ~ \Gone1  ~  \bigg( 10 ~ \bt0 ~ \z2 \bigg)      
+ \B3  ~  \bigg( 2 ~ \Lqr
- 2 ~ \Lfr \bigg)      
+ \B2  ~  \bigg( 
- 2 ~ \bt0 ~ \Lqr^2
+ 2 ~ \bt0 ~ \Lfr^2
+ 12 ~ \bt0 ~ \z2 \bigg)      
+ \B2 ~ \Gtone1  ~  \bigg( 4 ~ \Lqr
- 4 ~ \Lfr \bigg)      
+ \B2 ~ \Gone1  ~  \bigg( 4 ~ \Lqr
- 4 ~ \Lfr \bigg)      
+ \B1  ~  \bigg( 
- \bt1 ~ \Lqr^2
+ \bt1 ~ \Lfr^2
+ 6 ~ \bt1 ~ \z2
+ \frac{2}{3} ~ \bt0^2 ~ \Lqr^3
- \frac{2}{3} ~  \bt0^2 ~ \Lfr^3
- 12 ~ \bt0^2 ~ \z2 ~ \Lqr \bigg)      
+ \B1 ~ \Gttwo1  ~  \bigg( 2 ~ \Lqr
- 2 ~ \Lfr \bigg)      
+ \B1 ~ \Gtone2  ~  \bigg( 4 ~ \bt0 ~ \Lqr
- 4 ~ \bt0 ~ \Lfr \bigg)      
+ \B1 ~ \Gtone1  ~  \bigg( 
- 6 ~ \bt0 ~ \Lqr^2
+ 4 ~ \bt0 ~ \Lfr ~ \Lqr
+ 2 ~ \bt0 ~ \Lfr^2
+ 12 ~ \bt0 ~ \z2 \bigg)      
+ \B1 ~ \Gtone1^2  ~  \bigg( 4 ~ \Lqr
- 4 ~ \Lfr \bigg)      
+ \B1 ~ \Gtwo1  ~  \bigg( 2 ~ \Lqr
- 2 ~ \Lfr \bigg)      
+ \B1 ~ \Gone2  ~  \bigg( 4 ~ \bt0 ~ \Lqr
- 4 ~ \bt0 ~ \Lfr \bigg)      
+ \B1 ~ \Gone1  ~  \bigg( 
- 6 ~ \bt0 ~ \Lqr^2
+ 4 ~ \bt0 ~ \Lfr ~ \Lqr
+ 2 ~ \bt0 ~ \Lfr^2
+ 12 ~ \bt0 ~ \z2 \bigg)      
+ \B1 ~ \Gone1 ~ \Gtone1  ~  \bigg( 8 ~ \Lqr
- 8 ~ \Lfr \bigg)      
+ \B1 ~ \Gone1^2  ~  \bigg( 4 ~ \Lqr
- 4 ~ \Lfr \bigg)      
+ \B1 ~ \f1  ~  \bigg( 10 ~ \bt0 ~ \z2 ~ \Lqr
- 10 ~ \bt0 ~ \z2 ~ \Lfr \bigg)      
+ \B1 ~ \B2  ~  \bigg( 4 ~ \Lqr^2
- 8 ~ \Lfr ~ \Lqr
+ 4 ~ \Lfr^2 \bigg)      
+ \B1^2  ~  \bigg( 
- 2 ~ \bt0 ~ \Lqr^3
+ 2 ~ \bt0 ~ \Lfr ~ \Lqr^2
+ 2 ~ \bt0 ~ \Lfr^2 ~ \Lqr
- 2 ~ \bt0 ~  \Lfr^3
+ 12 ~ \bt0 ~ \z2 ~ \Lqr
- 12 ~ \bt0 ~ \z2 ~ \Lfr \bigg)      
+ \B1^2 ~ \Gtone1  ~  \bigg( 4 ~ \Lqr^2
- 8 ~ \Lfr ~ \Lqr
+ 4 ~ \Lfr^2 \bigg)      
+ \B1^2 ~ \Gone1  ~  \bigg( 4 ~ \Lqr^2
- 8 ~ \Lfr ~ \Lqr
+ 4 ~ \Lfr^2 \bigg)      
+ \B1^3  ~  \bigg( \frac{4}{3} ~ \Lqr^3
- 4 ~ \Lfr ~ \Lqr^2
+ 4 ~ \Lfr^2 ~ \Lqr
- \frac{4}{3} ~ \Lfr^3 \bigg)      
+ \A3  ~  \bigg( 5 ~ \z2 \bigg)      
+ \A2  ~  \bigg( \frac{16}{3} ~ \bt0 ~ \z3
- 10 ~ \bt0 ~ \z2 ~ \Lqr \bigg)      
+ \A2 ~ \Gtone1  ~  \bigg( 10 ~ \z2 \bigg)      
+ \A2 ~ \Gone1  ~  \bigg( 10 ~ \z2 \bigg)      
+ \A2 ~ \B1  ~  \bigg( 10 ~ \z2 ~ \Lqr
- 10 ~ \z2 ~ \Lfr \bigg)      
+ \A1  ~  \bigg( \frac{8}{3} ~ \bt1 ~ \z3
- 5 ~ \bt1 ~ \z2 ~ \Lqr
- \frac{16}{3} ~ \bt0^2 ~ \z3 ~ \Lqr
+ 5 ~ \bt0^2 ~ \z2 ~  \Lqr^2
+ \frac{21}{5} ~ \bt0^2 ~ \z2^2 \bigg)      
+ \A1 ~ \Gttwo1  ~  \bigg( 5 ~ \z2 \bigg)      
+ \A1 ~ \Gtone2  ~  \bigg( 10 ~ \bt0 ~ \z2 \bigg)      
+ \A1 ~ \Gtone1  ~  \bigg( \frac{16}{3} ~ \bt0 ~ \z3
- 20 ~ \bt0 ~ \z2 ~ \Lqr \bigg)      
+ \A1 ~ \Gtone1^2  ~  \bigg( 10 ~ \z2 \bigg)      
+ \A1 ~ \Gtwo1  ~  \bigg( 5 ~ \z2 \bigg)      
+ \A1 ~ \Gone2  ~  \bigg( 10 ~ \bt0 ~ \z2 \bigg)      
+ \A1 ~ \Gone1  ~  \bigg( \frac{16}{3} ~ \bt0 ~ \z3
- 20 ~ \bt0 ~ \z2 ~ \Lqr \bigg)      
+ \A1 ~ \Gone1 ~ \Gtone1  ~  \bigg( 20 ~ \z2 \bigg)      
+ \A1 ~ \Gone1^2  ~  \bigg( 10 ~ \z2 \bigg)      
+ \A1 ~ \f1  ~  \bigg( 25 ~ \bt0 ~ \z2^2 \bigg)      
+ \A1 ~ \B2  ~  \bigg( 10 ~ \z2 ~ \Lqr
- 10 ~ \z2 ~ \Lfr \bigg)      
+ \A1 ~ \B1  ~  \bigg( \frac{16}{3} ~ \bt0 ~ \z3 ~ \Lqr
- \frac{16}{3} ~ \bt0 ~ \z3 ~ \Lfr
- 15 ~ \bt0 ~ \z2 ~ \Lqr^2
+ 10 ~  \bt0 ~ \z2 ~ \Lfr ~ \Lqr
+ 5 ~ \bt0 ~ \z2 ~ \Lfr^2
+ 30 ~ \bt0 ~ \z2^2 \bigg)      
+ \A1 ~ \B1 ~ \Gtone1  ~  \bigg( 20 ~ \z2 ~ \Lqr
- 20 ~ \z2 ~ \Lfr \bigg)      
+ \A1 ~ \B1 ~ \Gone1  ~  \bigg( 20 ~ \z2 ~ \Lqr
- 20 ~ \z2 ~ \Lfr \bigg)      
+ \A1 ~ \B1^2  ~  \bigg( 10 ~ \z2 ~ \Lqr^2
- 20 ~ \z2 ~ \Lfr ~ \Lqr
+ 10 ~ \z2 ~ \Lfr^2 \bigg)      
+ \A1 ~ \A2  ~  \bigg( 25 ~ \z2^2 \bigg)      
+ \A1^2  ~  \bigg( \frac{40}{3} ~ \bt0 ~ \z2 ~ \z3
- 25 ~ \bt0 ~ \z2^2 ~ \Lqr \bigg)      
+ \A1^2 ~ \Gtone1  ~  \bigg( 25 ~ \z2^2 \bigg)      
+ \A1^2 ~ \Gone1  ~  \bigg( 25 ~ \z2^2 \bigg)      
+ \A1^2 ~ \B1  ~  \bigg( 25 ~ \z2^2 ~ \Lqr
- 25 ~ \z2^2 ~ \Lfr \bigg)      
+ \A1^3  ~  \bigg( \frac{125}{6} ~ \z2^3 \bigg)
\bigg], 
\end{autobreak} 
\end{align}\label{eq:g0b}

The resummed exponent as in Eq.\ (\ref{eq:resGNb}) is calculated to the N$^3$LL accuracy and collected below:

All the anomalous dimensions and constants can be found in Appendix \ref{app:B}
\subsection{Resummation ingredients for the Standard $N$ exponentiation}\label{app:Nres}
Below we present the resummed exponent for the Standard $N$-exponent as given in Eq.\ (\ref{eq:resN}).

\begin{align} 
\begin{autobreak} 
\gN1 = ~~~\bigg[ 
\Apo  ~  \bigg\{
+ \wp^{-1} ~ \fo  ~  \bigg( 2 \bigg)      
+ \fo  ~  \bigg( 
- 2 \bigg)      
+ 2 \bigg\} \bigg],  
\end{autobreak} 
\\ 
\begin{autobreak} 
\gN2 =  ~~~ \bigg[  
\Dpo  ~  \bigg\{ \fo  ~  \bigg( \frac{1}{2} \bigg) \bigg\} 
+ \Ap2  ~  \bigg\{ \fo  ~  \bigg( 
- 1 \bigg)      
+ \wp  ~  \bigg( 
- 1 \bigg)\bigg\}      
+ \Apo  ~  \bigg\{ \fo  ~  \bigg( \btp1
- 2 ~ \GE
+ \Lqr \bigg)      
+ \fo^2  ~  \bigg( \frac{1}{2} ~ \btp1 \bigg)      
+ \wp  ~  \bigg( \btp1
+ \Lfr \bigg) \bigg\}  \bigg], 
\end{autobreak} 
\\ 
\begin{autobreak} 
\gN3 = \beta_0 \bigg[  
\Dp2  ~  \bigg\{ \WpbWpp  ~  \bigg( 
- \frac{1}{2} \bigg) \bigg\} 
+ \Dpo  ~  \bigg\{ \fo\onemWpi  ~  \bigg( \frac{1}{2} ~ \btp1 \bigg)      
+ \WpbWpp  ~  \bigg( \frac{1}{2} ~ \btp1
- \GE
+ \frac{1}{2} ~ \Lqr \bigg) \bigg\}      
+ \Ap3  ~  \bigg\{ \varz  ~  \bigg( \frac{1}{2} \bigg) \bigg\}      
+ \Ap2  ~  \bigg\{ \WpbWpp  ~  \bigg( 
- \frac{3}{2} ~ \btp1
+ 2 ~ \GE
- \Lqr \bigg)      
+ \fo ~ \onemWpi  ~  \bigg( 
- \btp1 \bigg)      
+ \wp  ~  \bigg( \frac{1}{2} ~ \btp1
+ \Lfr \bigg) \bigg\}       
+ \Apo  ~  \bigg\{ \WpbWpp  ~  \bigg( \frac{1}{2} ~ \btp2
+ \frac{1}{2} ~ \btp1^2
- 2 ~ \GE ~ \btp1
+ 2 ~ \GE^2
+ \Lqr ~ \btp1
- 2 ~  \Lqr ~ \GE
+ \frac{1}{2} ~ \Lqr^2
+ 2 ~ \z2 \bigg)      
+ \fo  ~  \bigg( \btp2
- 2 ~ \GE ~ \btp1
+ \Lqr ~ \btp1 \bigg)      
+ \fo ~ \WpbWpp  ~  \bigg( \btp1^2
- 2 ~ \GE ~ \btp1
+ \Lqr ~ \btp1 \bigg)      
+ \fo^2 ~ \onemWpi  ~  \bigg( \frac{1}{2} ~ \btp1^2 \bigg)      
+ \wp  ~  \bigg( \frac{1}{2} ~ \btp2
- \frac{1}{2} ~ \btp1^2
- \frac{1}{2} ~ \Lfr^2 \bigg) \bigg\} \bigg], 
\end{autobreak} 
\\ 
\begin{autobreak} 
\gN4 =  \beta_0^2 \bigg[  
\Dp3  ~  \bigg\{ \varo  ~  \bigg( 
- \frac{1}{4} \bigg) \bigg\}      
+ \Dp2  ~  \bigg\{ \varo  ~  \bigg( \frac{1}{4} ~ \btp1
- \GE
+ \frac{1}{2} ~ \Lqr \bigg)      
+ \fo ~ \onemWpi^2  ~  \bigg( \frac{1}{2} ~ \btp1 \bigg) \bigg\}      
+ \Dpo  ~  \bigg\{ \varo  ~  \bigg( 
- \frac{1}{4} ~ \btp2
+ \frac{1}{4} ~ \btp1^2
- \GE^2
+ \Lqr ~ \GE
- \frac{1}{4} ~ \Lqr^2
- \z2 \bigg)      
+ \WpbWpp  ~  \bigg( \frac{1}{2} ~ \btp2
- \frac{1}{2} ~ \btp1^2 \bigg)      
+ \fo ~ \onemWpi^2  ~  \bigg( \GE ~ \btp1
- \frac{1}{2} ~ \Lqr ~ \btp1 \bigg)      
+ \fo^2 ~ \onemWpi^2  ~  \bigg( 
- \frac{1}{4} ~ \btp1^2 \bigg) \bigg\}      
+ \Ap4  ~  \bigg\{ \varo  ~  \bigg( \frac{1}{6} \bigg)      
+ \wp  ~  \bigg( 
- \frac{1}{3} \bigg)  \bigg\} 
+ \Ap3  ~  \bigg\{ \varo  ~  \bigg( 
- \frac{5}{12} ~ \btp1
+ \GE
- \frac{1}{2} ~ \Lqr \bigg)      
+ \fo ~ \onemWpi^2  ~  \bigg( 
- \frac{1}{2} ~ \btp1 \bigg)      
+ \wp  ~  \bigg( \frac{1}{3} ~ \btp1
+ \Lfr \bigg)  \bigg\}      
+ \Ap2  ~  \bigg\{ \varo  ~  \bigg( \frac{1}{3} ~ \btp2
- \frac{1}{12} ~ \btp1^2
- \GE ~ \btp1
+ 2 ~ \GE^2
+ \frac{1}{2} ~ \Lqr ~ \btp1
- 2  ~ \Lqr ~ \GE
+ \frac{1}{2} ~ \Lqr^2
+ 2 ~ \z2 \bigg)      
+ \WpbWpp  ~  \bigg( 
- \btp2
+ \btp1^2 \bigg)      
+ \fo ~ \onemWpi^2  ~  \bigg( \frac{1}{2} ~ \btp1^2
- 2 ~ \GE ~ \btp1
+ \Lqr ~ \btp1 \bigg)      
+ \fo^2 ~ \onemWpi^2  ~  \bigg( \frac{1}{2} ~ \btp1^2 \bigg)      
+ \wp  ~  \bigg( \frac{1}{3} ~ \btp2
- \frac{1}{3} ~ \btp1^2
- \Lfr^2 \bigg) \bigg\}      
+ \Apo  ~  \bigg\{ \varo  ~  \bigg( \frac{1}{12} ~ \btp3
- \frac{5}{12} ~ \btp1 ~ \btp2
+ \frac{1}{3} ~ \btp1^3
+ \GE ~ \btp2
- \GE ~  \btp1^2
+ \frac{4}{3} ~ \GE^3
- \frac{1}{2} ~ \Lqr ~ \btp2
+ \frac{1}{2} ~ \Lqr ~ \btp1^2
- 2 ~ \Lqr ~ \GE^2
+ \Lqr^2 ~ \GE
- \frac{1}{6} ~ \Lqr^3
+ \frac{8}{3} ~ \z3
+ 4 ~ \z2 ~ \GE
- 2 ~ \z2 ~ \Lqr \bigg)      
+ \WpbWpp  ~  \bigg( \btp1 ~ \btp2
- \btp1^3
- 2 ~ \GE ~ \btp2
+ 2 ~ \GE ~ \btp1^2
+ \Lqr ~ \btp2
- \Lqr ~ \btp1^2 \bigg)      
+ \fo  ~  \bigg( \frac{1}{2} ~ \btp3
- \frac{1}{2} ~ \btp1^3 \bigg)      
+ \fo ~ \WpbWpp  ~  \bigg( \btp1 ~ \btp2
- \btp1^3 \bigg)      
+ \fo ~ \onemWpi^2  ~  \bigg( 
- \frac{1}{2} ~ \btp1 ~ \btp2
+ \frac{1}{2} ~ \btp1^3
- 2 ~ \GE^2 ~ \btp1
+ 2 ~ \Lqr ~  \GE ~ \btp1
- \frac{1}{2} ~ \Lqr^2 ~ \btp1
- 2 ~ \z2 ~ \btp1 \bigg)      
+ \fo^2 ~ \onemWpi^2  ~  \bigg( \GE ~ \btp1^2
- \frac{1}{2} ~ \Lqr ~ \btp1^2 \bigg)      
+ \fo^3 ~ \onemWpi^2  ~  \bigg( 
- \frac{1}{6} ~ \btp1^3 \bigg)      
+ \wp  ~  \bigg( \frac{1}{3} ~ \btp3
- \frac{2}{3} ~ \btp1 ~ \btp2
+ \frac{1}{3} ~ \btp1^3
- \frac{1}{2} ~ \Lfr^2 ~ \btp1
+ \frac{1}{3} ~  \Lfr^3 \bigg) \bigg\} \bigg],
\end{autobreak} 
\end{align}

With $\lambda^{'} = (1- \lambda)$, where $\lambda = 2 a_s \btz \ln N $. $ \tilde{A}_i = A_i / \btz^i, \tilde{D}_i = D_i / \btz^i, \, \tilde{\beta}_i = \beta_i / \btz^{i+1}$.
The constants $g_0$ are given by

\begin{align} 
\begin{autobreak} 
\gZ1 =
\bigg[ \Gtone1  ~  \bigg( 2 \bigg)      
+ \Gone1  ~  \bigg( 2 \bigg)      
+ \f1  ~  \bigg( 2 ~ \GE \bigg)      
+ \B1  ~  \bigg( 2 ~ \Lqr
- 2 ~ \Lfr \bigg)      
+ \A1  ~  \bigg( 2 ~ \GE^2
- 2 ~ \Lqr ~ \GE
+ 2 ~ \Lfr ~ \GE
+ 5 ~ \z2 \bigg)
\bigg],   
\end{autobreak} 
\\ 
\begin{autobreak} 
\gZ2 =
\bigg[ \Gttwo1  ~  \bigg( 1 \bigg)      
+ \Gtone2  ~  \bigg( 2 ~ \bt0 \bigg)      
+ \Gtone1  ~  \bigg( 4 ~ \bt0 ~ \GE
- 2 ~ \bt0 ~ \Lqr \bigg)      
+ \Gtone1^2  ~  \bigg( 2 \bigg)      
+ \Gtwo1  ~  \bigg( 1 \bigg)      
+ \Gone2  ~  \bigg( 2 ~ \bt0 \bigg)      
+ \Gone1  ~  \bigg( 
- 2 ~ \bt0 ~ \Lqr \bigg)      
+ \Gone1 ~ \Gtone1  ~  \bigg( 4 \bigg)      
+ \Gone1^2  ~  \bigg( 2 \bigg)      
+ \f2  ~  \bigg( 2 ~ \GE \bigg)      
+ \f1  ~  \bigg( 2 ~ \bt0 ~ \GE^2
- 2 ~ \bt0 ~ \Lqr ~ \GE
+ 5 ~ \bt0 ~ \z2 \bigg)      
+ \f1 ~ \Gtone1  ~  \bigg( 4 ~ \GE \bigg)      
+ \f1 ~ \Gone1  ~  \bigg( 4 ~ \GE \bigg)      
+ \f1^2  ~  \bigg( 2 ~ \GE^2 \bigg)      
+ \B2  ~  \bigg( 2 ~ \Lqr
- 2 ~ \Lfr \bigg)      
+ \B1  ~  \bigg( 
- \bt0 ~ \Lqr^2
+ \bt0 ~ \Lfr^2
+ 6 ~ \bt0 ~ \z2 \bigg)      
+ \B1 ~ \Gtone1  ~  \bigg( 4 ~ \Lqr
- 4 ~ \Lfr \bigg)      
+ \B1 ~ \Gone1  ~  \bigg( 4 ~ \Lqr
- 4 ~ \Lfr \bigg)      
+ \B1 ~ \f1  ~  \bigg( 4 ~ \Lqr ~ \GE
- 4 ~ \Lfr ~ \GE \bigg)      
+ \B1^2  ~  \bigg( 2 ~ \Lqr^2
- 4 ~ \Lfr ~ \Lqr
+ 2 ~ \Lfr^2 \bigg)      
+ \A2  ~  \bigg( 2 ~ \GE^2
- 2 ~ \Lqr ~ \GE
+ 2 ~ \Lfr ~ \GE
+ 5 ~ \z2 \bigg)      
+ \A1  ~  \bigg( \frac{4}{3} ~ \bt0 ~ \GE^3
- 2 ~ \bt0 ~ \Lqr ~ \GE^2
+ \bt0 ~ \Lqr^2 ~ \GE
- \bt0 ~ \Lfr^2 ~ \GE
+ \frac{8}{3} ~ \bt0 ~ \z3
+ 4 ~ \bt0 ~ \z2 ~ \GE
- 5 ~ \bt0 ~ \z2 ~ \Lqr \bigg)      
+ \A1 ~ \Gtone1  ~  \bigg( 4 ~ \GE^2
- 4 ~ \Lqr ~ \GE
+ 4 ~ \Lfr ~ \GE
+ 10 ~ \z2 \bigg)      
+ \A1 ~ \Gone1  ~  \bigg( 4 ~ \GE^2
- 4 ~ \Lqr ~ \GE
+ 4 ~ \Lfr ~ \GE
+ 10 ~ \z2 \bigg)      
+ \A1 ~ \f1  ~  \bigg( 4 ~ \GE^3
- 4 ~ \Lqr ~ \GE^2
+ 4 ~ \Lfr ~ \GE^2
+ 10 ~ \z2 ~ \GE \bigg)      
+ \A1 ~ \B1  ~  \bigg( 4 ~ \Lqr ~ \GE^2
- 4 ~ \Lqr^2 ~ \GE
- 4 ~ \Lfr ~ \GE^2
+ 8 ~ \Lfr ~ \Lqr ~ \GE
- 4 ~  \Lfr^2 ~ \GE
+ 10 ~ \z2 ~ \Lqr
- 10 ~ \z2 ~ \Lfr \bigg)      
+ \A1^2  ~  \bigg( 2 ~ \GE^4
- 4 ~ \Lqr ~ \GE^3
+ 2 ~ \Lqr^2 ~ \GE^2
+ 4 ~ \Lfr ~ \GE^3
- 4 ~ \Lfr ~ \Lqr ~  \GE^2
+ 2 ~ \Lfr^2 ~ \GE^2
+ 10 ~ \z2 ~ \GE^2
- 10 ~ \z2 ~ \Lqr ~ \GE
+ 10 ~ \z2 ~ \Lfr ~ \GE
+ \frac{25}{2}  ~ \z2^2 \bigg)
\bigg],   
\end{autobreak} 
\\ 
\begin{autobreak} 
\gZ3 =
\bigg[ \Gtthree1  ~  \bigg( \frac{2}{3} \bigg)      
+ \Gttwo2  ~  \bigg( \frac{4}{3} ~ \bt0 \bigg)      
+ \Gttwo1  ~  \bigg( 4 ~ \bt0 ~ \GE
- 2 ~ \bt0 ~ \Lqr \bigg)      
+ \Gtone3  ~  \bigg( \frac{8}{3} ~ \bt0^2 \bigg)      
+ \Gtone2  ~  \bigg( \frac{4}{3} ~ \bt1
+ 8 ~ \bt0^2 ~ \GE
- 4 ~ \bt0^2 ~ \Lqr \bigg)      
+ \Gtone1  ~  \bigg( 4 ~ \bt1 ~ \GE
- 2 ~ \bt1 ~ \Lqr
+ 8 ~ \bt0^2 ~ \GE^2
- 8 ~ \bt0^2 ~ \Lqr ~ \GE
+ 2 ~  \bt0^2 ~ \Lqr^2
+ 8 ~ \bt0^2 ~ \z2 \bigg)      
+ \Gtone1 ~ \Gttwo1  ~  \bigg( 2 \bigg)      
+ \Gtone1 ~ \Gtone2  ~  \bigg( 4 ~ \bt0 \bigg)      
+ \Gtone1^2  ~  \bigg( 8 ~ \bt0 ~ \GE
- 4 ~ \bt0 ~ \Lqr \bigg)      
+ \Gtone1^3  ~  \bigg( \frac{4}{3} \bigg)      
+ \Gthree1  ~  \bigg( \frac{2}{3} \bigg)      
+ \Gtwo2  ~  \bigg( \frac{4}{3} ~ \bt0 \bigg)      
+ \Gtwo1  ~  \bigg( 
- 2 ~ \bt0 ~ \Lqr \bigg)      
+ \Gtwo1 ~ \Gtone1  ~  \bigg( 2 \bigg)      
+ \Gone3  ~  \bigg( \frac{8}{3} ~ \bt0^2 \bigg)      
+ \Gone2  ~  \bigg( \frac{4}{3} ~ \bt1
- 4 ~ \bt0^2 ~ \Lqr \bigg)      
+ \Gone2 ~ \Gtone1  ~  \bigg( 4 ~ \bt0 \bigg)      
+ \Gone1  ~  \bigg( 
- 2 ~ \bt1 ~ \Lqr
+ 2 ~ \bt0^2 ~ \Lqr^2
- 12 ~ \bt0^2 ~ \z2 \bigg)      
+ \Gone1 ~ \Gttwo1  ~  \bigg( 2 \bigg)      
+ \Gone1 ~ \Gtone2  ~  \bigg( 4 ~ \bt0 \bigg)      
+ \Gone1 ~ \Gtone1  ~  \bigg( 8 ~ \bt0 ~ \GE
- 8 ~ \bt0 ~ \Lqr \bigg)      
+ \Gone1 ~ \Gtone1^2  ~  \bigg( 4 \bigg)      
+ \Gone1 ~ \Gtwo1  ~  \bigg( 2 \bigg)      
+ \Gone1 ~ \Gone2  ~  \bigg( 4 ~ \bt0 \bigg)      
+ \Gone1^2  ~  \bigg( 
- 4 ~ \bt0 ~ \Lqr \bigg)      
+ \Gone1^2 ~ \Gtone1  ~  \bigg( 4 \bigg)      
+ \Gone1^3  ~  \bigg( \frac{4}{3} \bigg)      
+ \f3  ~  \bigg( 2 ~ \GE \bigg)      
+ \f2  ~  \bigg( 4 ~ \bt0 ~ \GE^2
- 4 ~ \bt0 ~ \Lqr ~ \GE
+ 10 ~ \bt0 ~ \z2 \bigg)      
+ \f2 ~ \Gtone1  ~  \bigg( 4 ~ \GE \bigg)      
+ \f2 ~ \Gone1  ~  \bigg( 4 ~ \GE \bigg)      
+ \f1  ~  \bigg( 2 ~ \bt1 ~ \GE^2
- 2 ~ \bt1 ~ \Lqr ~ \GE
+ 5 ~ \bt1 ~ \z2
+ \frac{8}{3} ~ \bt0^2 ~ \GE^3
- 4 ~  \bt0^2 ~ \Lqr ~ \GE^2
+ 2 ~ \bt0^2 ~ \Lqr^2 ~ \GE
+ \frac{16}{3} ~ \bt0^2 ~ \z3
+ 8 ~ \bt0^2 ~ \z2 ~ \GE
- 10 ~ \bt0^2 ~ \z2 ~ \Lqr \bigg)      
+ \f1 ~ \Gttwo1  ~  \bigg( 2 ~ \GE \bigg)      
+ \f1 ~ \Gtone2  ~  \bigg( 4 ~ \bt0 ~ \GE \bigg)      
+ \f1 ~ \Gtone1  ~  \bigg( 12 ~ \bt0 ~ \GE^2
- 8 ~ \bt0 ~ \Lqr ~ \GE
+ 10 ~ \bt0 ~ \z2 \bigg)      
+ \f1 ~ \Gtone1^2  ~  \bigg( 4 ~ \GE \bigg)      
+ \f1 ~ \Gtwo1  ~  \bigg( 2 ~ \GE \bigg)      
+ \f1 ~ \Gone2  ~  \bigg( 4 ~ \bt0 ~ \GE \bigg)      
+ \f1 ~ \Gone1  ~  \bigg( 4 ~ \bt0 ~ \GE^2
- 8 ~ \bt0 ~ \Lqr ~ \GE
+ 10 ~ \bt0 ~ \z2 \bigg)      
+ \f1 ~ \Gone1 ~ \Gtone1  ~  \bigg( 8 ~ \GE \bigg)      
+ \f1 ~ \Gone1^2  ~  \bigg( 4 ~ \GE \bigg)      
+ \f1 ~ \f2  ~  \bigg( 4 ~ \GE^2 \bigg)      
+ \f1^2  ~  \bigg( 4 ~ \bt0 ~ \GE^3
- 4 ~ \bt0 ~ \Lqr ~ \GE^2
+ 10 ~ \bt0 ~ \z2 ~ \GE \bigg)      
+ \f1^2 ~ \Gtone1  ~  \bigg( 4 ~ \GE^2 \bigg)      
+ \f1^2 ~ \Gone1  ~  \bigg( 4 ~ \GE^2 \bigg)      
+ \f1^3  ~  \bigg( \frac{4}{3} ~ \GE^3 \bigg)      
+ \B3  ~  \bigg( 2 ~ \Lqr
- 2 ~ \Lfr \bigg)      
+ \B2  ~  \bigg( 
- 2 ~ \bt0 ~ \Lqr^2
+ 2 ~ \bt0 ~ \Lfr^2
+ 12 ~ \bt0 ~ \z2 \bigg)      
+ \B2 ~ \Gtone1  ~  \bigg( 4 ~ \Lqr
- 4 ~ \Lfr \bigg)      
+ \B2 ~ \Gone1  ~  \bigg( 4 ~ \Lqr
- 4 ~ \Lfr \bigg)      
+ \B2 ~ \f1  ~  \bigg( 4 ~ \Lqr ~ \GE
- 4 ~ \Lfr ~ \GE \bigg)      
+ \B1  ~  \bigg( 
- \bt1 ~ \Lqr^2
+ \bt1 ~ \Lfr^2
+ 6 ~ \bt1 ~ \z2
+ \frac{2}{3} ~ \bt0^2 ~ \Lqr^3
- \frac{2}{3} ~  \bt0^2 ~ \Lfr^3
- 12 ~ \bt0^2 ~ \z2 ~ \Lqr \bigg)      
+ \B1 ~ \Gttwo1  ~  \bigg( 2 ~ \Lqr
- 2 ~ \Lfr \bigg)      
+ \B1 ~ \Gtone2  ~  \bigg( 4 ~ \bt0 ~ \Lqr
- 4 ~ \bt0 ~ \Lfr \bigg)      
+ \B1 ~ \Gtone1  ~  \bigg( 8 ~ \bt0 ~ \Lqr ~ \GE
- 6 ~ \bt0 ~ \Lqr^2
- 8 ~ \bt0 ~ \Lfr ~ \GE
+ 4 ~ \bt0 ~ \Lfr ~ \Lqr
+ 2 ~ \bt0 ~ \Lfr^2
+ 12 ~ \bt0 ~ \z2 \bigg)      
+ \B1 ~ \Gtone1^2  ~  \bigg( 4 ~ \Lqr
- 4 ~ \Lfr \bigg)      
+ \B1 ~ \Gtwo1  ~  \bigg( 2 ~ \Lqr
- 2 ~ \Lfr \bigg)      
+ \B1 ~ \Gone2  ~  \bigg( 4 ~ \bt0 ~ \Lqr
- 4 ~ \bt0 ~ \Lfr \bigg)      
+ \B1 ~ \Gone1  ~  \bigg( 
- 6 ~ \bt0 ~ \Lqr^2
+ 4 ~ \bt0 ~ \Lfr ~ \Lqr
+ 2 ~ \bt0 ~ \Lfr^2
+ 12 ~ \bt0 ~ \z2 \bigg)      
+ \B1 ~ \Gone1 ~ \Gtone1  ~  \bigg( 8 ~ \Lqr
- 8 ~ \Lfr \bigg)      
+ \B1 ~ \Gone1^2  ~  \bigg( 4 ~ \Lqr
- 4 ~ \Lfr \bigg)      
+ \B1 ~ \f2  ~  \bigg( 4 ~ \Lqr ~ \GE
- 4 ~ \Lfr ~ \GE \bigg)      
+ \B1 ~ \f1  ~  \bigg( 4 ~ \bt0 ~ \Lqr ~ \GE^2
- 6 ~ \bt0 ~ \Lqr^2 ~ \GE
- 4 ~ \bt0 ~ \Lfr ~ \GE^2
+ 4 ~ \bt0 ~  \Lfr ~ \Lqr ~ \GE
+ 2 ~ \bt0 ~ \Lfr^2 ~ \GE
+ 12 ~ \bt0 ~ \z2 ~ \GE
+ 10 ~ \bt0 ~ \z2 ~ \Lqr
- 10 ~ \bt0 ~  \z2 ~ \Lfr \bigg)      
+ \B1 ~ \f1 ~ \Gtone1  ~  \bigg( 8 ~ \Lqr ~ \GE
- 8 ~ \Lfr ~ \GE \bigg)      
+ \B1 ~ \f1 ~ \Gone1  ~  \bigg( 8 ~ \Lqr ~ \GE
- 8 ~ \Lfr ~ \GE \bigg)      
+ \B1 ~ \f1^2  ~  \bigg( 4 ~ \Lqr ~ \GE^2
- 4 ~ \Lfr ~ \GE^2 \bigg)      
+ \B1 ~ \B2  ~  \bigg( 4 ~ \Lqr^2
- 8 ~ \Lfr ~ \Lqr
+ 4 ~ \Lfr^2 \bigg)      
+ \B1^2  ~  \bigg( 
- 2 ~ \bt0 ~ \Lqr^3
+ 2 ~ \bt0 ~ \Lfr ~ \Lqr^2
+ 2 ~ \bt0 ~ \Lfr^2 ~ \Lqr
- 2 ~ \bt0 ~  \Lfr^3
+ 12 ~ \bt0 ~ \z2 ~ \Lqr
- 12 ~ \bt0 ~ \z2 ~ \Lfr \bigg)      
+ \B1^2 ~ \Gtone1  ~  \bigg( 4 ~ \Lqr^2
- 8 ~ \Lfr ~ \Lqr
+ 4 ~ \Lfr^2 \bigg)      
+ \B1^2 ~ \Gone1  ~  \bigg( 4 ~ \Lqr^2
- 8 ~ \Lfr ~ \Lqr
+ 4 ~ \Lfr^2 \bigg)      
+ \B1^2 ~ \f1  ~  \bigg( 4 ~ \Lqr^2 ~ \GE
- 8 ~ \Lfr ~ \Lqr ~ \GE
+ 4 ~ \Lfr^2 ~ \GE \bigg)      
+ \B1^3  ~  \bigg( \frac{4}{3} ~ \Lqr^3
- 4 ~ \Lfr ~ \Lqr^2
+ 4 ~ \Lfr^2 ~ \Lqr
- \frac{4}{3} ~ \Lfr^3 \bigg)      
+ \A3  ~  \bigg( 2 ~ \GE^2
- 2 ~ \Lqr ~ \GE
+ 2 ~ \Lfr ~ \GE
+ 5 ~ \z2 \bigg)      
+ \A2  ~  \bigg( \frac{8}{3} ~ \bt0 ~ \GE^3
- 4 ~ \bt0 ~ \Lqr ~ \GE^2
+ 2 ~ \bt0 ~ \Lqr^2 ~ \GE
- 2 ~ \bt0 ~ \Lfr^2 ~  \GE
+ \frac{16}{3} ~ \bt0 ~ \z3
+ 8 ~ \bt0 ~ \z2 ~ \GE
- 10 ~ \bt0 ~ \z2 ~ \Lqr \bigg)      
+ \A2 ~ \Gtone1  ~  \bigg( 4 ~ \GE^2
- 4 ~ \Lqr ~ \GE
+ 4 ~ \Lfr ~ \GE
+ 10 ~ \z2 \bigg)      
+ \A2 ~ \Gone1  ~  \bigg( 4 ~ \GE^2
- 4 ~ \Lqr ~ \GE
+ 4 ~ \Lfr ~ \GE
+ 10 ~ \z2 \bigg)      
+ \A2 ~ \f1  ~  \bigg( 4 ~ \GE^3
- 4 ~ \Lqr ~ \GE^2
+ 4 ~ \Lfr ~ \GE^2
+ 10 ~ \z2 ~ \GE \bigg)      
+ \A2 ~ \B1  ~  \bigg( 4 ~ \Lqr ~ \GE^2
- 4 ~ \Lqr^2 ~ \GE
- 4 ~ \Lfr ~ \GE^2
+ 8 ~ \Lfr ~ \Lqr ~ \GE
- 4 ~  \Lfr^2 ~ \GE
+ 10 ~ \z2 ~ \Lqr
- 10 ~ \z2 ~ \Lfr \bigg)      
+ \A1  ~  \bigg( \frac{4}{3} ~ \bt1 ~ \GE^3
- 2 ~ \bt1 ~ \Lqr ~ \GE^2
+ \bt1 ~ \Lqr^2 ~ \GE
- \bt1 ~ \Lfr^2 ~ \GE
+ \frac{8}{3} ~ \bt1 ~ \z3
+ 4 ~ \bt1 ~ \z2 ~ \GE
- 5 ~ \bt1 ~ \z2 ~ \Lqr
+ \frac{4}{3} ~ \bt0^2 ~ \GE^4
- \frac{8}{3} ~ \bt0^2 ~  \Lqr ~ \GE^3
+ 2 ~ \bt0^2 ~ \Lqr^2 ~ \GE^2
- \frac{2}{3} ~ \bt0^2 ~ \Lqr^3 ~ \GE
+ \frac{2}{3} ~ \bt0^2 ~ \Lfr^3 ~  \GE
+ \frac{32}{3} ~ \bt0^2 ~ \z3 ~ \GE
- \frac{16}{3} ~ \bt0^2 ~ \z3 ~ \Lqr
+ 8 ~ \bt0^2 ~ \z2 ~ \GE^2
- 8 ~ \bt0^2  ~ \z2 ~ \Lqr ~ \GE
+ 5 ~ \bt0^2 ~ \z2 ~ \Lqr^2
+ \frac{21}{5} ~ \bt0^2 ~ \z2^2 \bigg)      
+ \A1 ~ \Gttwo1  ~  \bigg( 2 ~ \GE^2
- 2 ~ \Lqr ~ \GE
+ 2 ~ \Lfr ~ \GE
+ 5 ~ \z2 \bigg)      
+ \A1 ~ \Gtone2  ~  \bigg( 4 ~ \bt0 ~ \GE^2
- 4 ~ \bt0 ~ \Lqr ~ \GE
+ 4 ~ \bt0 ~ \Lfr ~ \GE
+ 10 ~ \bt0 ~ \z2 \bigg)      
+ \A1 ~ \Gtone1  ~  \bigg( \frac{32}{3} ~ \bt0 ~ \GE^3
- 16 ~ \bt0 ~ \Lqr ~ \GE^2
+ 6 ~ \bt0 ~ \Lqr^2 ~ \GE
+ 8 ~ \bt0 ~  \Lfr ~ \GE^2
- 4 ~ \bt0 ~ \Lfr ~ \Lqr ~ \GE
- 2 ~ \bt0 ~ \Lfr^2 ~ \GE
+ \frac{16}{3} ~ \bt0 ~ \z3
+ 28 ~ \bt0 ~  \z2 ~ \GE
- 20 ~ \bt0 ~ \z2 ~ \Lqr \bigg)      
+ \A1 ~ \Gtone1^2  ~  \bigg( 4 ~ \GE^2
- 4 ~ \Lqr ~ \GE
+ 4 ~ \Lfr ~ \GE
+ 10 ~ \z2 \bigg)      
+ \A1 ~ \Gtwo1  ~  \bigg( 2 ~ \GE^2
- 2 ~ \Lqr ~ \GE
+ 2 ~ \Lfr ~ \GE
+ 5 ~ \z2 \bigg)      
+ \A1 ~ \Gone2  ~  \bigg( 4 ~ \bt0 ~ \GE^2
- 4 ~ \bt0 ~ \Lqr ~ \GE
+ 4 ~ \bt0 ~ \Lfr ~ \GE
+ 10 ~ \bt0 ~ \z2 \bigg)      
+ \A1 ~ \Gone1  ~  \bigg( \frac{8}{3} ~ \bt0 ~ \GE^3
- 8 ~ \bt0 ~ \Lqr ~ \GE^2
+ 6 ~ \bt0 ~ \Lqr^2 ~ \GE
- 4 ~ \bt0 ~ \Lfr  ~ \Lqr ~ \GE
- 2 ~ \bt0 ~ \Lfr^2 ~ \GE
+ \frac{16}{3} ~ \bt0 ~ \z3
+ 8 ~ \bt0 ~ \z2 ~ \GE
- 20 ~ \bt0 ~ \z2 ~ \Lqr \bigg)      
+ \A1 ~ \Gone1 ~ \Gtone1  ~  \bigg( 8 ~ \GE^2
- 8 ~ \Lqr ~ \GE
+ 8 ~ \Lfr ~ \GE
+ 20 ~ \z2 \bigg)      
+ \A1 ~ \Gone1^2  ~  \bigg( 4 ~ \GE^2
- 4 ~ \Lqr ~ \GE
+ 4 ~ \Lfr ~ \GE
+ 10 ~ \z2 \bigg)      
+ \A1 ~ \f2  ~  \bigg( 4 ~ \GE^3
- 4 ~ \Lqr ~ \GE^2
+ 4 ~ \Lfr ~ \GE^2
+ 10 ~ \z2 ~ \GE \bigg)      
+ \A1 ~ \f1  ~  \bigg( \frac{20}{3} ~ \bt0 ~ \GE^4
- 12 ~ \bt0 ~ \Lqr ~ \GE^3
+ 6 ~ \bt0 ~ \Lqr^2 ~ \GE^2
+ 4 ~ \bt0 ~  \Lfr ~ \GE^3
- 4 ~ \bt0 ~ \Lfr ~ \Lqr ~ \GE^2
- 2 ~ \bt0 ~ \Lfr^2 ~ \GE^2
+ \frac{16}{3} ~ \bt0 ~ \z3 ~ \GE
+ 28 ~ \bt0 ~ \z2 ~ \GE^2
- 30 ~ \bt0 ~ \z2 ~ \Lqr ~ \GE
+ 10 ~ \bt0 ~ \z2 ~ \Lfr ~ \GE
+ 25 ~ \bt0 ~ \z2^2 \bigg)      
+ \A1 ~ \f1 ~ \Gtone1  ~  \bigg( 8 ~ \GE^3
- 8 ~ \Lqr ~ \GE^2
+ 8 ~ \Lfr ~ \GE^2
+ 20 ~ \z2 ~ \GE \bigg)      
+ \A1 ~ \f1 ~ \Gone1  ~  \bigg( 8 ~ \GE^3
- 8 ~ \Lqr ~ \GE^2
+ 8 ~ \Lfr ~ \GE^2
+ 20 ~ \z2 ~ \GE \bigg)      
+ \A1 ~ \f1^2  ~  \bigg( 4 ~ \GE^4
- 4 ~ \Lqr ~ \GE^3
+ 4 ~ \Lfr ~ \GE^3
+ 10 ~ \z2 ~ \GE^2 \bigg)      
+ \A1 ~ \B2  ~  \bigg( 4 ~ \Lqr ~ \GE^2
- 4 ~ \Lqr^2 ~ \GE
- 4 ~ \Lfr ~ \GE^2
+ 8 ~ \Lfr ~ \Lqr ~ \GE
- 4 ~  \Lfr^2 ~ \GE
+ 10 ~ \z2 ~ \Lqr
- 10 ~ \z2 ~ \Lfr \bigg)      
+ \A1 ~ \B1  ~  \bigg( \frac{8}{3} ~ \bt0 ~ \Lqr ~ \GE^3
- 6 ~ \bt0 ~ \Lqr^2 ~ \GE^2
+ 4 ~ \bt0 ~ \Lqr^3 ~ \GE
- \frac{8}{3} ~  \bt0 ~ \Lfr ~ \GE^3
+ 4 ~ \bt0 ~ \Lfr ~ \Lqr ~ \GE^2
- 4 ~ \bt0 ~ \Lfr ~ \Lqr^2 ~ \GE
+ 2 ~ \bt0 ~ \Lfr^2 ~  \GE^2
- 4 ~ \bt0 ~ \Lfr^2 ~ \Lqr ~ \GE
+ 4 ~ \bt0 ~ \Lfr^3 ~ \GE
+ \frac{16}{3} ~ \bt0 ~ \z3 ~ \Lqr
- \frac{16}{3} ~  \bt0 ~ \z3 ~ \Lfr
+ 12 ~ \bt0 ~ \z2 ~ \GE^2
- 4 ~ \bt0 ~ \z2 ~ \Lqr ~ \GE
- 15 ~ \bt0 ~ \z2 ~ \Lqr^2
+ 4 ~  \bt0 ~ \z2 ~ \Lfr ~ \GE
+ 10 ~ \bt0 ~ \z2 ~ \Lfr ~ \Lqr
+ 5 ~ \bt0 ~ \z2 ~ \Lfr^2
+ 30 ~ \bt0 ~ \z2^2 \bigg)      
+ \A1 ~ \B1 ~ \Gtone1  ~  \bigg( 8 ~ \Lqr ~ \GE^2
- 8 ~ \Lqr^2 ~ \GE
- 8 ~ \Lfr ~ \GE^2
+ 16 ~ \Lfr ~ \Lqr ~ \GE
- 8 ~ \Lfr^2 ~ \GE
+ 20 ~ \z2 ~ \Lqr
- 20 ~ \z2 ~ \Lfr \bigg)      
+ \A1 ~ \B1 ~ \Gone1  ~  \bigg( 8 ~ \Lqr ~ \GE^2
- 8 ~ \Lqr^2 ~ \GE
- 8 ~ \Lfr ~ \GE^2
+ 16 ~ \Lfr ~ \Lqr ~ \GE
- 8 ~ \Lfr^2 ~ \GE
+ 20 ~ \z2 ~ \Lqr
- 20 ~ \z2 ~ \Lfr \bigg)      
+ \A1 ~ \B1 ~ \f1  ~  \bigg( 8 ~ \Lqr ~ \GE^3
- 8 ~ \Lqr^2 ~ \GE^2
- 8 ~ \Lfr ~ \GE^3
+ 16 ~ \Lfr ~ \Lqr ~ \GE^2
- 8 ~ \Lfr^2 ~ \GE^2
+ 20 ~ \z2 ~ \Lqr ~ \GE
- 20 ~ \z2 ~ \Lfr ~ \GE \bigg)      
+ \A1 ~ \B1^2  ~  \bigg( 4 ~ \Lqr^2 ~ \GE^2
- 4 ~ \Lqr^3 ~ \GE
- 8 ~ \Lfr ~ \Lqr ~ \GE^2
+ 12 ~ \Lfr ~ \Lqr^2  ~ \GE
+ 4 ~ \Lfr^2 ~ \GE^2
- 12 ~ \Lfr^2 ~ \Lqr ~ \GE
+ 4 ~ \Lfr^3 ~ \GE
+ 10 ~ \z2 ~ \Lqr^2
- 20 ~  \z2 ~ \Lfr ~ \Lqr
+ 10 ~ \z2 ~ \Lfr^2 \bigg)      
+ \A1 ~ \A2  ~  \bigg( 4 ~ \GE^4
- 8 ~ \Lqr ~ \GE^3
+ 4 ~ \Lqr^2 ~ \GE^2
+ 8 ~ \Lfr ~ \GE^3
- 8 ~ \Lfr ~ \Lqr  ~ \GE^2
+ 4 ~ \Lfr^2 ~ \GE^2
+ 20 ~ \z2 ~ \GE^2
- 20 ~ \z2 ~ \Lqr ~ \GE
+ 20 ~ \z2 ~ \Lfr ~ \GE
+ 25 ~  \z2^2 \bigg)      
+ \A1^2  ~  \bigg( \frac{8}{3} ~ \bt0 ~ \GE^5
- \frac{20}{3} ~ \bt0 ~ \Lqr ~ \GE^4
+ 6 ~ \bt0 ~ \Lqr^2 ~ \GE^3
- 2 ~ \bt0 ~  \Lqr^3 ~ \GE^2
+ \frac{8}{3} ~ \bt0 ~ \Lfr ~ \GE^4
- 4 ~ \bt0 ~ \Lfr ~ \Lqr ~ \GE^3
+ 2 ~ \bt0 ~ \Lfr ~ \Lqr^2 ~  \GE^2
- 2 ~ \bt0 ~ \Lfr^2 ~ \GE^3
+ 2 ~ \bt0 ~ \Lfr^2 ~ \Lqr ~ \GE^2
- 2 ~ \bt0 ~ \Lfr^3 ~ \GE^2
+ \frac{16}{3} ~ \bt0 ~ \z3 ~ \GE^2
- \frac{16}{3} ~ \bt0 ~ \z3 ~ \Lqr ~ \GE
+ \frac{16}{3} ~ \bt0 ~ \z3 ~ \Lfr ~ \GE
+ \frac{44}{3} ~ \bt0  ~ \z2 ~ \GE^3
- 28 ~ \bt0 ~ \z2 ~ \Lqr ~ \GE^2
+ 15 ~ \bt0 ~ \z2 ~ \Lqr^2 ~ \GE
+ 8 ~ \bt0 ~ \z2 ~ \Lfr ~  \GE^2
- 10 ~ \bt0 ~ \z2 ~ \Lfr ~ \Lqr ~ \GE
- 5 ~ \bt0 ~ \z2 ~ \Lfr^2 ~ \GE
+ \frac{40}{3} ~ \bt0 ~ \z2 ~ \z3
+ 20  ~ \bt0 ~ \z2^2 ~ \GE
- 25 ~ \bt0 ~ \z2^2 ~ \Lqr \bigg)      
+ \A1^2 ~ \Gtone1  ~  \bigg( 4 ~ \GE^4
- 8 ~ \Lqr ~ \GE^3
+ 4 ~ \Lqr^2 ~ \GE^2
+ 8 ~ \Lfr ~ \GE^3
- 8 ~ \Lfr  ~ \Lqr ~ \GE^2
+ 4 ~ \Lfr^2 ~ \GE^2
+ 20 ~ \z2 ~ \GE^2
- 20 ~ \z2 ~ \Lqr ~ \GE
+ 20 ~ \z2 ~ \Lfr ~ \GE
+ 25 ~ \z2^2 \bigg)      
+ \A1^2 ~ \Gone1  ~  \bigg( 4 ~ \GE^4
- 8 ~ \Lqr ~ \GE^3
+ 4 ~ \Lqr^2 ~ \GE^2
+ 8 ~ \Lfr ~ \GE^3
- 8 ~ \Lfr ~  \Lqr ~ \GE^2
+ 4 ~ \Lfr^2 ~ \GE^2
+ 20 ~ \z2 ~ \GE^2
- 20 ~ \z2 ~ \Lqr ~ \GE
+ 20 ~ \z2 ~ \Lfr ~ \GE
+ 25 ~ \z2^2 \bigg)      
+ \A1^2 ~ \f1  ~  \bigg( 4 ~ \GE^5
- 8 ~ \Lqr ~ \GE^4
+ 4 ~ \Lqr^2 ~ \GE^3
+ 8 ~ \Lfr ~ \GE^4
- 8 ~ \Lfr ~  \Lqr ~ \GE^3
+ 4 ~ \Lfr^2 ~ \GE^3
+ 20 ~ \z2 ~ \GE^3
- 20 ~ \z2 ~ \Lqr ~ \GE^2
+ 20 ~ \z2 ~ \Lfr ~  \GE^2
+ 25 ~ \z2^2 ~ \GE \bigg)      
+ \A1^2 ~ \B1  ~  \bigg( 4 ~ \Lqr ~ \GE^4
- 8 ~ \Lqr^2 ~ \GE^3
+ 4 ~ \Lqr^3 ~ \GE^2
- 4 ~ \Lfr ~ \GE^4
+ 16 ~ \Lfr ~ \Lqr ~ \GE^3
- 12 ~ \Lfr ~ \Lqr^2 ~ \GE^2
- 8 ~ \Lfr^2 ~ \GE^3
+ 12 ~ \Lfr^2 ~ \Lqr ~  \GE^2
- 4 ~ \Lfr^3 ~ \GE^2
+ 20 ~ \z2 ~ \Lqr ~ \GE^2
- 20 ~ \z2 ~ \Lqr^2 ~ \GE
- 20 ~ \z2 ~ \Lfr ~  \GE^2
+ 40 ~ \z2 ~ \Lfr ~ \Lqr ~ \GE
- 20 ~ \z2 ~ \Lfr^2 ~ \GE
+ 25 ~ \z2^2 ~ \Lqr
- 25 ~ \z2^2 ~ \Lfr \bigg)      
+ \A1^3  ~  \bigg( \frac{4}{3} ~ \GE^6
- 4 ~ \Lqr ~ \GE^5
+ 4 ~ \Lqr^2 ~ \GE^4
- \frac{4}{3} ~ \Lqr^3 ~ \GE^3
+ 4 ~  \Lfr ~ \GE^5
- 8 ~ \Lfr ~ \Lqr ~ \GE^4
+ 4 ~ \Lfr ~ \Lqr^2 ~ \GE^3
+ 4 ~ \Lfr^2 ~ \GE^4
- 4 ~ \Lfr^2  ~ \Lqr ~ \GE^3
+ \frac{4}{3} ~ \Lfr^3 ~ \GE^3
+ 10 ~ \z2 ~ \GE^4
- 20 ~ \z2 ~ \Lqr ~ \GE^3
+ 10 ~ \z2 ~  \Lqr^2 ~ \GE^2
+ 20 ~ \z2 ~ \Lfr ~ \GE^3
- 20 ~ \z2 ~ \Lfr ~ \Lqr ~ \GE^2
+ 10 ~ \z2 ~ \Lfr^2 ~ \GE^2
+ 25 ~ \z2^2 ~ \GE^2
- 25 ~ \z2^2 ~ \Lqr ~ \GE
+ 25 ~ \z2^2 ~ \Lfr ~ \GE
+ \frac{125}{6} ~ \z2^3 \bigg)
\bigg], 
\end{autobreak} 
\end{align}

\subsection{Resummation ingredients for the Soft exponentiation}\label{app:softres}
In the case for `Soft exponentiation', all the terms coming from the soft function are exponentiated and hence this means all the contribution to the finite (N-independent) piece from the soft function is also being exponentiated. This renders the $g_0$ coefficients of the Standard $\Nb$ threshold and changes also the resumed exponent. We write these changes below in terms of the Standard $\Nb$ threshold exponent and pre-factor, 
\begin{align}
g_1^{\rm Soft} &= \bg_1 \,,  \nn\\
g_2^{\rm Soft} &= \bg_2 + \as ~\Delta^{\rm Soft}_{g_2} \,,  \nn\\
g_3^{\rm Soft} &= \bg_3 + \as^2~ \Delta^{\rm Soft}_{g_3} \,,  \nn\\
g_4^{\rm Soft} &= \bg_4 + \as^3 ~\Delta^{\rm Soft}_{g_4} \,,  \nn\\
\end{align}
where the coefficients $\Delta^{\rm Soft}_{g_i}$ are given as, 

\begin{align} 
\begin{autobreak} 
\gNS1 =
\bigg[ \Gtone1  ~  \bigg( 2 \bigg)      
+ \f1  ~  \bigg( 
- \Lqr \bigg)      
+ \A1  ~  \bigg( \frac{1}{2} ~ \Lqr^2
+ 2 ~ \z2 \bigg)
\bigg],   
\end{autobreak} 
\\ 
\begin{autobreak} 
\gNS2 =
\bigg[ \Gttwo1  ~  \bigg( 1 \bigg)      
+ \Gtone2  ~  \bigg( 2 ~ \bt0 \bigg)      
+ \Gtone1  ~  \bigg( 
- 2 ~ \bt0 ~ \Lqr \bigg)      
+ \f2  ~  \bigg( 
- \Lqr \bigg)      
+ \f1  ~  \bigg( \frac{1}{2} ~ \bt0 ~ \Lqr^2
+ 2 ~ \bt0 ~ \z2 \bigg)      
+ \A2  ~  \bigg( \frac{1}{2} ~ \Lqr^2
+ 2 ~ \z2 \bigg)      
+ \A1  ~  \bigg( 
- \frac{1}{6} ~ \bt0 ~ \Lqr^3
+ \frac{8}{3} ~ \bt0 ~ \z3
- 2 ~ \bt0 ~ \z2 ~ \Lqr \bigg)
\bigg],   
\end{autobreak} 
\\ 
\begin{autobreak} 
\gNS3 =
\bigg[ \Gtthree1  ~  \bigg( \frac{2}{3} \bigg)      
+ \Gttwo2  ~  \bigg( \frac{4}{3} ~ \bt0 \bigg)      
+ \Gttwo1  ~  \bigg( 
- 2 ~ \bt0 ~ \Lqr \bigg)      
+ \Gtone3  ~  \bigg( \frac{8}{3} ~ \bt0^2 \bigg)      
+ \Gtone2  ~  \bigg( \frac{4}{3} ~ \bt1
- 4 ~ \bt0^2 ~ \Lqr \bigg)      
+ \Gtone1  ~  \bigg( 
- 2 ~ \bt1 ~ \Lqr
+ 2 ~ \bt0^2 ~ \Lqr^2
+ 8 ~ \bt0^2 ~ \z2 \bigg)      
+ \f3  ~  \bigg( 
- \Lqr \bigg)      
+ \f2  ~  \bigg( \bt0 ~ \Lqr^2
+ 4 ~ \bt0 ~ \z2 \bigg)      
+ \f1  ~  \bigg( \frac{1}{2} ~ \bt1 ~ \Lqr^2
+ 2 ~ \bt1 ~ \z2
- \frac{1}{3} ~ \bt0^2 ~ \Lqr^3
+ \frac{16}{3} ~ \bt0^2 ~ \z3
- 4  ~ \bt0^2 ~ \z2 ~ \Lqr \bigg)      
+ \A3  ~  \bigg( \frac{1}{2} ~ \Lqr^2
+ 2 ~ \z2 \bigg)      
+ \A2  ~  \bigg( 
- \frac{1}{3} ~ \bt0 ~ \Lqr^3
+ \frac{16}{3} ~ \bt0 ~ \z3
- 4 ~ \bt0 ~ \z2 ~ \Lqr \bigg)      
+ \A1  ~  \bigg( 
- \frac{1}{6} ~ \bt1 ~ \Lqr^3
+ \frac{8}{3} ~ \bt1 ~ \z3
- 2 ~ \bt1 ~ \z2 ~ \Lqr
+ \frac{1}{12} ~ \bt0^2 ~  \Lqr^4
- \frac{16}{3} ~ \bt0^2 ~ \z3 ~ \Lqr
+ 2 ~ \bt0^2 ~ \z2 ~ \Lqr^2
+ \frac{36}{5} ~ \bt0^2 ~ \z2^2 \bigg)
\bigg] \,.
\end{autobreak} 
\end{align}
The $N$-independent constants in the case can be put in the following form:
\begin{align}
g_{01}^{\rm Soft} &= \bg_{01} + \as ~\Delta^{\rm Soft}_{g_{01}} \,,  \nn\\
g_{02}^{\rm Soft} &= \bg_{02} + \as^2~ \Delta^{\rm Soft}_{g_{02}} \,,  \nn\\
g_{03}^{\rm Soft} &= \bg_{03} + \as^3 ~\Delta^{\rm Soft}_{g_{03}} \,,  \nn\\
\end{align}
where the coefficients $\Delta^{\rm Soft}_{g_{0i}}$ are given by,

\begin{align} 
\begin{autobreak} 
\gSOFT1 =
\bigg[ \Gtone1  ~  \bigg( 
- 2 \bigg)      
+ \f1  ~  \bigg( \Lqr \bigg)      
+ \A1  ~  \bigg( 
- \frac{1}{2} ~ \Lqr^2
- 2 ~ \z2 \bigg)
\bigg],   
\end{autobreak} 
\\ 
\begin{autobreak} 
\gSOFT2 =
\bigg[ \Gttwo1  ~  \bigg( 
- 1 \bigg)      
+ \Gtone2  ~  \bigg( 
- 2 ~ \bt0 \bigg)      
+ \Gtone1  ~  \bigg( 2 ~ \bt0 ~ \Lqr \bigg)      
+ \Gtone1^2  ~  \bigg( 
- 2 \bigg)      
+ \Gone1 ~ \Gtone1  ~  \bigg( 
- 4 \bigg)      
+ \f2  ~  \bigg( \Lqr \bigg)      
+ \f1  ~  \bigg( 
- \frac{1}{2} ~ \bt0 ~ \Lqr^2
- 2 ~ \bt0 ~ \z2 \bigg)      
+ \f1 ~ \Gone1  ~  \bigg( 2 ~ \Lqr \bigg)      
+ \f1^2  ~  \bigg( \frac{1}{2} ~ \Lqr^2 \bigg)      
+ \B1 ~ \Gtone1  ~  \bigg( 
- 4 ~ \Lqr
+ 4 ~ \Lfr \bigg)      
+ \B1 ~ \f1  ~  \bigg( 2 ~ \Lqr^2
- 2 ~ \Lfr ~ \Lqr \bigg)      
+ \A2  ~  \bigg( 
- \frac{1}{2} ~ \Lqr^2
- 2 ~ \z2 \bigg)      
+ \A1  ~  \bigg( \frac{1}{6} ~ \bt0 ~ \Lqr^3
- \frac{8}{3} ~ \bt0 ~ \z3
+ 2 ~ \bt0 ~ \z2 ~ \Lqr \bigg)      
+ \A1 ~ \Gtone1  ~  \bigg( 
- 10 ~ \z2 \bigg)      
+ \A1 ~ \Gone1  ~  \bigg( 
- \Lqr^2
- 4 ~ \z2 \bigg)      
+ \A1 ~ \f1  ~  \bigg( 
- \frac{1}{2} ~ \Lqr^3
+ 3 ~ \z2 ~ \Lqr \bigg)      
+ \A1 ~ \B1  ~  \bigg( 
- \Lqr^3
+ \Lfr ~ \Lqr^2
- 4 ~ \z2 ~ \Lqr
+ 4 ~ \z2 ~ \Lfr \bigg)      
+ \A1^2  ~  \bigg( \frac{1}{8} ~ \Lqr^4
- \frac{3}{2} ~ \z2 ~ \Lqr^2
- 8 ~ \z2^2 \bigg)
\bigg],   
\end{autobreak} 
\\ 
\begin{autobreak} 
\gSOFT3 =
\bigg[ \Gtthree1  ~  \bigg( 
- \frac{2}{3} \bigg)      
+ \Gttwo2  ~  \bigg( 
- \frac{4}{3} ~ \bt0 \bigg)      
+ \Gttwo1  ~  \bigg( 2 ~ \bt0 ~ \Lqr \bigg)      
+ \Gtone3  ~  \bigg( 
- \frac{8}{3} ~ \bt0^2 \bigg)      
+ \Gtone2  ~  \bigg( 
- \frac{4}{3} ~ \bt1
+ 4 ~ \bt0^2 ~ \Lqr \bigg)      
+ \Gtone1  ~  \bigg( 2 ~ \bt1 ~ \Lqr
- 2 ~ \bt0^2 ~ \Lqr^2
- 8 ~ \bt0^2 ~ \z2 \bigg)      
+ \Gtone1 ~ \Gttwo1  ~  \bigg( 
- 2 \bigg)      
+ \Gtone1 ~ \Gtone2  ~  \bigg( 
- 4 ~ \bt0 \bigg)      
+ \Gtone1^2  ~  \bigg( 4 ~ \bt0 ~ \Lqr \bigg)      
+ \Gtone1^3  ~  \bigg( 
- \frac{4}{3} \bigg)      
+ \Gtwo1 ~ \Gtone1  ~  \bigg( 
- 2 \bigg)      
+ \Gone2 ~ \Gtone1  ~  \bigg( 
- 4 ~ \bt0 \bigg)      
+ \Gone1 ~ \Gttwo1  ~  \bigg( 
- 2 \bigg)      
+ \Gone1 ~ \Gtone2  ~  \bigg( 
- 4 ~ \bt0 \bigg)      
+ \Gone1 ~ \Gtone1  ~  \bigg( 8 ~ \bt0 ~ \Lqr \bigg)      
+ \Gone1 ~ \Gtone1^2  ~  \bigg( 
- 4 \bigg)      
+ \Gone1^2 ~ \Gtone1  ~  \bigg( 
- 4 \bigg)      
+ \f3  ~  \bigg( \Lqr \bigg)      
+ \f2  ~  \bigg( 
- \bt0 ~ \Lqr^2
- 4 ~ \bt0 ~ \z2 \bigg)      
+ \f2 ~ \Gone1  ~  \bigg( 2 ~ \Lqr \bigg)      
+ \f1  ~  \bigg( 
- \frac{1}{2} ~ \bt1 ~ \Lqr^2
- 2 ~ \bt1 ~ \z2
+ \frac{1}{3} ~ \bt0^2 ~ \Lqr^3
- \frac{16}{3} ~ \bt0^2 ~ \z3
+ 4 ~ \bt0^2 ~ \z2 ~ \Lqr \bigg)      
+ \f1 ~ \Gtone1  ~  \bigg( 
- 10 ~ \bt0 ~ \z2 \bigg)      
+ \f1 ~ \Gtwo1  ~  \bigg( \Lqr \bigg)      
+ \f1 ~ \Gone2  ~  \bigg( 2 ~ \bt0 ~ \Lqr \bigg)      
+ \f1 ~ \Gone1  ~  \bigg( 
- 3 ~ \bt0 ~ \Lqr^2
- 4 ~ \bt0 ~ \z2 \bigg)      
+ \f1 ~ \Gone1^2  ~  \bigg( 2 ~ \Lqr \bigg)      
+ \f1 ~ \f2  ~  \bigg( \Lqr^2 \bigg)      
+ \f1^2  ~  \bigg( 
- \frac{1}{2} ~ \bt0 ~ \Lqr^3
+ 3 ~ \bt0 ~ \z2 ~ \Lqr \bigg)      
+ \f1^2 ~ \Gone1  ~  \bigg( \Lqr^2 \bigg)      
+ \f1^3  ~  \bigg( \frac{1}{6} ~ \Lqr^3 \bigg)      
+ \B2 ~ \Gtone1  ~  \bigg( 
- 4 ~ \Lqr
+ 4 ~ \Lfr \bigg)      
+ \B2 ~ \f1  ~  \bigg( 2 ~ \Lqr^2
- 2 ~ \Lfr ~ \Lqr \bigg)      
+ \B1 ~ \Gttwo1  ~  \bigg( 
- 2 ~ \Lqr
+ 2 ~ \Lfr \bigg)      
+ \B1 ~ \Gtone2  ~  \bigg( 
- 4 ~ \bt0 ~ \Lqr
+ 4 ~ \bt0 ~ \Lfr \bigg)      
+ \B1 ~ \Gtone1  ~  \bigg( 6 ~ \bt0 ~ \Lqr^2
- 4 ~ \bt0 ~ \Lfr ~ \Lqr
- 2 ~ \bt0 ~ \Lfr^2
- 12 ~ \bt0 ~ \z2 \bigg)      
+ \B1 ~ \Gtone1^2  ~  \bigg( 
- 4 ~ \Lqr
+ 4 ~ \Lfr \bigg)      
+ \B1 ~ \Gone1 ~ \Gtone1  ~  \bigg( 
- 8 ~ \Lqr
+ 8 ~ \Lfr \bigg)      
+ \B1 ~ \f2  ~  \bigg( 2 ~ \Lqr^2
- 2 ~ \Lfr ~ \Lqr \bigg)      
+ \B1 ~ \f1  ~  \bigg( 
- 2 ~ \bt0 ~ \Lqr^3
+ \bt0 ~ \Lfr ~ \Lqr^2
+ \bt0 ~ \Lfr^2 ~ \Lqr
+ 2 ~ \bt0 ~ \z2 ~  \Lqr
+ 4 ~ \bt0 ~ \z2 ~ \Lfr \bigg)      
+ \B1 ~ \f1 ~ \Gone1  ~  \bigg( 4 ~ \Lqr^2
- 4 ~ \Lfr ~ \Lqr \bigg)      
+ \B1 ~ \f1^2  ~  \bigg( \Lqr^3
- \Lfr ~ \Lqr^2 \bigg)      
+ \B1^2 ~ \Gtone1  ~  \bigg( 
- 4 ~ \Lqr^2
+ 8 ~ \Lfr ~ \Lqr
- 4 ~ \Lfr^2 \bigg)      
+ \B1^2 ~ \f1  ~  \bigg( 2 ~ \Lqr^3
- 4 ~ \Lfr ~ \Lqr^2
+ 2 ~ \Lfr^2 ~ \Lqr \bigg)      
+ \A3  ~  \bigg( 
- \frac{1}{2} ~ \Lqr^2
- 2 ~ \z2 \bigg)      
+ \A2  ~  \bigg( \frac{1}{3} ~ \bt0 ~ \Lqr^3
- \frac{16}{3} ~ \bt0 ~ \z3
+ 4 ~ \bt0 ~ \z2 ~ \Lqr \bigg)      
+ \A2 ~ \Gtone1  ~  \bigg( 
- 10 ~ \z2 \bigg)      
+ \A2 ~ \Gone1  ~  \bigg( 
- \Lqr^2
- 4 ~ \z2 \bigg)      
+ \A2 ~ \f1  ~  \bigg( 
- \frac{1}{2} ~ \Lqr^3
+ 3 ~ \z2 ~ \Lqr \bigg)      
+ \A2 ~ \B1  ~  \bigg( 
- \Lqr^3
+ \Lfr ~ \Lqr^2
- 4 ~ \z2 ~ \Lqr
+ 4 ~ \z2 ~ \Lfr \bigg)      
+ \A1  ~  \bigg( \frac{1}{6} ~ \bt1 ~ \Lqr^3
- \frac{8}{3} ~ \bt1 ~ \z3
+ 2 ~ \bt1 ~ \z2 ~ \Lqr
- \frac{1}{12} ~ \bt0^2 ~ \Lqr^4
+ \frac{16}{3} ~ \bt0^2 ~ \z3 ~ \Lqr
- 2 ~ \bt0^2 ~ \z2 ~ \Lqr^2
- \frac{36}{5} ~ \bt0^2 ~ \z2^2 \bigg)      
+ \A1 ~ \Gttwo1  ~  \bigg( 
- 5 ~ \z2 \bigg)      
+ \A1 ~ \Gtone2  ~  \bigg( 
- 10 ~ \bt0 ~ \z2 \bigg)      
+ \A1 ~ \Gtone1  ~  \bigg( 
- \frac{16}{3} ~ \bt0 ~ \z3
+ 20 ~ \bt0 ~ \z2 ~ \Lqr \bigg)      
+ \A1 ~ \Gtone1^2  ~  \bigg( 
- 10 ~ \z2 \bigg)      
+ \A1 ~ \Gtwo1  ~  \bigg( 
- \frac{1}{2} ~ \Lqr^2
- 2 ~ \z2 \bigg)      
+ \A1 ~ \Gone2  ~  \bigg( 
- \bt0 ~ \Lqr^2
- 4 ~ \bt0 ~ \z2 \bigg)      
+ \A1 ~ \Gone1  ~  \bigg( \frac{4}{3} ~ \bt0 ~ \Lqr^3
- \frac{16}{3} ~ \bt0 ~ \z3
+ 8 ~ \bt0 ~ \z2 ~ \Lqr \bigg)      
+ \A1 ~ \Gone1 ~ \Gtone1  ~  \bigg( 
- 20 ~ \z2 \bigg)      
+ \A1 ~ \Gone1^2  ~  \bigg( 
- \Lqr^2
- 4 ~ \z2 \bigg)      
+ \A1 ~ \f2  ~  \bigg( 
- \frac{1}{2} ~ \Lqr^3
+ 3 ~ \z2 ~ \Lqr \bigg)      
+ \A1 ~ \f1  ~  \bigg( \frac{5}{12} ~ \bt0 ~ \Lqr^4
- 6 ~ \bt0 ~ \z2 ~ \Lqr^2
- 16 ~ \bt0 ~ \z2^2 \bigg)      
+ \A1 ~ \f1 ~ \Gone1  ~  \bigg( 
- \Lqr^3
+ 6 ~ \z2 ~ \Lqr \bigg)      
+ \A1 ~ \f1^2  ~  \bigg( 
- \frac{1}{4} ~ \Lqr^4
+ \frac{3}{2} ~ \z2 ~ \Lqr^2 \bigg)      
+ \A1 ~ \B2  ~  \bigg( 
- \Lqr^3
+ \Lfr ~ \Lqr^2
- 4 ~ \z2 ~ \Lqr
+ 4 ~ \z2 ~ \Lfr \bigg)      
+ \A1 ~ \B1  ~  \bigg( \frac{5}{6} ~ \bt0 ~ \Lqr^4
- \frac{1}{3} ~ \bt0 ~ \Lfr ~ \Lqr^3
- \frac{1}{2} ~ \bt0 ~ \Lfr^2 ~ \Lqr^2
- \frac{16}{3} ~ \bt0 ~ \z3 ~ \Lqr
+ \frac{16}{3} ~ \bt0 ~ \z3 ~ \Lfr
+ 3 ~ \bt0 ~ \z2 ~ \Lqr^2
- 4 ~ \bt0 ~ \z2 ~ \Lfr ~ \Lqr
- 2 ~ \bt0 ~ \z2 ~ \Lfr^2
- 12 ~ \bt0 ~ \z2^2 \bigg)      
+ \A1 ~ \B1 ~ \Gtone1  ~  \bigg( 
- 20 ~ \z2 ~ \Lqr
+ 20 ~ \z2 ~ \Lfr \bigg)      
+ \A1 ~ \B1 ~ \Gone1  ~  \bigg( 
- 2 ~ \Lqr^3
+ 2 ~ \Lfr ~ \Lqr^2
- 8 ~ \z2 ~ \Lqr
+ 8 ~ \z2 ~ \Lfr \bigg)      
+ \A1 ~ \B1 ~ \f1  ~  \bigg( 
- \Lqr^4
+ \Lfr ~ \Lqr^3
+ 6 ~ \z2 ~ \Lqr^2
- 6 ~ \z2 ~ \Lfr ~ \Lqr \bigg)      
+ \A1 ~ \B1^2  ~  \bigg( 
- \Lqr^4
+ 2 ~ \Lfr ~ \Lqr^3
- \Lfr^2 ~ \Lqr^2
- 4 ~ \z2 ~ \Lqr^2
+ 8 ~ \z2 ~  \Lfr ~ \Lqr
- 4 ~ \z2 ~ \Lfr^2 \bigg)      
+ \A1 ~ \A2  ~  \bigg( \frac{1}{4} ~ \Lqr^4
- 3 ~ \z2 ~ \Lqr^2
- 16 ~ \z2^2 \bigg)      
+ \A1^2  ~  \bigg( 
- \frac{1}{12} ~ \bt0 ~ \Lqr^5
+ 2 ~ \bt0 ~ \z2 ~ \Lqr^3
- \frac{40}{3} ~ \bt0 ~ \z2 ~ \z3
+ 16 ~ \bt0  ~ \z2^2 ~ \Lqr \bigg)      
+ \A1^2 ~ \Gtone1  ~  \bigg( 
- 25 ~ \z2^2 \bigg)      
+ \A1^2 ~ \Gone1  ~  \bigg( \frac{1}{4} ~ \Lqr^4
- 3 ~ \z2 ~ \Lqr^2
- 16 ~ \z2^2 \bigg)      
+ \A1^2 ~ \f1  ~  \bigg( \frac{1}{8} ~ \Lqr^5
- \frac{3}{2} ~ \z2 ~ \Lqr^3
+ \frac{9}{2} ~ \z2^2 ~ \Lqr \bigg)      
+ \A1^2 ~ \B1  ~  \bigg( \frac{1}{4} ~ \Lqr^5
- \frac{1}{4} ~ \Lfr ~ \Lqr^4
- 3 ~ \z2 ~ \Lqr^3
+ 3 ~ \z2 ~ \Lfr ~ \Lqr^2
- 16 ~ \z2^2 ~ \Lqr
+ 16 ~ \z2^2 ~ \Lfr \bigg)      
+ \A1^3  ~  \bigg( 
- \frac{1}{48} ~ \Lqr^6
+ \frac{3}{8} ~ \z2 ~ \Lqr^4
- \frac{9}{4} ~ \z2^2 ~ \Lqr^2
- \frac{49}{3} ~ \z2^3 \bigg)
\bigg]\,. 
\end{autobreak} 
\end{align}

\subsection{Resummation ingredients for the All exponentiation}\label{app:allres}
In the case for `All exponentiation', the complete $g_0$ is being exponentiated along with the large-$N$ pieces. This brings into modification only for the resummed exponent compared to the `Standard $\overline{N}$ exponentiation'. We write the resummed exponent in this case in terms of $\overline{N}$ exponents as,
\begin{align}
g_1^{\rm All} &= \gb_1 \,,  \nn\\
g_2^{\rm All} &= \gb_2 +  \as~ \Delta^{\rm All}_{g_2} \,,  \nn\\
g_3^{\rm All} &= \gb_3 +  \as^2~ \Delta^{\rm All}_{g_3} \,,  \nn\\
g_4^{\rm All} &= \gb_4 +  \as^3 ~\Delta^{\rm All}_{g_4} \,,  \nn\\
\end{align}
where $\Delta^{\rm All}_{g_i}$ terms are found from exponentiating also the complete $g_0$ prefactor and they are given as,

\begin{align}
 \Delta^{\rm All}_{g_2} &= \gb_{01}^{} \,, \nn\\
  \Delta^{\rm All}_{g_2} &=  \bigg(-\frac{\gb_{01}^2}{2} +  \gb_{02}^{} \bigg) \,, \nn\\
  \Delta^{\rm All}_{g_2} &= \bigg(\frac{\gb_{01}^3}{3} - \gb_{01}^{} \gb_{02}^{} +  \gb_{03}^{} \bigg) \,,
\end{align}
where the coefficients $\gb_{0i}$ are given in \ref{eq:g0b}.

\section{Anomalous dimensions} \label{app:anodim}

Here we present all the anomalous dimensions used in performing the resummation.\\
The cusp anomalous dimensions are given as 
 
\begin{align} 
\begin{autobreak} 
\A1 =
\bigg[ \CF  ~  \bigg( 4 \bigg)
\bigg],   
\end{autobreak} 
\\ 
\begin{autobreak} 
\A2 =
\bigg[ \CF ~ \NF  ~  \bigg( 
- \frac{40}{9} \bigg)      
+ \CF ~ \CA  ~  \bigg( \frac{268}{9}
- 8 ~ \z2 \bigg)
\bigg],   
\end{autobreak} 
\\ 
\begin{autobreak} 
\A3 =
\bigg[ \CF ~ \NF^2  ~  \bigg( 
- \frac{16}{27} \bigg)      
+ \CF ~ \CA ~ \NF  ~  \bigg( 
- \frac{836}{27}
- \frac{112}{3} ~ \z3
+ \frac{160}{9} ~ \z2 \bigg)      
+ \CF ~ \CA^2  ~  \bigg( \frac{490}{3}
+ \frac{88}{3} ~ \z3
- \frac{1072}{9} ~ \z2
+ \frac{176}{5} ~ \z2^2 \bigg)      
+ \CF^2 ~ \NF  ~  \bigg( 
- \frac{110}{3}
+ 32 ~ \z3 \bigg)
\bigg], 
\end{autobreak} 
\\
\begin{autobreak} 
\A4 =
\bigg[ \CF  \dFAoNA    \bigg( \frac{7040}{3}  \z5
+ \frac{256}{3}  \z3
- 768  \z3^2
- 256  \z2
- \frac{15872}{35}  \z2^3 \bigg)      
+ \CF  \NF  \dFFoNA    \bigg( 
- \frac{2560}{3}  \z5
- \frac{512}{3}  \z3
+ 512  \z2 \bigg)      
+ \CF  \NF^3    \bigg( 
- \frac{32}{81}
+ \frac{64}{27}  \z3 \bigg)      
+ \CF^2  \NF^2    \bigg( \frac{2392}{81}
- \frac{640}{9}  \z3
+ \frac{64}{5}  \z2^2 \bigg)      
+ \CF^3  \NF    \bigg( \frac{572}{9}
- 320  \z5
+ \frac{592}{3}  \z3 \bigg)      
+ \CA  \CF  \NF^2    \bigg( \frac{923}{81}
+ \frac{2240}{27}  \z3
- \frac{608}{81}  \z2
- \frac{224}{15}  \z2^2 \bigg)      
+ \CA  \CF^2  \NF    \bigg( 
- \frac{34066}{81}
+ 160  \z5
+ \frac{3712}{9}  \z3
+ \frac{440}{3}  \z2
- 128  \z2  \z3
- \frac{352}{5}  \z2^2 \bigg)      
+ \CA^2  \CF  \NF    \bigg( 
- \frac{24137}{81}
+ \frac{2096}{9}  \z5
- \frac{23104}{27}  \z3
+ \frac{20320}{81}  \z2
+ \frac{448}{3}  \z2  \z3
- \frac{352}{15}  \z2^2 \bigg)      
+ \CA^3  \CF    \bigg( \frac{84278}{81}
- \frac{3608}{9}  \z5
+ \frac{20944}{27}  \z3
- 16  \z3^2
- \frac{88400}{81}  \z2
- \frac{352}{3}  \z2  \z3
+ \frac{3608}{5}  \z2^2
- \frac{20032}{105}  \z2^3 \bigg)
\bigg], 
\end{autobreak} 
\end{align}

The quartic casimirs are given by
\begin{align}
\frac{d_F^{abcd}d_A^{abcd}}{N_A} &=  \frac{n_c (n_c^2+6)}{48}, \qquad
\frac{d_F^{abcd}d_F^{abcd}}{N_A} = \frac{(n_c^4 - 6 n_c^2 + 18)}{96n_c^2},
\end{align} 
with $N_A = n_c^2 -1$ and $N_F = n_c$ where $n_c = 3$ for QCD.\\

The universal $D$ coefficients are given as,

\begin{align} 
\begin{autobreak} 
D_1 = {C}_{F}
\bigg[0
\bigg],   
\end{autobreak} 
\\ 
\begin{autobreak} 
D_2 = {C}_{F}
\bigg[ \nf    \bigg( \frac{224}{27}
- \frac{32}{3}  \z2 \bigg)      
+ \Ca    \bigg( 
- \frac{1616}{27}
+ 56  \z3
+ \frac{176}{3}  \z2 \bigg)
\bigg],   
\end{autobreak} 
\\ 
\begin{autobreak} 
D_3 = { C}_{F}
\bigg[ \nf^2    \bigg( 
- \frac{3712}{729}
+ \frac{320}{27}  \z3
+ \frac{640}{27}  \z2 \bigg)      
+ \Cf  \nf    \bigg( \frac{3422}{27}
- \frac{608}{9}  \z3
- 32  \z2
- \frac{64}{5}  \z2^2 \bigg)      
+ \Ca  \nf    \bigg( \frac{125252}{729}
- \frac{2480}{9}  \z3
- \frac{29392}{81}  \z2
+ \frac{736}{15}  \z2^2 \bigg)      
+ \Ca^2    \bigg( 
- \frac{594058}{729}
- 384  \z5
+ \frac{40144}{27}  \z3
+ \frac{98224}{81}  \z2
- \frac{352}{3}   \z2  \z3
- \frac{2992}{15}  \z2^2 \bigg)
\bigg]\,.
\end{autobreak} 
\end{align}

The coefficients $ B$ are given as 

\begin{align} 
\begin{autobreak} 
\B1 =
\bigg[ \CF  ~  \bigg( 3 \bigg)
\bigg],   
\end{autobreak} 
\\ 
\begin{autobreak} 
\B2 =
\bigg[ \CF ~ \NF  ~  \bigg( 
- \frac{1}{3}
- \frac{8}{3} ~ \z2 \bigg)      
+ \CF ~ \CA  ~  \bigg( \frac{17}{6}
- 12 ~ \z3
+ \frac{44}{3} ~ \z2 \bigg)      
+ \CF^2  ~  \bigg( \frac{3}{2}
+ 24 ~ \z3
- 12 ~ \z2 \bigg)
\bigg],   
\end{autobreak} 
\\ 
\begin{autobreak} 
\B3 =
\bigg[ \CF ~ \NF^2  ~  \bigg( 
- \frac{17}{9}
- \frac{16}{9} ~ \z3
+ \frac{80}{27} ~ \z2 \bigg)      
+ \CF ~ \CA ~ \NF  ~  \bigg( 20
+ \frac{200}{9} ~ \z3
- \frac{1336}{27} ~ \z2
+ \frac{4}{5} ~ \z2^2 \bigg)      
+ \CF ~ \CA^2  ~  \bigg( 
- \frac{1657}{36}
+ 40 ~ \z5
- \frac{1552}{9} ~ \z3
+ \frac{4496}{27} ~ \z2
- 2 ~ \z2^2 \bigg)      
+ \CF^2 ~ \NF  ~  \bigg( 
- 23
- \frac{136}{3} ~ \z3
+ \frac{20}{3} ~ \z2
+ \frac{232}{15} ~ \z2^2 \bigg)      
+ \CF^2 ~ \CA  ~  \bigg( \frac{151}{4}
+ 120 ~ \z5
+ \frac{844}{3} ~ \z3
- \frac{410}{3} ~ \z2
+ 16 ~ \z2 ~ \z3
- \frac{988}{15} ~  \z2^2 \bigg)      
+ \CF^3  ~  \bigg( \frac{29}{2}
- 240 ~ \z5
+ 68 ~ \z3
+ 18 ~ \z2
- 32 ~ \z2 ~ \z3
+ \frac{288}{5} ~ \z2^2 \bigg)
\bigg], 
\end{autobreak} 
\end{align}
The anomalous dimensions $f$ are given as

\begin{align} 
\begin{autobreak} 
\f1 =  
\bigg[
  0
\bigg],   
\end{autobreak} 
\\ 
\begin{autobreak} 
\f2 =
\bigg[ \CF ~ \NF  ~  \bigg( 
- \frac{112}{27}
+ \frac{4}{3} ~ \z2 \bigg)      
+ \CF ~ \CA  ~  \bigg( \frac{808}{27}
- 28 ~ \z3
- \frac{22}{3} ~ \z2 \bigg)
\bigg],   
\end{autobreak} 
\\ 
\begin{autobreak} 
\f3 =
\bigg[ \CF ~ \NF^2  ~  \bigg( 
- \frac{2080}{729}
+ \frac{112}{27} ~ \z3
- \frac{40}{27} ~ \z2 \bigg)      
+ \CF ~ \CA ~ \NF  ~  \bigg( 
- \frac{11842}{729}
+ \frac{728}{27} ~ \z3
+ \frac{2828}{81} ~ \z2
- \frac{96}{5} ~ \z2^2 \bigg)      
+ \CF ~ \CA^2  ~  \bigg( \frac{136781}{729}
+ 192 ~ \z5
- \frac{1316}{3} ~ \z3
- \frac{12650}{81} ~ \z2
+ \frac{176}{3} ~ \z2 ~  \z3
+ \frac{352}{5} ~ \z2^2 \bigg)      
+ \CF^2 ~ \NF  ~  \bigg( 
- \frac{1711}{27}
+ \frac{304}{9} ~ \z3
+ 4 ~ \z2
+ \frac{32}{5} ~ \z2^2 \bigg)
\bigg], 
\end{autobreak} 
\end{align}
The finite $G$ coefficients coming from the explicit calculation of the  form factor are given as

\begin{align} 
\begin{autobreak} 
\GGO1 =
\bigg[ \CF  ~  \bigg( 
- 8
+ \z2 \bigg)
\bigg],   
\end{autobreak} 
\\ 
\begin{autobreak} 
\GGO2 =
\bigg[ \CF  ~  \bigg( 8
- \frac{7}{3} ~ \z3
- \frac{3}{4} ~ \z2 \bigg)
\bigg],   
\end{autobreak} 
\\ 
\begin{autobreak} 
\GGO3 =
\bigg[ \CF  ~  \bigg( 
- 8
+ \frac{7}{4} ~ \z3
+ \z2
+ \frac{47}{80} ~ \z2^2 \bigg)
\bigg],   
\end{autobreak} 
\\ 
\begin{autobreak} 
\GGTW1 =
\bigg[ \CF ~ \NF  ~  \bigg( \frac{5813}{162}
- \frac{8}{3} ~ \z3
+ \frac{37}{9} ~ \z2 \bigg)      
+ \CF ~ \CA  ~  \bigg( 
- \frac{70165}{324}
+ \frac{260}{3} ~ \z3
- \frac{575}{18} ~ \z2
+ \frac{88}{5} ~ \z2^2 \bigg)      
+ \CF^2  ~  \bigg( 
- \frac{1}{4}
- 60 ~ \z3
+ 58 ~ \z2
- \frac{88}{5} ~ \z2^2 \bigg)
\bigg],   
\end{autobreak} 
\\ 
\begin{autobreak} 
\GGTW2 =
\bigg[ \CF ~ \NF  ~  \bigg( 
- \frac{129389}{1944}
+ \frac{301}{27} ~ \z3
- \frac{425}{54} ~ \z2
+ \frac{7}{12} ~ \z2^2 \bigg)      
+ \CF ~ \CA  ~  \bigg( \frac{1547797}{3888}
- 51 ~ \z5
- \frac{12479}{54} ~ \z3
+ \frac{7297}{108} ~ \z2
+ \frac{89}{3} ~ \z2 ~  \z3
- \frac{653}{24} ~ \z2^2 \bigg)      
+ \CF^2  ~  \bigg( 
- \frac{109}{16}
+ 12 ~ \z5
+ 184 ~ \z3
- \frac{437}{4} ~ \z2
- 28 ~ \z2 ~ \z3
+ \frac{108}{5} ~  \z2^2 \bigg)
\bigg],   
\end{autobreak} 
\\ 
\begin{autobreak} 
\GGTH1 =
\bigg[ \CF ~ \Nfour ~ \nfv  ~  \bigg( 12
- 80 ~ \z5
+ 14 ~ \z3
+ 30 ~ \z2
- \frac{6}{5} ~ \z2^2 \bigg)      
+ \CF ~ \NF^2  ~  \bigg( 
- \frac{258445}{2187}
+ \frac{536}{81} ~ \z3
- \frac{3466}{81} ~ \z2
- \frac{40}{9} ~ \z2^2 \bigg)      
+ \CF ~ \CA ~ \NF  ~  \bigg( \frac{3702974}{2187}
- 72 ~ \z5
- \frac{68660}{81} ~ \z3
+ \frac{155008}{243} ~ \z2
+ \frac{392}{9} ~ \z2 ~ \z3
- \frac{1298}{45} ~ \z2^2 \bigg)      
+ \CF ~ \CA^2  ~  \bigg( 
- \frac{48902713}{8748}
+ \frac{688}{3} ~ \z5
+ \frac{85883}{18} ~ \z3
- \frac{1136}{3} ~ \z3^2
- \frac{1083305}{486} ~ \z2
+ \frac{1786}{9} ~ \z2 ~ \z3
+ \frac{37271}{90} ~ \z2^2
- \frac{6152}{63} ~ \z2^3 \bigg)      
+ \CF^2 ~ \NF  ~  \bigg( \frac{73271}{162}
- \frac{368}{3} ~ \z5
+ \frac{19700}{27} ~ \z3
- \frac{7541}{18} ~ \z2
- \frac{152}{3} ~  \z2 ~ \z3
- \frac{704}{45} ~ \z2^2 \bigg)      
+ \CF^2 ~ \CA  ~  \bigg( \frac{230}{3}
- \frac{3020}{3} ~ \z5
- \frac{23402}{9} ~ \z3
+ 296 ~ \z3^2
+ \frac{55499}{18} ~ \z2
- \frac{3448}{3} ~ \z2 ~ \z3
+ \frac{2432}{45} ~ \z2^2
- \frac{15448}{105} ~ \z2^3 \bigg)      
+ \CF^3  ~  \bigg( 
- \frac{1527}{4}
+ 1992 ~ \z5
- 2130 ~ \z3
+ 48 ~ \z3^2
- 206 ~ \z2
+ 840 ~ \z2 ~ \z3
- 534 ~ \z2^2
+ \frac{21584}{105} ~ \z2^3 \bigg)
\bigg], 
\end{autobreak} 
\end{align}
Here $N_4 = (n_c^2-4)/n_c$ and $n_{fv}$ is proportional to the charge weighted sum of the quark flavors \cite{Gehrmann:2010ue}.
The finite $\tilde{G}$ coefficients are found to be

\begin{align} 
\begin{autobreak} 
\GtGO1 =
\bigg[ \CF  ~  \bigg( 
- 3 ~ \z2 \bigg)
\bigg],   
\end{autobreak} 
\\ 
\begin{autobreak} 
\GtGO2 =
\bigg[ \CF  ~  \bigg( \frac{7}{3} ~ \z3 \bigg)
\bigg],   
\end{autobreak} 
\\ 
\begin{autobreak} 
\GtGO3 =
\bigg[ \CF  ~  \bigg( 
- \frac{3}{16} ~ \z2^2 \bigg)
\bigg],   
\end{autobreak} 
\\ 
\begin{autobreak} 
\GtGTW1 =
\bigg[ \CF ~ \NF  ~  \bigg( 
- \frac{328}{81}
+ \frac{32}{3} ~ \z3
+ \frac{70}{9} ~ \z2 \bigg)      
+ \CF ~ \CA  ~  \bigg( \frac{2428}{81}
- \frac{176}{3} ~ \z3
- \frac{469}{9} ~ \z2
+ 4 ~ \z2^2 \bigg)
\bigg],   
\end{autobreak} 
\\ 
\begin{autobreak} 
\GtGTW2 =
\bigg[ \CF ~ \NF  ~  \bigg( \frac{976}{243}
- \frac{310}{27} ~ \z3
- \frac{196}{27} ~ \z2
- \frac{1}{20} ~ \z2^2 \bigg)      
+ \CF ~ \CA  ~  \bigg( 
- \frac{7288}{243}
+ 43 ~ \z5
+ \frac{2077}{27} ~ \z3
+ \frac{1414}{27} ~ \z2
- \frac{203}{3} ~ \z2 ~ \z3
+ \frac{11}{40} ~ \z2^2 \bigg)
\bigg],   
\end{autobreak} 
\\ 
\begin{autobreak} 
\GtGTH1 =
\bigg[ \CF ~ \NF^2  ~  \bigg( \frac{11584}{2187}
- \frac{2720}{81} ~ \z3
- \frac{1996}{81} ~ \z2
+ \frac{32}{9} ~ \z2^2 \bigg)      
+ \CF ~ \CA ~ \NF  ~  \bigg( 
- \frac{716509}{4374}
+ \frac{148}{3} ~ \z5
+ \frac{45956}{81} ~ \z3
+ \frac{105059}{243} ~ \z2
- \frac{1208}{9} ~ \z2 ~ \z3
- \frac{532}{9} ~ \z2^2 \bigg)      
+ \CF ~ \CA^2  ~  \bigg( \frac{7135981}{8748}
- \frac{1430}{3} ~ \z5
- \frac{59648}{27} ~ \z3
+ \frac{536}{3} ~ \z3^2
- \frac{765127}{486} ~ \z2
+ \frac{11000}{9} ~ \z2 ~ \z3
+ \frac{1964}{9} ~ \z2^2
+ \frac{152}{63} ~ \z2^3 \bigg)      
+ \CF^2 ~ \NF  ~  \bigg( 
- \frac{42727}{324}
+ \frac{112}{3} ~ \z5
+ \frac{2536}{27} ~ \z3
+ \frac{605}{6} ~ \z2
- 88 ~ \z2 ~  \z3
+ \frac{152}{15} ~ \z2^2 \bigg)
\bigg]\,.
\end{autobreak} 
\end{align}
\bibliographystyle{JHEP}
\bibliography{references}
\end{document}